\documentclass[prd,preprint,superscriptaddress,
  tightenlines,nofootinbib, eqsecnum
  ]{revtex4}

\usepackage{amsmath}
\usepackage{amsfonts}
\usepackage{amssymb}
\usepackage{bm}
\usepackage{hyperref}
\usepackage{mathrsfs}
\usepackage{graphicx}

\usepackage{empheq}

\usepackage{ulem}
\usepackage[usenames]{color}

\definecolor{darkgreen}{rgb}{0,0.5,0}

\hypersetup{
    bookmarks=true,         
    unicode=false,          
    pdftoolbar=true,        
    pdfmenubar=true,        
    pdffitwindow=false,     
    pdfstartview={FitH},    
    pdftitle={My title},    
    pdfauthor={Author},     
    pdfsubject={Subject},   
    pdfcreator={Creator},   
    pdfproducer={Producer}, 
    pdfkeywords={keyword1} {key2} {key3}, 
    pdfnewwindow=true,      
    colorlinks=true,       
    linkcolor=red,          
    citecolor=cyan,        
    filecolor=magenta,      
    urlcolor=darkgreen,           
    linktocpage=true
}

\allowdisplaybreaks

\DeclareSymbolFontAlphabet{\mathrsfs}{rsfs}
\DeclareMathAlphabet{\mathcal}{OMS}{cmsy}{m}{n}

\newcommand{\beq}{\begin{equation}}
\newcommand{\eeq}{\end{equation}} 

\newcommand{\dd}{\mathrm{d}}
 
\newcommand{\de}{\mathrm{e}} 

\begin{document}

\title{The mass quadrupole moment of compact binary systems \\at the fourth post-Newtonian order}

\author{Tanguy Marchand}\email{Tanguy.Marchand@ens.fr}
\affiliation{$\mathcal{G}\mathbb{R}\varepsilon{\mathbb{C}}\mathcal{O}$,
  Institut d'Astrophysique de Paris,\\ UMR 7095, CNRS, Sorbonne
  Universit{\'e}s \& UPMC Univ Paris 6,\\ 98\textsuperscript{bis}
  boulevard Arago, 75014 Paris, France}
\affiliation{Laboratoire APC -- Astroparticule et Cosmologie, \\
Universit{\'e} Paris Diderot Paris 7, 75013 Paris, France}

\author{Quentin Henry}\email{henry@iap.fr}
\affiliation{$\mathcal{G}\mathbb{R}\varepsilon{\mathbb{C}}\mathcal{O}$,
  Institut d'Astrophysique de Paris,\\ UMR 7095, CNRS, Sorbonne
  Universit{\'e}s \& UPMC Univ Paris 6,\\ 98\textsuperscript{bis}
  boulevard Arago, 75014 Paris, France}

\author{Fran\c{c}ois Larrouturou}\email{francois.larrouturou@iap.fr}
\affiliation{$\mathcal{G}\mathbb{R}\varepsilon{\mathbb{C}}\mathcal{O}$,
  Institut d'Astrophysique de Paris,\\ UMR 7095, CNRS, Sorbonne
  Universit{\'e}s \& UPMC Univ Paris 6,\\ 98\textsuperscript{bis}
  boulevard Arago, 75014 Paris, France}

\author{Sylvain Marsat}\email{marsat@apc.in2p3.fr}
\affiliation{Laboratoire APC -- Astroparticule et Cosmologie, \\
Universit{\'e} Paris Diderot Paris 7, 75013 Paris, France}

\author{Guillaume Faye}\email{guillaume.faye@iap.fr}
\affiliation{$\mathcal{G}\mathbb{R}\varepsilon{\mathbb{C}}\mathcal{O}$,
  Institut d'Astrophysique de Paris,\\ UMR 7095, CNRS, Sorbonne
  Universit{\'e}s \& UPMC Univ Paris 6,\\ 98\textsuperscript{bis}
  boulevard Arago, 75014 Paris, France}

\author{Luc Blanchet}\email{luc.blanchet@iap.fr}
\affiliation{$\mathcal{G}\mathbb{R}\varepsilon{\mathbb{C}}\mathcal{O}$,
  Institut d'Astrophysique de Paris,\\ UMR 7095, CNRS, Sorbonne
  Universit{\'e}s \& UPMC Univ Paris 6,\\ 98\textsuperscript{bis}
  boulevard Arago, 75014 Paris, France}

\date{\today}

\begin{abstract}
The mass-type quadrupole moment of inspiralling compact binaries (without spins) is computed at the fourth post-Newtonian (4PN) approximation of general relativity. The multipole moments are defined by matching between the field in the exterior zone of the matter system and the PN field in the near zone, following the multipolar-post-Minkowskian (MPM)-PN formalism. The matching implies a specific regularization for handling infra-red (IR) divergences of the multipole moments at infinity, based on the Hadamard finite part procedure. On the other hand, the calculation entails ultra-violet (UV) divergences due to the modelling of compact objects by delta-functions, that are treated with dimensional regularization (DR). In future work we intend to systematically study the IR divergences by means of dimensional regularization as well. Our result constitutes an important step in the goal of obtaining the gravitational wave templates of inspiralling compact binary systems with 4PN/4.5PN accuracy. 
\end{abstract}


\maketitle

\section{Introduction} 
\label{sec:intro}

The post-Newtonian (PN) approximation has played a crucial role in the data analysis of the recent discovery by the LIGO-Virgo detectors of gravitational waves (GW) generated by the coalescence of two neutron stars~\cite{GW170817}. It also constitutes an important input for validating the early inspiral phase of the coalescence of two black holes~\cite{GW150914, LIGOrun1}. In this paper, motivated by the need of improving the accuracy of template waveforms generated by inspiralling compact binaries (neutron stars or black holes), notably in view of the future LISA mission, we tackle the computation of the mass-type quadrupole moment of compact binary systems (without spins) at the high 4PN approximation level.\footnote{Here 4PN refers to the terms of order $1/c^8$ in the waveform and energy flux, \textit{i.e.} beyond the Einstein quadrupole formula. Seen as a small dissipative radiation reaction effect, this corresponds to the order 6.5PN or $1/c^{13}$ in the equations of motion, beyond the Newtonian acceleration.} This calculation is part of our current program to obtain the orbital phase evolution and waveform of inspiralling binaries up to the 4.5PN approximation. 

So far the main results achieved in this program are:\footnote{See Ref.~\cite{BlanchetLR} for a review of previous (pre-4PN) works.}
\begin{enumerate}
\item The non-linear effects in the GW propagation (so-called tails-of-tails-of-tails) that are responsible for the 4.5PN coefficient in the energy flux for circular orbits~\cite{MBF16}. This result has been confirmed by an independent PN reexpansion of resummed waveforms~\cite{MNagar17};
\item The conservative part of the equations of motion up to the 4PN order, obtained from the Fokker Lagrangian in harmonic coordinates~\cite{BBBFMa, BBBFMb, BBBFMc, MBBF17, BBFM17}. This result has also been obtained by means of the canonical Hamiltonian formalism of general relativity~\cite{JaraS13, JaraS15, DJS14, DJS15eob, DJS16}, and by the effective field theory (EFT)~\cite{GR06,FS4PN, FStail, GLPR16,Po16, FS19, FPRS19, Blumlein20}.
\end{enumerate}

The next steps are the computation of the multipole moments of the compact binary source to consistent order. To control the GW energy flux at the 4PN order, we essentially need the mass-type quadrupole moment at the 4PN order (it is currently known at the 3.5PN order~\cite{BIJ02, BI04mult, BFIS08, FMBI12}) and the current-type quadrupole moment at the 3PN order; indeed the other moments are already known, see for instance~\cite{BFIS08}. In particular, the mass-type octupole moment at the 3PN order has been computed in~\cite{FBI15}. To control the waveform and polarisation modes at the 4PN order we need more precision, for instance the current quadrupole and mass octupole moments would be needed up to 3.5PN order.

The multipole moments are defined within the so-called MPM-PN formalism, which consists in computing the external field of the source by means of a multipolar-post-Minkowskian (MPM) expansion~\cite{BD86, B87, BD92, BDI95}, which is matched to the inner field obtained by direct PN iteration~\cite{B95, B98mult, PB02, BFN05}. Two possible alternative approaches also able to compute multipole moments at high PN orders, are the direct integration of the relaxed field equations (DIRE)~\cite{WW96} and the EFT~\cite{LMRY19}, both being currently developed at the 2PN order. See also Ref.~\cite{compere18} for a discussion on various alternative definitions of multipole moments for radiating space-times.

In the present paper we are devoted to the (long and demanding) computation of the mass quadrupole moment of compact binary systems at the 4PN order. As in the problem of the equations of motion, a very important aspect of the calculation is the proper use of regularizations. At the 3PN order, it was found that ultra-violet (UV) divergences occur, due to the  model of point particles, and have to be dealt with dimensional regularization (DR)~\cite{DJSdim, BDEI04}. At the 4PN order in the equations of motion, not only are there UV divergences, but also infra-red (IR) ones, linked to the presence of tails at 4PN order, and those must also be cured by means of DR~\cite{BBBFMc}. In particular it was shown that DR completely resolves the problem of ambiguity parameters in the 4PN equations of motion, including the ones associated with IR divergences~\cite{MBBF17}. Such feature of DR was also pointed out in the EFT approach~\cite{PR17}, and used to obtain equivalent ambiguity-free results~\cite{FS19, FPRS19, Blumlein20}.

In the present computation we shall use DR for the UV divergences but leave open the problem of the IR divergences. Indeed this problem requires a separate analysis and is independent from the long calculations we perform in this paper. Therefore we shall continue to use, as in all our previous works~\cite{BlanchetLR}, the Hadamard partie finie procedure consisting of using the natural regulator $r^B=\vert\mathbf{x}\vert^B$ in the multipole moments and selecting the finite part in the expansion when $B\to 0$ (thus throwing away the poles $1/B^n$). This specific procedure comes as a direct consequence of the matching between the MPM exterior field and the near zone PN metric in 3 dimensions~\cite{B95, B98mult, PB02, BFN05}. 

However the recent work on the equations of motion showed that a combination of the Hadamard procedure together with DR is required in the presence of IR divergences. This is the so-called ``$\eta\varepsilon$'' regularization scheme introduced in Ref.~\cite{MBBF17}, which we shall better rename here the ``$B\varepsilon$'' regularization since the extra parameter $\eta$ (in addition to $\varepsilon=d-3$) is nothing but $B$ in the Hadamard partie finie process. In the $B\varepsilon$ regularization the limit $B\to 0$ is considered first and shown to be finite (\textit{i.e.}, no poles $\propto 1/B$) for generic non-integer values of $\varepsilon$~\cite{MBBF17}. Then the limit $\varepsilon\to 0$ is applied and this reduces to the standard DR; in particular, the poles $\propto 1/\varepsilon$ are renormalized by appropriate shifts of trajectories. Again we postpone to future work the task of understanding whether the Hadamard partie finie IR regularization should be replaced by the ``$B\varepsilon$'' regularization.

At the 4PN order for non-spinning black-hole binaries (and at 2.5PN order for spinning black holes) the GW flux and phasing will require inclusion of absorption effects by the black-hole horizons \cite{PS95,TMT97,Alvi01,Po08,Chatz12}. The MPM-PN formalism developed for point-particle binaries is not suitable to compute these effects. As usual, they have to be added to the results computed here.

The plan of this paper is as follows. In Sec.~\ref{sec:mult} we recall from previous works the definitions of the general mass multipole moments (of order $\ell$) in $d$ dimensions. In Sec.~\ref{sec:MQPot} we express the 4PN quadrupole moment in terms of elementary potentials, and find that all the terms can be computed, thanks in particular to the techniques of super-potentials and of surface-integrals. The problem of the dimensional regularization of UV divergences (including distributional derivatives of singular functions) is dealt with in Sec.~\ref{sec:dimregUVgen}, and the computation of the various types of potentials and terms is done in Sec.~\ref{sec:ComputePot}. Crucial to this calculation is the application of certain UV shifts of the trajectories determined in previous works on the 4PN equations of motion. Finally we present in Sec.~\ref{sec:resultMQ} the 4PN mass quadrupole moment in the case of circular orbit. The complete expression of the 4PN metric in $d$ dimensions is relagated in Appendix~\ref{app:PNpotentials}; the shifts are given in Appendix~\ref{app:shift}; finally we give in Appendix~\ref{app:MQAsPot} the exhaustive list of all the terms composing the 4PN quadrupole moment in terms of elementary potentials.

\section{The mass-type multipole moments}
\label{sec:mult}

We provide first the expression of the $\ell$-th order mass-type multipole moments of a general isolated source in 3 dimensions \cite{B98mult}:\footnote{The notation is: $L = i_1\cdots i_\ell$ for a multi-index composed of $\ell$ multipolar indices $i_1, \cdots, i_\ell$; $iL = i i_1\cdots i_\ell$ for a multi-index with $\ell+1$ indices; $x_L = x_{i_1}\cdots x_{i_\ell}$ for the product of $\ell$ spatial vectors $x^i = x_i$. The symmetric-trace-free (STF) projection is denoted by $\hat{x}_L = \mathrm{STF}(x_{i_1}\cdots x_{i_\ell})$, or sometimes using brackets surrounding the indices, for instance $x_{\langle L}v_{P\rangle}$. Similarly, $\partial_L = \partial_{i_1}\cdots \partial_{i_\ell}$ for the product of $\ell$ partial derivatives $\partial_i=\partial/\partial x^i$, and $\hat{\partial}_L = \mathrm{STF}( \partial_{i_1}\cdots \partial_{i_\ell})$. In the case of summed-up (dummy) multi-indices $L$, we do not write the $\ell$ summations from 1 to 3 over their indices. The superscript $(n)$ denotes $n$ time derivatives, and an overbar indicates a PN-expanded quantity.}
\begin{align}
\label{ValueILGeneral}
I_L(t)&= \mathop{\mathrm{FP}}_{B=0} \int \dd^3\mathbf{x} \left(\frac{r}{r_0}\right)^B \int^1_{-1} \dd z\biggl\{ \delta_\ell(z)\,\hat{x}_L\,\overline{\Sigma} -\frac{4(2\ell+1)}{c^2(\ell+1)(2\ell+3)} \,\delta_{\ell+1}(z) \,\hat{x}_{iL} \,\overline{\Sigma}^{(1)}_i \nonumber\\
 & \qquad \qquad \qquad \qquad + \frac{2(2\ell+1)}{c^4(\ell+1)(\ell+2)(2\ell+5)}\,\delta_{\ell+2}(z) \,\hat{x}_{ijL} \,\overline{\Sigma}^{(2)}_{ij} \biggr\} (\mathbf{x}, t+z r/c) \,.
\end{align}
The source terms are defined from the PN expansion of the components of the pseudo stress-energy tensor of the matter system in harmonic coordinates, \textit{i.e.} $\overline{\tau}^{\mu\nu}$ where the overbar means the PN expansion, as (with $\overline{\tau}^{ii}=\delta_{ij}\overline{\tau}^{ij}$)
\begin{equation}\label{Sigma}
\overline{\Sigma} = \frac{\overline{\tau}^{00}+\overline\tau^{ii}}{c^2}\,,\qquad\overline{\Sigma}_i = \frac{\overline{\tau}^{0i}}{c}\,, \qquad\overline{\Sigma}_{ij} = \overline{\tau}^{ij}\,.
\end{equation}
The pseudo stress-energy tensor is defined from the gauge-fixed Einstein field equations as
\begin{equation}\label{EFE}
\Box h^{\mu\nu}=\frac{16\pi G}{c^4}\tau^{\mu\nu}\,,
\end{equation}
where $\Box\equiv\Box_\eta$ is the flat d'Alembertian operator and the field variable is defined by $h^{\mu\nu} = \sqrt{-g}\,g^{\mu\nu}-\eta^{\mu\nu}$, with $g=\text{det}(g_{\rho\sigma})$ and $\eta^{\mu\nu}=\text{diag}(-1,1,1,1)$. It obeys the usual harmonic coordinates condition $\partial_\nu h^{\mu\nu}=0$. The pseudo tensor is composed of a matter part and a gravitational part that we denote as
\begin{equation}\label{tau}
\tau^{\mu\nu} = \vert g\vert T^{\mu\nu}+\frac{c^4}{16\pi
G}\,\Lambda^{\mu\nu}\,,
\end{equation}
where $T^{\mu\nu}$ is the matter stress-energy tensor, and $\Lambda^{\mu\nu}$ represents the gravitational source term which is a complicated non-linear, at least quadratic, functional of $h^{\rho\sigma}$ and its first and second space-time derivatives.

An important feature of Eq.~\eqref{ValueILGeneral} is the presence of the Hadamard finite part (FP) when $B\rightarrow 0$, with regularization factor $(r/r_0)^B$ where $r=\vert \mathbf{x} \vert$ and $r_0$ is an arbitrary length scale. The role of this finite part operation is to deal with the IR divergences initially introduced in the multipole moments by the fact that the PN-expanded integrand of the multipole moments is valid in the near zone, and typically diverges at spatial infinity (when $r\rightarrow +\infty$). This specific regularization of the multipole moments is actually imposed by the matching between the inner PN field and the outer MPM field, given the particular way the MPM metric is generated at each post-Minkowskian order~\cite{B98mult}. However, as mentioned in Sec.~\ref{sec:intro}, more work is required to investigate whether the IR divergences should be treated instead by a variant of DR (\textit{i.e.}, the $B\varepsilon$ regularization of Ref.~\cite{MBBF17}).

Concerning the UV divergences, they may be cured using the Hadamard partie finie regularization up to the 2PN order. As is well known, this technique fails at the 3PN and 4PN orders where DR has to be systematically used. In our practical calculation, a first evaluation using the Hadamard partie finie for the UV divergences is done, and then we systematically add the appropriate correction accounting for the DR. 

The functions $\overline{\Sigma}$, $\overline{\Sigma}_i$ and $\overline{\Sigma}_{ij}$ in the integrand of Eq.~\eqref{ValueILGeneral} are evaluated at the spatial point $\mathbf{x}$ and at time $t+z r/c$ where $r=\vert\mathbf{x}\vert$. In addition there is the extra integration variable $z$ entering the auxiliary function 
\begin{equation}\label{delta}
\delta_\ell(z) = \frac{(2\ell+1)!!}{2^{\ell+1}\ell!}(1-z^2)^\ell\,, \qquad
\int_{-1}^{1} \dd z\,\delta_\ell(z) = 1\,.
\end{equation}
But in fact, as we are going to compute the PN expansion of $I_L$, the integral over $z$ is given as an explicit asymptotic PN series using the property of the functions $\delta_\ell(z)$ that
\begin{subequations}
\label{integralOverZ}
\begin{align}
\int_{-1}^1\dd z \,\delta_\ell(z) \,\overline{\Sigma} (\mathbf{x}, t+ z r/c) &= \sum_{k=0}^{+\infty} \alpha_{k,\ell}\left(\frac{r}{c}\frac{\partial}{\partial t}\right)^{2k} \overline{\Sigma}(\mathbf{x}, t)\,,\\
\text{with}\quad\alpha_{k,\ell} &\equiv \frac{(2\ell+1)!!}{(2k)!! (2\ell + 2k +1)!!}\,.
\end{align}
\end{subequations}
Therefore, $I_L(t)$ is just a sum of time derivatives of 3-dimensional integrals of PN quantities depending on the pseudo stress-energy tensor $\overline{\tau}^{\mu\nu}(\mathbf{x}, t)$.

In the case of a compact binary source we use the usual point-particle stress-energy tensor for the matter. We write the components of the matter tensor for two particles as
\begin{equation}\label{Tmunu}
T^{\mu\nu} = \mu_1 \,v_1^\mu v_1^\nu\,\delta(\mathbf{x} - \bm{y}_1) + 1 \leftrightarrow 2 \,, 
\end{equation}
where $v_1^\mu=\dd y_1^\mu/\dd t$ is the coordinate velocity of the particle, $v_1^\mu = (c, v_1^i)$, and $\delta$ is the 3-dimensional Dirac function. We have (with $m_1$ the constant PN mass)
\begin{equation}\label{mu}
\mu_1(t) = \frac{1}{\sqrt{-(g)_1}}\frac{m_1}{\sqrt{-(g_{\mu\nu})_1 \frac{v_1^\mu v_1^\nu}{c^2}}} \,,
\end{equation}
where the index $1$ indicates that the metric has to be evaluated at the location $\mathbf{x} = \bm{y}_1$; in the self-gravitating case, self-field divergences
are removed by means of DR.

As we shall use DR for the UV divergences, we require the generalization of the expression of the multipole moments to $d$ dimensions. The formula for the general $\ell$-th mass-type multipole moment reads~\cite{BDEI04}
\begin{eqnarray}\label{ILexpr2}
I_L(t)&=&\frac{d-1}{2(d-2)}\mathop{\mathrm{FP}}_{B=0}\int \dd^d \mathbf{x}\left(\frac{r}{r_0}\right)^B \biggr\{\hat{x}_L\,\overline{ \Sigma}_{[\ell]} 
\nonumber\\
&&\qquad\qquad -\frac{4(d+2\ell-2)} {c^2(d+\ell-2)(d+2\ell)}\,\hat{x}_{iL}\, \overline{\Sigma}^{(1)}_{i[\ell+1]} \nonumber\\
&&\qquad\qquad +\frac{2(d+2\ell-2)} {c^4(d+\ell-1)(d+\ell-2)(d+2\ell+2)} \,\hat{x}_{ijL}\, \overline{\Sigma}^{(2)}_{ij[\ell+2]} \nonumber\\
&&\qquad\qquad - \frac{4(d-3)(d+2\ell-2)}{c^2(d-1)(d+\ell-2)(d+2\ell)} B \,\hat{x}_{iL}\,\frac{x_j}{r^2} \,\overline{\Sigma}_{ij[\ell+1]} \biggr\}(\mathbf{x},t)\,.
\end{eqnarray}
The overall $d$-dependent factor in front is such that~\eqref{ILexpr2} reduces to the usual Newtonian-looking expression of the multipole moments in the Newtonian approximation, given by $I_L = m_1 \,\hat{y}_1^{L} + 1 \leftrightarrow 2 + \mathcal{O}(c^{-2})$. The last term of~\eqref{ILexpr2} will in fact not contribute because of the $B$ and the $d-3$ factors appearing simultaneously. To see this one splits the integral into a near-zone contribution $r<\mathcal{R}$ and a far-zone one $r>\mathcal{R}$. In the near zone integral one has to apply the limit $B\to 0$ since there are no IR divergences (hence no poles $\propto 1/B$), while the far zone one which is UV-finite is to be computed with $d=3$, so both integrals are separately zero. In future work we shall investigate the fate of the last term in Eq.~\eqref{ILexpr2} within the $B\varepsilon$ regularization.

As before the source terms are defined from the PN expansion of the pseudo stress-energy tensor in $d$ dimensions, $\overline{\tau}^{\mu\nu}$, defined by the Einstein field equations in harmonic coordinates, which take the same form as in Eqs.~\eqref{EFE}--\eqref{tau}, except that the Newton constant there reads $G=\ell_0^{d-3}G_\text{N}$, where $\ell_0$ is the characteristic length scale associated with DR and $G_\text{N}$ is the Newton constant in 3 dimensions. We have
\begin{equation}\label{Sigmad}
\overline{\Sigma} = \frac{2}{d-1}\frac{(d-2)\overline{\tau}^{00}+\overline{\tau}^{ii}}{c^2}\,,
\end{equation}
while $\overline{\Sigma}_i$ and $\overline{\Sigma}_{ij}$ take the same form as in~\eqref{Sigma}. Of course the matter terms are still given by the point mass expressions~\eqref{Tmunu}--\eqref{mu} but with now $\delta=\delta^{(d)}$, the Dirac function in $d$ dimensions. The generalization of Eqs.~\eqref{delta}--\eqref{integralOverZ} to $d$ dimensions reads as
\begin{equation}\label{Sellcompact}
\overline{\Sigma}_\ell(\mathbf{x},t)=\int_{-1}^1 \dd z
\,\delta_\ell^{(\varepsilon)} (z)
\,\overline{\Sigma}(\mathbf{x},t+zr/c)\,,
\end{equation}
where we have posed $\varepsilon = d-3$, and
\begin{equation}\label{deltal}
\delta_\ell^{(\varepsilon)} (z) \equiv
\frac{\Gamma\left(\ell+\frac{3}{2}+\frac{\varepsilon}{2}\right)}{
\Gamma\left(\frac{1}{2}\right)\Gamma
\left(\ell+1+\frac{\varepsilon}{2}\right)}
\,(1-z^2)^{\ell+\frac{\varepsilon}{2}},
\qquad\int_{-1}^{1}
\dd z\,\delta_\ell^{(\varepsilon)}(z) = 1\,.
\end{equation}
In practice we only need the formal PN expansion
\begin{equation}\label{series}
\overline{\Sigma}_{[\ell]}(\mathbf{x},t)
= \sum_{k=0}^{+\infty}\alpha_\ell^k
\left(\frac{r}{c}\frac{\partial}{\partial
t}\right)^{2k}\overline\Sigma(\mathbf{x},t)\,,
\end{equation}
with the numerical coefficients now being given by
\begin{equation}\label{coeffs}
\alpha_\ell^k = \frac{1}{2^{2k} k!}\frac{\Gamma(\ell+\frac{d}{2})}{\Gamma(\ell+\frac{d}{2}+k)}\,.
\end{equation}
It is very useful to define for the matter stress-energy tensor the following matter currents 
\begin{equation}\label{sigma}
\sigma = \frac{2}{d-1}\frac{(d-2)T^{00}+T^{ii}}{c^2}\,,\qquad\sigma_i = \frac{T^{0i}}{c}\,, \qquad \sigma_{ij} = T^{ij}\,,
\end{equation}
that are given in the case of compact binary systems by
\begin{subequations}\label{mattercurrent}
\begin{align}
\sigma &= \tilde{\mu}_1 \,\delta (\mathbf{x} - \bm{y}_1) + 1 \leftrightarrow 2\,, \\
\sigma_i &= \mu_1 v_1^i \,\delta (\mathbf{x} -  \bm{y}_1) + 1 \leftrightarrow 2\,, \\
\sigma_{ij} &=\mu_1 v_1^i v_1^j \,\delta (\mathbf{x} -  \bm{y}_1) + 1 \leftrightarrow 2\,,
\end{align}
\end{subequations}
where $\bm{y}_1=(y_1^i)$ is the particle's position, $\bm{v}_1=\dd\bm{y}_1/\dd t=(v_1^i)$ the coordinate velocity, and we have introduced, besides $\mu_1$ which keeps the same expression as in 3 dimensions, see Eq.~\eqref{mu}, the useful tilded version 
\begin{equation}\label{mutilde}
\tilde{\mu}_1 = \frac{2}{d-1}\left(d-2 + \frac{v_1^2}{c^2}\right) \mu_1\,.
\end{equation}
The matter current densities~\eqref{mattercurrent} generate all the compact-support terms in the expression of the quadrupole moment. 

\section{The quadrupole moment as a function of potentials}
\label{sec:MQPot}

\subsection{The elementary potentials}
\label{sec:pot}

To start our derivation of the quadrupole moment we need to inject into~\eqref{ILexpr2} the PN metric $\overline{h}^{\mu\nu}$ which is an explicit solution of the Einstein field equations~\eqref{EFE} valid in the near zone (we recall that the
overbar means PN expansion). In our recent computation of the equations of motion by means of the Fokker Lagrangian, the order to which the PN metric had to be expanded was given by the so-called ``$n+2$'' method~\cite{BBBFMa}. In the case of the mass quadrupole moment, such a method does not exist, and $\overline{h}^{\mu\nu}$ has to be expanded to higher PN order. We find that the metric components $\overline{h}^{00}$, $\overline{h}^{0i}$ and $\overline{h}^{ij}$ are respectively to be expanded up to orders $c^{-8}$, $c^{-7}$ and $c^{-8}$ included. Thus, we need $\overline{h}^{00}$ and $\overline{h}^{0i}$ at 3PN order (remind that $c^{-8}$ in $\overline{h}^{00}$ actually corresponds to 3PN), and $\overline{h}^{ij}$ at the 4PN order.

Building up on previous works~\cite{BIJ02, BI04mult, BDEI05dr, FBI15} we parametrize the metric with appropriate PN elementary retarded potentials, namely scalar potentials $V$, $K$, $\hat{X}$ and $\hat{T}$, vector potentials $V_i$, $\hat{R}_i$ and $\hat{Y}_i$, and tensor ones $\hat{W}_{ij}$, $\hat{Z}_{ij}$ and $\hat{M}_{ij}$. The structure of our parametrization in 3 dimensions is 
\begin{subequations}\label{metric4PN3}
\begin{align}
\overline{h}^{00} &= - \frac{4V}{c^{2}} - \frac{2}{c^{4}} \left( \hat{W} + 4 V^2\right) - \frac{8}{c^{6}} \left( \hat{X} + \cdots \right) - \frac{64}{c^{8}} \left( \hat{T} + \cdots \right) + \mathcal{O}(c^{-10})\,, \\
\overline{h}^{0i} &=  \frac{4V_{i}}{c^{3}} + \frac{8}{c^{5}} \left( \hat{R}_{i} + V_{i} V\right) + \frac{16}{c^7}\left( \hat{Y}_{i} + \cdots\right) + \mathcal{O}(c^{-9})\,,\\
\overline{h}^{ij} &= - \frac{4}{c^{4}} \Bigl(\hat{W}_{ij} - \frac{1}{2}\delta_{ij} \hat{W}\Bigr) - \frac{16}{c^{4}} \Bigl(\hat{Z}_{ij} - \frac{1}{2}\delta_{ij} \hat{Z}\Bigr) - \frac{32}{c^8}\Bigl( \hat{M}_{ij} + \cdots \Bigr) + \mathcal{O}(c^{-10})\,.
\end{align}
\end{subequations}
The ellipsis symbolizes non-linear products of these elementary potentials. The complete expression of the metric at 4PN order in $d$ dimensions is given in Eqs.~\eqref{metric4PN} of Appendix~\ref{app:PNpotentials}. The potentials obey some imbricated flat space-time wave equations. Some have a compact support like
\begin{equation}\label{V3}
\Box V = - 4 \pi G\, \sigma \,,\qquad
\Box V_{i} = - 4 \pi G\, \sigma_{i} \,,
\end{equation}
while there are many quadratic non-linear terms (sometimes called ``$\partial V\partial V$'') such as in
\begin{equation}\label{W3}
\Box\hat{W}_{ij} = -4\pi G\bigl(\sigma_{ij}
-\delta_{ij}\,\sigma_{kk} \bigr) - \partial_i V \partial_j V\,,
\end{equation}
and higher order terms (called ``non-compact'') such as the cubic term $\hat{W}_{ij}\,\partial_{ij}V$ in
\begin{equation}\label{X3}
\Box\hat{X} = - 4 \pi G V \sigma_{ii}+\hat{W}_{ij} \,\partial_{ij} V+2 V_i \partial_t \partial_i V+ V \partial_t^2 V + \frac{3}{2} (\partial_t V)^2 - 2 \partial_i V_j \partial_j V_i\,.
\end{equation}
See Eqs.~\eqref{defpotentials}--\eqref{defpotentials4PN} for thorough definitions of all these potentials in $d$ dimensions. Among these note that the only purely 4PN potential which is needed for the 4PN quadrupole moment is $\hat{M}_{ij}$ (new with the present paper), which obeys the equation in 3 dimensions:
\begin{align}\label{Mij3d}
\Box\hat{M}_{ij} ={}& G \pi \Bigl[\Bigl(-4 V_{i} V_{j}
 + \delta^{ij} (2 V_{a} V_{a}
 + \hat{X})\Bigr) \sigma
 + 4\bigl(\hat{R}_{(i}
 + V V_{(i}\bigr) \sigma_{j)}
 - 2 \hat{W}_{a(i} \sigma_{j) a} \\
&\quad
 - 4 V^2 \sigma_{ij}
 + \delta^{ij} \Bigl(-2 \hat{R}_{b} \sigma_{b}
 - 2 V V_{m} \sigma_{m}
 + \frac{1}{2} \hat{W}_{kl} \sigma_{kl}
 + V^2 \sigma_{p p} -  \frac{1}{2} \hat{W} \sigma_{q q}\Bigr)\Bigr]
 \nonumber\\
& -  \partial_t V_{i} \partial_t V_{j}
 + V_{a} \partial_t \partial_{a}\hat{W}_{ij}
 + 2 V_{(i} \partial_{a}V_{j)} \partial_{a}V
 -  \partial_t \hat{W}_{(i}{}_{a} \partial_{a}V_{j)}\nonumber\\
& + \frac{1}{2} \hat{W}_{ab} \partial_{ab} \hat{W}_{ij}
 -  \frac{1}{2} \partial_{a}\hat{W}_{i b} \partial_{b}\hat{W}_{ja}
 + \frac{1}{2} \hat{W}_{(i}{}_{a} \partial_{a}V \partial_{j)}V -  \frac{1}{4} \partial_{i}\hat{W}_{ab} \partial_{j}\hat{W}_{ab} \nonumber\\
& - 2\partial_{a}V_{(i} \bigl( \partial_{j)}\hat{R}_{a}
 + V_{a} \partial_{j)}V\bigr)
 -  2 \partial_{a}\hat{R}_{(i} \partial_{j)}V_{a}
 + \partial_t \hat{W}_{(i}{}_{a} \partial_{j)}V_{a}
 + \partial_{b}\hat{W}_{a(i} \partial_{j)}\hat{W}_{ab}\nonumber\\
& + 2\partial_{(i}V_{a} \partial_{j)}\hat{R}_{a}
 + \Bigl(- 2\partial_t \hat{R}_{(i}
 + \frac{1}{2} V_{(i} \partial_t V
 -  \partial_{(i}\hat{X}\Bigr) \partial_{j)}V
 + V \Bigl(\frac{1}{2} \partial_t^{2} \hat{W}_{ij}
 - 2 \partial_t V_{(i} \partial_{j)}V\Bigr)\nonumber\\
& + \delta_{ij} \Bigl[ - \frac{1}{2} V_{a} \partial_t \partial_{a}\hat{W}
 -  \frac{1}{4} V^2 \partial_t^2 V
 + \partial_t \hat{R}_{a} \partial_{a}V
 -  \frac{1}{4} \hat{W}_{ab} \partial_{ab}\hat{W} + \partial_{a}V_{b} \partial_{b}\hat{R}_{a}\nonumber\\
&\quad
 -  \frac{1}{8} \hat{W}_{ab} \partial_{a}V \partial_{b}V
 -  \frac{1}{2} \partial_t \hat{W}_{ab} \partial_{a}V_{b} -  \frac{1}{4} \partial_{a}\hat{W}_{bc} \partial_{c}\hat{W}_{ab} -  \frac{1}{4} V_{a} \partial_t V \partial_{a}V \nonumber\\
&\quad 
 + V \Bigl(\frac{3}{8} (\partial_t V)^2
 -  \frac{1}{2} V_{a} \partial_t \partial_{a}V
 -  \frac{1}{4} \partial_t^{2} \hat{W} + \partial_t V_{a} \partial_{a}V
 -  \frac{1}{4} \hat{W}_{ab} \partial_{ab}V
 + \frac{1}{2} \partial_{a}V_{b} \partial_{b}V_{a}\Bigr) \Bigr]\,.\nonumber
\end{align}

Inserting the PN metric into the mass quadrupole moment, we obtain the full expression in terms of the latter PN potentials. However we do not know explicitly all the potentials (either in $3$ or $d$ dimensions), since they are solutions of complicated wave equations such as~\eqref{Mij3d}. Thus, crucial simplifications of the result have to be performed first, in order to put the expression into computable form; see the complete result for all the terms in Appendix~\ref{app:MQAsPot}. 

\subsection{The method of super-potentials}
\label{sec:superpot}

The first technique we have used in order to be able to compute all the terms at the 4PN order is the method of ``super-potentials''. Many of the most difficult terms are of the form $\phi\,P$ where $\phi$ is a simple potential or derivative of a potential, and $P$ is a complicated potential whose expression in the whole space is not known. For instance $P$ could be the 4PN potential $\hat{M}_{ij}$ entering the spatial components of the metric and obeying the equation~\eqref{Mij3d}. On the other hand, in our case $\phi$ is one of the following potentials: $\partial_{ij} V$, $\partial_{t}\partial_i V$ or $\partial_j V_i$. 

To compute the integral $\int \dd^{3} \mathbf{x} \,r^B\,\hat{x}_L \,\phi \,P$ (in 3 or $d$ dimensions) we notice that $\hat{x}_L \,\phi$ may be recast in the form of a Laplace operator acting on some solution $\Psi^{\phi}_L$:
\begin{equation}\label{DeltaPsiL}
\Delta \Psi^{\phi}_L = \hat{x}_L \phi\,.
\end{equation}
Assuming that $\Psi^{\phi}_L$ can be constructed analytically, a mere integration by part yields a volume integral whose source is known explicitly, namely $-\int \dd^{3} \mathbf{x} \,r^B\, \Psi^{\phi}_L \,\Delta P$, plus terms that are essentially surface integrals at infinity when the Hadamard finite part is applied.

Now, as it turns out, it is possible to construct the solution $\Psi^{\phi}_L$ by defining the super-potentials of $\phi$ as the hierarchy of solutions $\phi_{2k}$ of the sequence of Poisson equations
\begin{equation}\label{Deltaphi2k}
\Delta\phi_{2k+2}=\phi_{2k}\,,
\end{equation}
together with $\phi_0=\phi$. We thus have $\Delta^k\phi_{2k}=\phi$. The solution of Eq.~\eqref{DeltaPsiL} is then given in analytic closed form as~\cite{BFW14b}
\begin{equation}\label{PsiL}
\Psi^{\phi}_L = \Delta^{-1} \bigl(\hat{x}_{L}\,\phi\bigr) =
\sum_{k=0}^{\ell}\frac{(-2)^k\ell!}{(\ell-k)!}\,x_{\langle
	L-K}\partial_{K\rangle}\phi_{2k+2}\,.
\end{equation}
This formula has been derived by induction in $3$ dimensions in~\cite{BFW14b} but the proof works as well in $d$ dimensions and no extra factor needs to be added. The precise choice of the Poisson solutions involved in the above algorithms is irrelevant for this particular problem, hence the operator
$\Delta^{-1}$ has not been precisely defined in Eq.~\eqref{PsiL}. However, it is convenient in practice to take $\Delta^{-1}=\widetilde{\Delta^{-1}}$, where $\widetilde{\Delta^{-1}}=\mathrm{FP}_{B=0}\Delta^{-1}(r/r_0)^B$ represents the Poisson integral regularized at infinity by means of the Hadamard finite part prescription.

With this tool in hands, we can thus transform the integral we were looking for into the much more tractable form
\begin{equation}\label{tractable}
\int \dd^{3} \mathbf{x} \,r^B\,\hat{x}_L \,\phi P = \int \dd^{3} \mathbf{x} \,r^B\Bigl(\Psi^{\phi}_L \Delta P + \partial_i\Bigl[ \partial_i \Psi^{\phi}_L P - \Psi^{\phi}_L \partial_i P\Bigr] \Bigr)\,.
\end{equation}
The first term involves the source of the potential $P$, and is therefore computable, while the second one is a surface term, which is also computable [see Sec.~\ref{sec:intpart}]. For instance, in the case where $P=\hat{M}_{ij}$, with $\hat{M}_{ij}$ being the 4PN tensor potential, we will replace $\Delta\hat{M}_{ij}$ by the source given explicitly by Eq.~\eqref{Mij3d}, which is correct since we are already at the maximal 4PN order and thus $\hat{M}_{ij}$ is merely Newtonian.

In general $\phi$ will be equal to some derivative of a compact-support potential, such as $\phi=\partial_{ab}V$ where $V$ is given in~\eqref{V3}, but let us illustrate the computation of the super-potentials with the simpler case $\phi=V$. The compact source of this potential is $\sigma=\tilde{\mu}_1 \delta (\mathbf{x}-\mathbf{y_1})+1\leftrightarrow 2$ where $\tilde{\mu}_1$ is a function of time defined by Eq.~\eqref{mutilde}. Using the symmetric propagator (we neglect the odd dissipative effects which do not impact our calculation here\footnote{We have checked explicitly that dissipative contributions that are even in powers of $1/c$, due for instance to the coupling of two odd terms, never arise in the mass quadrupole at the 4PN order.}) we have
\begin{subequations}\label{formV}
\begin{align}
V &= \sum_{k=0}^{+\infty} \left(\frac{\partial}{c\partial t}\right)^{2k} U_{2k}\,,\\
\text{where}\quad U_{2k} &= - 4 \pi G \,\Delta^{-k-1} \bigl[ \tilde{\mu}_1 \delta (\mathbf{x}-\mathbf{y_1}) \bigr] + 1\leftrightarrow 2 \,.
\end{align}
\end{subequations}
From that definition we see that $U_{2k}$ is the super-potential of the Newtonian potential $U$ obeying the Poisson equation $\Delta U = -4\pi G \sigma$. It is straightforward to compute the functions $U_{2k}$ using the Appendix B of~\cite{BDE04} and we find
\begin{equation}\label{U2k}
U_{2k} = \frac{G\,\tilde{k}}{2^k (2k)!!} \,\frac{\Gamma(2-\frac{d}{2})}{\Gamma(k+2-\frac{d}{2})} \,\tilde{\mu}_1 \,r_1^{2k+2-d} + 1\leftrightarrow 2 \,,
\end{equation}
generalizing Eq.~(4.18) in~\cite{BFW14b} to $d$ dimensions. Finally, from Eqs.~\eqref{formV} we see that the super-potentials of $V$ are given in terms of those of $U$ by
\begin{equation}\label{V2k}
V_{2k} = \sum_{j=0}^{+\infty} \left(\frac{\partial}{c\partial t}\right)^{2j} U_{2k+2j}\,.
\end{equation}
By inserting~\eqref{V2k} into~\eqref{PsiL} we can obtain at each PN order an explicit expression for the superpotential $\Psi_L^V$. To compute that quantity but for some space-time derivative of a potential, for instance $\Psi_L^{\partial_{ab}V}$ or $\Psi_L^{\partial_t\partial_{a}V}$, we proceed in a similar way as all the space and time derivatives commute. Finally we are able to compute in a straightforward way all the super-potentials we need [see Table~\ref{tab:listPotOrder}].  

\subsection{Integrations by part and surface terms}
\label{sec:intpart}

The second technique to simplify the expression of the quadrupole moment is to integrate some terms by part and transform volume integrals into simpler surface integrals at infinity. For instance, we systematically rewrite integrals involving the double gradient of a simple compact-support potential like $V$, defined in~\eqref{V3}, and a difficult one $P$ (with non-compact support) as
\begin{equation}
\label{intpart}
\partial_i V \partial_i P = \frac{1}{2} \Bigl[\Delta (V P) - V \Delta P - P \Delta V\Bigr]\,.
\end{equation}
The second term in~\eqref{intpart} is much simpler because it contains (modulo higher PN corrections) the source of the potentials $P$, \textit{i.e.} $\Delta P = S + \mathcal{O}(c^{-2})$. The third term is also easy to evaluate because it depends only on the value of the potential $P$ at the location of the particles, since $V$ has a compact support: $\Delta V = - 4 \pi G \sigma + \mathcal{O}(c^{-2})$. 

As for the first term in~\eqref{intpart}, it yields an example of a so-called ``Laplacian term'', coming after integration by parts from the derivation of the regularization factor $(r/r_0)^B$. It is made of a surface integral at infinity, plus a possible volume integral whose expression in terms of the potentials is significantly simpler than the original one, see below. The surface integrals of the Laplacian terms, as well as the analogous so-called ``divergence terms'', are very easy to integrate within the Hadamard finite part prescription. Therefore, we keep as much as possible the terms into Laplacian or divergence form.

Following~\cite{BI04mult} the generic Laplacian term, \textit{e.g.}, coming from the first term in the right side of~\eqref{intpart} with $G=\frac{1}{2}V P$, reads
\begin{equation}\label{LaplacianTerm1}
T_L = \mathop{\mathrm{FP}}_{B=0} \int \dd^3 \mathbf{x} \left( \frac{r}{r_0}\right)^B \!\!\hat{x}_L\, r^{2k} \Delta G\,,
\end{equation}
where the factor $r^{2k}$ (with $k\in \mathbb{N}$) arises when applying the formula~\eqref{series} that implements the integration with respect to $z$. Integrating the Laplacian by parts, we obtain
\begin{align}\label{LaplacianTerm2}
T_L &= 2k(2k+2\ell+1) \mathop{\mathrm{FP}}_{B=0} \int \dd^3 \mathbf{x} \left( \frac{r}{r_0}\right)^B\!\! \hat{x}_L\, r^{2k-2} G \nonumber \\ &+ \mathop{\mathrm{FP}}_{B=0} B (B + 4k+2\ell +1) \,r_{0}^{-B}\int_{r > \mathcal{R}} \dd^3 \mathbf{x} \,r^{B-2} \hat{x}_L \, r^{2k} G \,.
\end{align}
Thanks to the prefactor $B$, the second integral can be restricted to the far zone: $r>\mathcal{R}$, where $\mathcal{R}$ is an arbitrary length. Indeed, the matching equation which leads to the expressions of the multipole moments is originally applied for smooth matter distributions, so that the metric is smooth everywhere in the near zone. As such, $G$ is also smooth there and, due to the factor $B$, the near zone contribution is zero after the FP procedure. (In practice, the point-particle approximation leads to UV divergences, but these are separately treated by DR, in which the FP plays no role.)

Because of the $B$ prefactor in~\eqref{LaplacianTerm2}, we need to look at the $1/B$ pole in the integral. This pole can only come from a radial integral of the form $\int_\mathcal{R}^{+\infty} \dd r \,r^{B-1} = - \mathcal{R}^B/B$. Thus, considering the asymptotic expansion of $G$ when $r \to \infty$, which we denote by $\mathcal{M}(G)$ as it is identical to the multipole expansion, we find that the pole comes only from the term of order $r^{-2k-\ell -1}$ in that expansion. At the 4PN order, we also obtain a logarithmic dependence in the asymptotic expansion of some of the potentials. Hence, if we define $X_p(\mathbf{n})$ and $X_p^{\ln} (\mathbf{n})$ to be the coefficients of $r^{-p -1}$ and $r^{-p -1}\ln r$ in the multipole expansion, we have
\begin{equation}\label{expG}
\mathcal{M}(G) = \dots + \frac{1}{r^{2k+\ell+1}}\left[X_{2k+\ell}(\mathbf{n}) + X_{2k+\ell}^{\ln}(\mathbf{n}) \ln \left(\frac{r}{r_0}\right)\right]+ o\left(r^{-2k-\ell-1}\right)\,,
\end{equation}
so that we finally obtain (applying the definition of the FP)
\begin{align}\label{JL}
T_L &= 2k(2k+2\ell+1) \mathop{\mathrm{FP}}_{B=0} \int \dd^3 \mathbf{x} \left( \frac{r}{r_0}\right)^B\!\! \hat{x}_L\, r^{2k-2} G \nonumber \\ &+ \int \dd \Omega \,\hat{n}_L \Bigl[-(4k+2\ell +1)X_{2k+\ell}(\bm{n})+ X_{2k+\ell}^{\ln} (\mathbf{n})\Bigr]\,.
\end{align}
The first integral is still of the form \eqref{LaplacianTerm1} except that the Laplacian factor $\Delta G$ has been replaced by $G$ itself. Since the latter quantity is typically the product of several potentials (or their derivatives), the integrand has actually been simplified. As for the second integral, it is a surface contribution that depends on neither scale $\mathcal{R}$ nor $r_0$.
We have presented in~\eqref{JL} a general formula; however with our conventions for the simplification of the 4PN quadrupole moment, we will only need the case $k=0$. In such case the coefficient of the first term in Eq.~\eqref{JL} vanishes and the only task consists in evaluating the angular integral involving the $1/r^{\ell+1}$ coefficients of $G$. In particular, there is then no need to know (and in general we do not know) $G$ everywhere but just its asymptotic expansion, which is computed from the known source of the potential [see Sec.~\ref{sec:PotAtInf}].

The other surface integrals occurring are the ``divergence terms'', for instance the second term in the right side of~\eqref{tractable}. Indeed, most of these terms come from the method of super-potentials. They are of the form
\begin{equation}
K = \mathop{\mathrm{FP}}_{B=0} \int \dd^3 \mathbf{x} \left(\frac{r}{r_0}\right)^B \partial_i H_i\,.
\end{equation}
A similar reasoning to the one before shows that they depend only on the $1/r^2$ coefficient, say $Y_i(\mathbf{n})$, in the asymptotic or multipole expansion $\mathcal{M}(H_i)$ when $r\to\infty$:
\begin{equation}\label{expHi}
\mathcal{M}(H_i) = \dots + \frac{1}{r^2}\left[Y_i(\mathbf{n}) + Y_i^{\ln}(\mathbf{n}) \ln \left(\frac{r}{r_0}\right)\right] + o\left(r^{-2}\right)\,,
\end{equation}
and the FP procedure yields simply 
\begin{equation}
K = \int \dd \Omega \,n_i Y_i(\mathbf{n})\,.
\end{equation}
We find that there in no contribution from the logarithm in~\eqref{expHi}.

By the previous method, we have to obtain the asymptotic expansion when $r\to+\infty$ of $G$ or $H_i$ [Eqs.~\eqref{expG} or~\eqref{expHi}], where $G$ and $H_i$ are made of products of derivatives of potentials, involving in general one potential which is not known in the whole space, for instance the 4PN potential $P=\hat{M}_{ij}$ or its trace. To find the expansion when $r\to+\infty$ of such potential, we rely on the method explained in Sec.~\ref{sec:PotAtInf}.

Once the latter two techniques --- super-potentials and surface integrals --- have been applied, one obtains an extremely long expression for the quadrupole moment as a function of potentials and super-potentials, where all terms can be explicitly integrated. The full result is presented in the Appendix~\ref{app:MQAsPot}. 

\section{Dimensional regularization of UV divergences}\label{sec:dimregUVgen}

Our approach being based on an effective representation of the bodies by Dirac distributions, it relies on DR to deal with the
divergences related to the point-like character of the object description. As stressed in Sec.~\ref{sec:intro}, this regularization was shown, both by traditional PN methods and by the EFT, to be able to tackle this issue properly and without ambiguities at the 3PN order and beyond.

\subsection{Regularization of potentials and volume integrals}\label{sec:dimregUV}

The computation of the volume integrals that remain in the quadrupole after the treatment of Laplacian and divergence terms requires the potentials in $d$ dimensions but only in the form of a local expansion around the particles $\bm{y}_{1,2}$ (third column in the list of potential presented in Table~\ref{tab:listPotOrder}), since it is ultimately the difference between the 3-dimensional potential computed by means of Hadamard's regularization and its $d$-dimensional counterpart that we really need to control, and that the parts outside the singularities cancel in the limit $d\to 3$. The compact parts of potentials pose no problem as we just have to iterate the symmetric propagator with the Green function of the Laplace operator in $d$ dimensions. Notably, the Poisson solution of $\Delta u_1 = -4\pi \delta^{(d)}(\mathbf{x}-\bm{y}_1)$ reads\footnote{See the Appendix B in~\cite{BDE04} for a compendium of useful formulas in $d$ dimensions.}
\begin{equation}\label{greend}
u_1 = \tilde{k}\,r_1^{2-d}\,,\qquad\tilde{k} = \frac{\Gamma(\frac{d-2}{2})}{\pi^{\frac{d-2}{2}}}\,.
\end{equation}
To compute potentials in $d$ dimensions one needs in principle the generalization of functions such as $g$ given by~\eqref{Fock} in $d$ dimensions. This is known in the case of $g$, but it is rather cumbersome and needed only if one wants to compute the potentials in $d$ dimensions in the whole space. However, the expansion of $g$ in $d$ dimensions around $\bm{y}_{1,2}$ is quite handy, and given by the Appendix B of~\cite{BBBFMa}. Therefore, while some potentials cannot be easily computed for any $\mathbf{x} \in \mathbb{R}^d$, we can compute them around the singularities $\bm{y}_{1,2}$, which is actually sufficient to apply the DR.

In Hadamard's regularization, the $3$-dimensional spatial integral is defined by the \textit{partie finie} (Pf) prescription, depending on two constants $s_1$ and $s_2$ associated with logarithmic divergences at the two singular points (see Ref.~\cite{BFreg} for a review of this regularization), say
\begin{equation}\label{IHad}
I_{\mathcal{R}}  = \mathop{\mathrm{Pf}}_{s_1,s_2}\int_{r<\mathcal{R}}
\dd^3\mathbf{x}\,S(\mathbf{x})\, ,
\end{equation}
where $S$ denotes the non-compact support integrand and we limit the integration to a spherical ball $r<\mathcal{R}$ in order to keep only the UV divergences (see~\cite{BFeom} for a review). On the other hand, in DR, the integral is automatically regularized by means of the analytic continuation in the dimension $d$, so that
\begin{equation}\label{Id}
I^{(d)}_{\mathcal{R}} = \int_{r<\mathcal{R}}
\dd^d\mathbf{x}\,S^{(d)}(\mathbf{x})\,.
\end{equation}

We assume that we can compute the expansion of the source in $d$ dimensions in the vicinity of the particles, say $\bm{y}_1$. In 3 dimensions, we have the expansion around $\bm{y}_1$, 
\begin{equation}\label{expS}
S(\mathbf{x}) = \sum_{p_0 \leqslant p \leqslant N} r_1^p \mathop{\sigma}_1{}_{\!p} (\mathbf{n}_1) + o(r_1^N) \,,
\end{equation}
where $r_1=\vert\mathbf{x}-\bm{y}_1\vert$ and the coefficients $\mathop{\sigma}_{1p}$ depend on the unit direction $\mathbf{n}_1=(\mathbf{x}-\bm{y}_1)/r_1$ of approach to the singularity. In $d$ dimensions, we have a similar albeit more complicated expansion (where we pose $\varepsilon=d-3$)
\begin{equation}\label{expSd}
S^{(d)}(\mathbf{x}) = \sum_{\substack{p_0\leqslant p\leqslant N\\ q_0\leqslant q\leqslant q_1}} r_1^{p+q\varepsilon}  \mathop{\sigma}_1{}^{(\varepsilon)}_{\!p,q} (\mathbf{n}_1) + o(r_1^N) \,,
\end{equation}
where the coefficients now depend on an extra integer $q$ reflecting the  more complicated structure of the expansion involving powers $p+q\varepsilon$ (here both $p$, $q \in\mathbb{Z}$). Since the two expansions~\eqref{expS} and~\eqref{expSd} must agree in the limit $\varepsilon\to 0$, the relation
\begin{equation}\label{eps0}
\sum_{q_0\leqslant q\leqslant q_1} \mathop{\sigma}_1{}^{(0)}_{\!p,q} = \mathop{\sigma}_1{}_{\!p}\,,
\end{equation}
must hold for any $p$.

Now, we are interested in the \textit{difference} between DR and the Hadamard partie finie, because this is precisely what we have to add to the Hadamard result~\eqref{IHad} in order to get the correct $d$-dimensional result~\eqref{Id}. This difference is
\begin{equation}\label{DI}
\mathcal{D}I = I^{(d)}_{\mathcal{R}} - I_{\mathcal{R}} \,,
\end{equation}
of which we merely compute the pole part $\propto 1/\varepsilon$ and the finite part $\propto\varepsilon^0$ in the Laurent expansion when $\varepsilon \rightarrow 0$, the other terms vanishing in that limit. The key point is that the difference~\eqref{DI} does not depend on $\mathcal{R}$ but only on the coefficients of the expansion around the two singularities as defined by~\eqref{expSd}, modulo neglected $\mathcal{O}(\varepsilon)$ terms. We have~\cite{BDEI05dr}
\begin{align}\label{IDiffHadDR}
\mathcal{D}I =& \frac{\Omega_{d-1}}{\varepsilon}\sum_{q_0\leqslant q\leqslant q_1} \left[\frac{1}{q+1} + \varepsilon \ln \left(\frac{s_1}{\ell_0}\right) \right]\langle\mathop{\sigma}_1{}^{(\varepsilon)}_{\!-3,q}\rangle + 1\leftrightarrow 2\,,
\end{align}
where it is crucial that the angular average be performed in $d$ dimensions, \textit{i.e.},
\begin{equation}\label{IntAngul}
\langle\mathop{\sigma}_1{}_{\!p,q}^{(\varepsilon)}\rangle
 = \int\frac{\dd\Omega_{d-1}}{\Omega_{d-1}}
\mathop{\sigma}_1{}_{\!p,q}^{(\varepsilon)}(\mathbf{n}_1)\,,\qquad \Omega_{d-1}= \frac{2\pi^{\frac{d}{2}}}{\Gamma\left(\frac{d}{2}\right)}\,.
\end{equation}
Here, $\dd \Omega_{d-1}(\mathbf{n}_1)$ is the solid angle on the $(d-1)$-dimensional sphere and $\Omega_{d-1}=\int\dd \Omega_{d-1}$. In actual calculations, we have verified that there is no problem with the value $q=-1$ in~\eqref{IDiffHadDR} since we always have $\langle{}_1\sigma_{-3,-1}^{(\varepsilon)}\rangle=0$. 

As we see from~\eqref{IDiffHadDR}, the calculation generates many UV-type poles $1/\varepsilon$, but we shall prove that all the poles can be removed by the specific shift determined from the 4PN equations of motion in Refs.~\cite{BBBFMc,MBBF17}. However, before we do so, let us discuss that at 4PN order all the poles are not of the form~\eqref{IDiffHadDR}, but there are other poles coming directly from the calculation of potentials at the location of the particles with DR. This more involved case is detailed now.

\subsection{Regularization of the potentials at the locations of the particles}\label{sec:potpart}

For the integrals with compact support, only the values of the potentials at $\mathbf{x} = \bm{y}_{1,2}$ are required. Obviously, when the potential is known in $d$ dimensions, we can directly deduce its value at the particles $\bm{y}_{1,2}$. This is however not the case for some of the most difficult potentials. In particular, we have to use another method to compute the values of $\hat{X}$, $\hat{Y}_i$, $\hat{M}=\hat{M}_{ii}$ and $\hat{T}$ at $\bm{y}_{1,2}$ in $d$ dimensions. From the fourth column of Table~\ref{tab:listPotOrder}, we see that these values are required at Newtonian order, except for the case of $\hat{X}$, needed at 1PN order.

Let $S$ be the source of a potential $P$ that we want to compute at the particle positions $\mathbf{x} = \bm{y}_{1,2}$. At Newtonian order for $P$, we simply have $\Delta P = S$. The source admits some expansions around the singularities in 3 dimensions and $d$ dimensions given by Eqs.~\eqref{expS} and~\eqref{expSd} above. As usual, we proceed in two steps, Hadamard's regularization, to which we add the corrections due to DR. 

We thus compute first the value of the potential $P$ at point 1 in $3$ dimensions using the Hadamard partie finie. In that case, the potential at any field point $\mathbf{x}'$ is given by the Poisson integral of the singular source $S(\mathbf{x})$, and defined in the sense of the Hadamard partie finie,
\begin{equation}\label{Px}
P_{\mathcal{R}}(\mathbf{x}')=-\frac{1}{4\pi}\,\mathop{\mathrm{Pf}}_{s_1,s_2}
\int_{r<\mathcal{R}}\frac{\dd^3\mathbf{x}}{\vert\mathbf{x}-
\mathbf{x}'\vert}\,S(\mathbf{x})\,,
\end{equation}
where $s_1$ and $s_2$ are the two associated arbitrary constants. Again, we restrict the integration volume to a spherical ball $r<\mathcal{R}$ in order to focus attention on UV divergences. Now, it has been shown in~\cite{BFreg} that the Hadamard partie finie of the potential $P$, or rather $P_{\mathcal{R}}$ when the IR part of the integral is ignored, at the location of the singular point 1, reads
\begin{equation}\label{P1Had}
(P_{\mathcal{R}})_1 = - \frac{1}{4\pi} \,\mathop{\mathrm{Pf}}_{s_1,s_2} \int_{r<\mathcal{R}} \frac{\dd^3\mathbf{x}}{r_1} \,S(\mathbf{x}) + \left[\ln\left(\frac{r_1'}{s_1}\right) - 1\right] \langle\mathop{\sigma}_1{}_{\!-2}(\mathbf{n}_1)\rangle\,.
\end{equation}
The first term corresponds to the naive replacement of $\mathbf{x}'$ by the source point $\bm{y}_1$ in~\eqref{Px}, while the second term accounts for the presence of the logarithmic divergence $\ln r_1'=\ln\vert\mathbf{x}'-\bm{y}_1\vert$ in the limit $\mathbf{x}'\to\bm{y}_1$, the formally infinite contribution $\ln r_1'$ therein being considered to be a constant (see~\cite{BFreg} for more details). Note that the constant $s_1$ cancels out between the two terms in~\eqref{P1Had} so that $(P)_1$ depends only on $s_2$ and $r_1'$. The second term in~\eqref{P1Had} contains the usual angle average in 3 dimensions,
\begin{equation}\label{angle}
\langle\mathop{\sigma}_1{}_{\!-2}\rangle = \int \frac{\dd \Omega_1}{4\pi} \mathop{\sigma}_1{}_{\!-2}(\mathbf{n}_1)\,.
\end{equation}
After having computed the Hadamard value $(P)_1$ in this way, we correct it so that in corresponds to DR. 

In DR the value of the potential $P^{(d)}_{\mathcal{R}}$ at the point $\bm{y}_1$ is simply obtained by replacing $\mathbf{x}'$ by $\bm{y}_1$ inside the Poisson integral in $d$ dimension, since the regularization is taken care of by the analytic continuation in $d$. Hence
\begin{equation}\label{P1dimreg}
P^{(d)}_{\mathcal{R}}(\bm{y}_1) = -\frac{\tilde{k}}{4\pi}\,\int_{r< \mathcal{R}}\frac{\dd^d\mathbf{x}}{r_1^{d-2}}\,S^{(d)}(\mathbf{x})\,,
\end{equation}
where $\tilde{k}$ has been defined in Eq.~\eqref{greend}. Given the results $(P)_1$ and $P^{(d)}(\bm{y}_1)$ of the two regularizations we define their UV difference as
\begin{equation}\label{DP1}
\mathcal{D}P(1) = P^{(d)}_{\mathcal{R}}(\bm{y}_1)-(P_{\mathcal{R}})_1\,,
\end{equation}
which is independent of $\mathcal{R}$. We only compute the pole part followed by the finite part when $\varepsilon \rightarrow 0$. The difference depends again only on the coefficients of the expansion around the two singularities as given by~\eqref{expSd}, but the formula is more involved than in~\eqref{IDiffHadDR}. We have~\cite{BDE04}
\begin{align}\label{P1DiffHadDR}
\mathcal{D}P(1) =& -\frac{1}{\varepsilon (1+\varepsilon)}\sum_{q_0\leqslant q\leqslant q_1} \left(\frac{1}{q} + \varepsilon \left[\ln \left(\frac{r_1'}{\ell_0}\right) -1\right]\right)\langle\mathop{\sigma}_1{}^{(\varepsilon)}_{\!-2,q}\rangle\nonumber \\
& -\frac{1}{\varepsilon (1+\varepsilon)} \sum_{q_0\leqslant q\leqslant q_1} \left(\frac{1}{q+1} + \varepsilon \ln \left(\frac{s_2}{\ell_0}\right)\right) \sum_{\ell = 0}^{+\infty} \frac{(-)^\ell}{\ell!}\,\partial_L\left(\frac{1}{r_{12}^{1+\varepsilon}}\right)\langle n_2^L\mathop{\sigma}_2{}^{(\varepsilon)}_{\!-\ell-3,q}\rangle\,.
\end{align}
In the second term the sum over $\ell$ is actually finite since there is a maximal order of the singularity, bounded by a negative integer $p_0$ in~\eqref{expSd}.

After consistently correcting the results obtained in Hadamard's regularization by means of Eqs.~\eqref{IDiffHadDR} and~\eqref{P1DiffHadDR}, the constants $s_1$, $s_2$, $r'_1$ and $r'_2$ must individually cancel out since they are actually absent in $d$ dimensions. Instead, poles associated with logarithms of the characteristic length scale $\ell_0$ may arise. They do at the 3PN order and beyond, in both the gravitational field and the accelerations~\cite{DJSdim, BDE04, BDEI05dr}. In previous works, those entering the accelerations have been conveniently traded, by applying an unphysical shift of the worldlines, for two logarithmic constants, denoted as $\ln r'_1$, $\ln r'_2$~\cite{BDE04, BBBFMa}. Indeed, the ensuing equations of motion have then the same form as the ones derived from a purely 3-dimensional calculation based on Hadamard's treatment. In particular, $\ln r'_1$ and $\ln r'_2$ play the role of trackers for the UV divergences. A different, simpler, but very close, choice of shift will be made in section~\ref{sec:resultMQ} by taking $r'_1=r'_2=r_0'$.

\subsection{Distributional derivatives and the Gel'fand-Shilov formula}

Finally, we need to take care of the compact support contributions that are generated by the purely distributional part of the derivatives of potentials appearing in the non-compact terms in $d$ dimensions. For that purpose, we use the Schwartz distributional derivative~\cite{Schwartz} or, equivalently, the Gel'fand-Shilov formula~\cite{gelfand}. The distributional derivatives are imperatively to be applied in $d$ dimensions in order to be well defined and avoid the appearance of undesirable products of Dirac distributions with functions that are singular on their support, like $r_1^{-1}\delta^{(3)}(\mathbf{x}-\bm{y}_1)$. Let $P$ be one of our elementary potentials $(V, V_i, \hat{W}_{ij}, \cdots)$ presented in Appendix~\ref{app:PNpotentials}. Around the two singularities, it admits an expansion similar to~\eqref{expSd},
namely
\begin{equation}\label{expPd}
P = \sum_{\substack{p_0\leqslant p\leqslant N\\ q_0\leqslant q\leqslant q_1}} r_1^{p+q\varepsilon}  \mathop{f}_1{}^{(\varepsilon)}_{\!p,q} (\mathbf{n}_1) + o(r_1^N) \,,
\end{equation}
where, in particular, the maximal divergence corresponds to the generally negative power $p_0\in\mathbb{Z}$. Then, the distributional derivative of this potential is given by
\begin{equation}\label{distrderiv}
\partial_{i} P = (\partial_{i} P)_\text{ord} + D_{i}[P]\,,
\end{equation}
where the first term represents the ``ordinary'' piece of the derivative, while the purely distributional part reads
\begin{equation}\label{distrpart}
D_{i}[P] = \Omega_{d-1} \sum_{\ell=0}^{+\infty} \frac{(-)^\ell}{\ell!} \,\partial_L\delta^{(d)}(\mathbf{x}-\bm{y}_1)\,\langle n_1^{iL} \mathop{f}_1{}_{\!-\ell-2,-1} \rangle + 1 \leftrightarrow 2\,,
\end{equation}
which is a generalized version of the Gel'fand-Shilov formula~\cite{gelfand}. We use here the notation~\eqref{IntAngul} for the angular average in $d$ dimensions, and denote by $L$ the multi-index $i_1\cdots i_\ell$ with $\ell$ indices (and $n_1^{iL}=n_1^{i}n_1^{i_1}\cdots n_1^{i_\ell}$). The only contributions to $D_{i}[P]$ come thus from the singular terms with powers $p = -\ell -2$ and with $q=-1$. Moreover, as in~\eqref{P1DiffHadDR} the sum in the right-hand side of~\eqref{distrpart} is actually finite since we have $\ell \leqslant -2 - p_0$.

Typically, distributional spatial derivatives will contribute when computing the second derivative of a potential, say $\partial_{ij} P$. In that case, we shall have $\partial_{ij} P = (\partial_{ij} P)_\text{ord} + D_{ij}[P]$ with the distributional term given by (see~\cite{BFreg} for a review)
\begin{equation}\label{distrpartij}
D_{ij}[P] = D_{i}[\partial_j P] + \partial_i D_{j}[P]\,,
\end{equation}
where $D_i$ represents the distributional derivative operator defined by~\eqref{distrpart}. Only terms linear in the Dirac functions $\delta(\mathbf{x}-\bm{y}_{1})$ or $\delta(\mathbf{x}-\bm{y}_{2})$ are to be kept since the product of two delta-functions, or derivatives of delta-functions, is always zero in DR. So, the partial derivatives in Eq.~\eqref{distrpartij} are to be taken as ordinary.

To compute the distributional time derivative, one first define the partial derivative with respect to the source points $\bm{y}_{1,2}$ as
\begin{equation}\label{distrpart1}
\mathop{D}_1{}_{i}[P] = - \Omega_{d-1} \sum_{\ell=0}^{+\infty} \frac{(-)^\ell}{\ell!} \,\partial_L\delta^{(d)}(\mathbf{x}-\bm{y}_1)\,\langle n_1^{iL} \mathop{f}_1{}_{\!-\ell-2,-1} \rangle \quad\text{and}\quad 1\leftrightarrow 2\,.
\end{equation}
This definition is consistent with the translational invariance of $r_1=|\mathbf{x}-\bm{y}_1|$, which is nothing but the small expansion parameter of the potentials near the singularity $\mathbf{x}=\bm{y}_1$. It also implies that $\mathop{D}_{i}[P]+\mathop{D}_{1i}[P]+\mathop{D}_{2i}[P]=0$, which ensures that the partial derivatives obey $\partial_{i} P+\partial_{1i} P+\partial_{2i} P=0$; this is the consequence of the fact that the potential, as a function, depends on the trajectories only through the two distances to the field point $\mathbf{r}_1 = \mathbf{x}-\bm{y}_{1}$ and $\mathbf{r}_2 = \mathbf{x}-\bm{y}_{2}$. Then, the distributional time derivative is naturally obtained as $\partial_{t} P = (\partial_{t} P)_\text{ord} + D_{t}[P]$, where
\begin{equation}\label{distrpartt}
D_{t}[P] = v_1^i\mathop{D}_1{}_{i}[P] + v_2^i\mathop{D}_1{}_{2}[P]\,.
\end{equation}
Mixed time-space or second time derivatives are computed using 
\begin{subequations}\label{distrpartttit}
\begin{align}
D_{it}[P] &= D_{i}[\partial_t P] + \partial_i D_{t}[P]\,,\\
D_{tt}[P] &= D_{t}[\partial_t P] + \partial_t D_{t}[P]\,.
\end{align}
\end{subequations}
We observe from~\eqref{distrpartij} and~\eqref{distrpartttit} that the operations of applying successive distributional derivatives do not \textit{a priori} commute. Fortunately, we have checked that this non-commutation does not affect our computation of the quadrupole moment up to 4PN order.

Let us finally mention an important point. At the 4PN order, there are terms involving the spatial derivative $\partial_i\hat{X}$ of the non-linear potential $\hat{X}$, defined by~\eqref{X3} in 3 dimensions and provided in Appendix~\ref{app:PNpotentials} in $d$ dimensions. This potential involves some cubic ``self'' terms that diverge like $1/r_1^3$ in 3 dimensions, and like $1/r_1^{3d-6}$ in $d$ dimensions, when $r_1\to 0$, as they essentially arise from the product of three Newtonian-like potentials. Now, if we were to apply the formula~\eqref{distrpart} in 3 dimensions, we would find that the derivative $\partial_i\hat{X}$ does contain a distributional term, proportional to the derivative of the Dirac function in 3 dimensions. However, in $d$ dimensions, the situation is different. Indeed, we see from~\eqref{distrpart} that $\partial_i (1/r_1^{3d-6})$ does not generate any distributional term, because the singularity corresponds to the value $q=-3$ while one needs $q=-1$ for the singularity to contribute in Eq.~\eqref{distrpart}. Therefore, the derivative $\partial_i\hat{X}$ should just be considered as ordinary. This is why we had to be careful at computing distributional derivatives only in $d$ dimensions. We found in our calculations that only second derivatives of potentials, like $\partial_{ij} P$ or $\partial_{t}^2 P$, yield distributional contributions.

\section{Computation of the various types of terms}
\label{sec:ComputePot}

Based on the computational and regularization methods explained above, from the sum of terms composing the mass quadrupole provided in the Appendix~\ref{app:MQAsPot}, we can list all the potentials required to control the mass quadrupole at 4PN. These potentials can be required either everywhere when they enter non-compact support integrals, or only at the location of particles $\mathbf{x} = \bm{y}_{1,2}$ when they enter compact-support terms, as was seen in Sec.~\ref{sec:dimregUVgen}. For instance, the potentials at $\mathbf{x} = \bm{y}_{1}$ (excluding the superpotentials) are those that enter $\tilde{\mu}_1$ at 4PN order and $\mu_1$ at 3PN order [see Appendix~\ref{app:PNpotentials}]. Besides, the potentials that enter non-compact integrals are needed in $d$ dimensions in a neighborhood of the particles $\bm{y}_{1,2}$ in order to implement the UV DR. Finally, in order to compute the surface terms (either Laplacian or divergence terms), we need to know their explicit expressions when $r\to\infty$ in 3 dimensions only. 

The techniques we employ to compute these potentials are well documented elsewhere~\cite{BDI95, BFP98, BFeom, BIJ02, BI04mult, FBI15}. We provide below a short summary of those and outline their most salient features. The Table~\ref{tab:listPotOrder} gives the different PN orders to which the various types of potentials are required for this computation.
\begin{table}[h]
\begin{center}
\begin{tabular}{| c || c | c | c | c |}
\hline
 Potential & 3-dim whole space & $d$-dim near $\bm{y}_{1,2}$ & ~at $\bm{y}_{1,2}$~ & $r\to +\infty$ \tabularnewline
    \hline \hline
    $V$ & 2PN & 2PN & 3PN & 3PN \tabularnewline
    \hline
      $V_i$ & 2PN & 2PN & 2PN & 2PN \tabularnewline
    \hline
      $K$ & $\times$ & 1PN & $\times$ & $\times$ \tabularnewline
    \hline
     $\Psi_{ij}^{\partial_{ab} V}$ &2PN&2PN&2PN&2PN \tabularnewline     
       \hline
     $\Psi_{ij}^{\partial_a V_b}$ &1PN&1PN&1PN&1PN \tabularnewline     
        \hline
     $\Psi_{ij}^{\partial_t \partial_a V}$ &1PN&1PN&1PN&1PN \tabularnewline     
        \hline
     $\Psi_{ijk}^{\partial_a V}$ &1PN&1PN&1PN&1PN \tabularnewline 
     \hline
      $\hat{W}_{ij}$ &1PN & 1PN &1PN & 2PN \tabularnewline
    \hline
     $\hat{W}$ & 1PN & 1PN & 2PN &2PN \tabularnewline
     \hline
      $\hat{Z}_{ij}$ &N & N & N& 1PN \tabularnewline
    \hline
     $\hat{Z}$ &N& N& N &1PN \tabularnewline
     \hline
      $\hat{R}_{i}$ &N& N& 1PN& 1PN \tabularnewline
    \hline
     $\hat{Y}_i$ &$\times$& $\times$&N&N \tabularnewline
     \hline
      $\hat{X}$ &N& N&1PN&1PN \tabularnewline
    \hline
     $\hat{M}_{ij}$ &$\times$&$\times$&$\times$&N \tabularnewline     
    \hline
     $\hat{M}$ &$\times$&$\times$&N&N \tabularnewline    
     \hline
  $\hat{T}$ &$\times$&$\times$&N&N \tabularnewline    
     \hline
 \end{tabular}
\caption{List of the PN orders required for the different potentials and super-potentials to control the 4PN mass quadrupole. The notation for the super-potential associated with the potential $\phi$ at multipolar order $\ell$ is $\Psi^{\phi}_L$, as defined by Eqs.~\eqref{Deltaphi2k}--\eqref{PsiL}. The second column corresponds to the potentials computed in $3$ dimensions in the whole space (for all $\mathbf{x}\in\mathbb{R}^3$). They are required for performing the volume integrals of the non-compact support terms. The third column corresponds to the potentials computed in $d$ dimensions but in the form of an expansion around the particles ($\bm{y}_1$ and $\bm{y}_2$). These expansions are inserted into the ``difference'' due to the DR of UV divergences. The next column is the value of the potentials at the location of particles ($\bm{y}_1$ or $\bm{y}_2$) needed for the compact-support terms, while the last one corresponds to the potentials computed in the form of an expansion at infinity, needed for evaluation of the surface integrals.
Note the particular case of the potential $K$, which always appears combined with $d-3$ factor, thus playing no role except for dimensional regularization.}
\label{tab:listPotOrder}
 \end{center}
\end{table}

\subsection{Compact-support potentials}\label{sec:compact}

The first potentials to compute are the compact support potentials, \textit{i.e.} $V$, $V_i$ and $K$ in Table~\ref{tab:listPotOrder}, and the compact-support parts of more complicated potentials. The d'Alembertian of these potentials is proportional to the matter source densities $\sigma$, $\sigma_i$ or $\sigma_{ij}$ defined in Eq.~\eqref{sigma}. We need the compact-support potentials only in 3 dimensions (none of them develop a pole). They are computed using the symmetric propagator, by iteration of the Green's function of the Laplace operator, say 
\begin{equation}\label{compact} 
P = \Box^{-1}_\text{sym}\delta^{(3)} = -\frac{1}{4\pi}\left[\frac{1}{r} + \frac{1}{c^2}\partial_t^2\left(\frac{r}{2}\right) + \frac{1}{c^4}\partial_t^4\left(\frac{r^3}{24}\right) + \cdots\right]\,.
\end{equation}
We do not include here the ``odd'' parity (dissipative) terms at orders 2.5PN and 3.5PN as they are already known~\cite{BIJ02}.

The values of $\hat{Y}_i$, $\hat{M}_{ij}$ and its trace $\hat{M}$, and $\hat{T}$ have been computed at $\mathbf{x} = \bm{y}_1$ by applying successively Eqs.~\eqref{P1Had} and~\eqref{P1DiffHadDR}. For the 1PN potential $\hat{X}$ the procedure is a little more complex but is fully explained in Sec.~IV of~\cite{BDE04} [see in particular Eq.~(4.30a) there]. In fact, the values of $\hat{X}$ and $\hat{T}$ at point $\bm{y}_1$ have already been computed in Ref.~\cite{BDLW10a} and we used (and recomputed) those results.

Note that some of the potentials, computable in the whole space in $d$ dimensions, are finite in the ``bulk'', \textit{i.e.} outside the singularities, but develop a pole when computed at the points $\bm{y}_{1,2}$. This is the case of the potentials $\hat{W}\equiv\hat{W}_{ii}$ and $\hat{Z}$ which, according to the Table~\ref{tab:listPotOrder}, are respectively needed at 2PN and 1PN orders at the points $\bm{y}_{1,2}$, in order to obtain the effective mass $\tilde{\mu}_1$ at the 4PN order given by~\eqref{mutilde1pot}. The values of these potentials have been computed directly from their expression in the whole space as well as by the application of the procedure relying on Eqs.~\eqref{P1Had} and~\eqref{P1DiffHadDR}. In the case of the 2PN potential $\hat{W}_{ij}$, we have obtained it in whole space and at the points $\bm{y}_{1,2}$ following the regularization [see Sec.~\ref{sec:Wij2PN}]; but we have also computed the trace $\hat{W}$ by solving the equivalent more convenient equation [see Eq.~\eqref{W3}]
\begin{equation}\label{eqWii}
\Box\Bigl(\hat{W}+\frac{V^2}{2}\Bigr) = 8\pi G\Bigl(\sigma_{ii}-\frac{1}{2}\sigma\,V\Bigr) - \frac{1}{c^2}(\partial_t V)^2\,.
\end{equation}
The advantage of this alternative formulation is that the non-compact support term is to be computed at relative 1PN order only. This permits us to apply the machinery of Eqs.~\eqref{P1Had} and~\eqref{P1DiffHadDR}, and its generalization at 1PN described in Ref.~\cite{BDE04}. As already said, the end result for these potentials computed at point 1 in DR includes a pole $1/\varepsilon$. For instance the relevant combination of potentials $\hat{W}$ and $\hat{Z}$ entering the 4PN effective mass $\tilde{\mu}_1$ [see Eq.~\eqref{mutilde1pot}] contains the pole:
\begin{equation}\label{WZpole}
(\hat{W})_1 + \frac{4}{c^2}(\hat{Z})_1 = -\frac{5G^4 m_{1}^2 m_{2}^2}{3 c^4 \varepsilon r_{12}^4} + \mathcal{O}(\varepsilon^0)\,.
\end{equation}
The effective mass $\tilde{\mu}_1$ itself will thus also contain this pole, together with several others due to the non-compact potentials $\hat{X}$, $\hat{T}$, $\hat{Y}_i$ and $\hat{M}$ computed at 1 [see Table~\ref{tab:listPotOrder}]. We find
\begin{equation}\label{mutildepole}
\tilde{\mu}_1 = \frac{G^3 m_1^2 m_2}{c^8 \varepsilon r_{12}^3}\left[ \frac{53}{15} m_1 \left( 3(n_{12}v_{12})^2 - v_{12}^2 + \frac{G m_1}{r_{12}}\right) + \frac{G m_2}{r_{12}}\left(\frac{79}{5}m_1 - \frac{1}{3} m_2\right)\right] + \mathcal{O}(\varepsilon^0)\,.
\end{equation}
We mention also the tricky case of the potential $\hat{R}_{i}$ at the 1PN order which, when computed at point 1, contains a ``cancelled'' pole, in the sense that the (1PN generalisation of the) formula~\eqref{P1DiffHadDR} has no pole, nevertheless it contains a finite non-zero term $\mathcal{O}(\varepsilon^0)$ which does contribute to the effective mass $\tilde{\mu}_1$ at 4PN order.

In Sec.~\ref{sec:resultMQ}, we shall show that all the UV poles coming from volume integrals on non-compact support terms, according to Eq.~\eqref{IDiffHadDR}, and the values at 1 of non-compact potentials, after Eq.~\eqref{P1DiffHadDR}, are removed by applying the \textit{same} specific shift as the one computed from the 4PN equations of motion in Refs.~\cite{BBBFMa, MBBF17}. This is an important verification of the present calculation, showing the consistency with the calculation of the equations of motion.

\subsection{Non compact-support potentials}\label{sec:noncompact}

The non-compact support terms are non-linear terms of basically two types. One is a rather simple type called ``$\partial V \partial V$'', made of a quadratic product of derivatives of compact support potentials, for instance $V$, $V_i$ and $K$, or the compact parts of other potentials. To the lowest PN order, we compute these terms at any field point $\mathbf{x}$ in analytic closed form using the elementary solution $g$ of the Poisson equation (see~\cite{BFeom} for more details)
\begin{equation}\label{geq}
\Delta g = \frac{1}{r_1 r_2}\,,
\end{equation}
where $r_1=\vert\mathbf{x}-\bm{y}_1\vert$ and $r_2=\vert\mathbf{x}-\bm{y}_2\vert$. The function $g$ solving~\eqref{geq}, in the sense of distributions in 3 dimensions, is the Fock function~\cite{Fock} 
\begin{equation}\label{Fock}
g^\text{Fock} = \ln S \equiv \ln \bigl(r_1 + r_2 + r_{12}\bigr)\,,
\end{equation}
where $r_{12}=\vert\bm{y}_1-\bm{y}_2\vert$. When handling PN corrections, it is also necessary to consider the solutions of the iterated Poisson equations and more complicated elementary functions. These are fully reviewed in Sec.~\ref{sec:Wij2PN}, and in addition, a subtle point about the use of those elementary functions, namely that they must be correctly matched to the exterior zone~\cite{BFeom}. As we shall see the general way to proceed in order to ensure this matching is to use Eqs.~\eqref{eq:kernels_matching}--\eqref{eq:source_matching} below.

The only potential which is needed in the whole space in 3 dimensions and that involves a piece more complicated than the $\partial V \partial V$ terms is $\hat{X}$. The more involved structure of the latter piece corresponds to the cubic non-linear source term $\hat{W}_{ij}\partial_{ij}V$ in Eq.~\eqref{X3}. Fortunately, from Table~\ref{tab:listPotOrder}, this term is needed at Newtonian order only and, as it turns out, we know it in closed analytic form at that order. As shown in~\cite{BDI95, BFP98}, in order to construct it, it suffices to find the solutions of the two Poisson equations (together with $1\leftrightarrow 2$)
\begin{subequations}\label{KH}
\begin{align}
\Delta K_1 &= 2~\partial_{ij}\left({1\over r_2}\right)
\partial_{ij}\ln r_1  \,,\\
\Delta H_1 &= 2~\partial_{ij}\left({1\over r_1}\right) \frac{\partial^2 g}{\partial y_1^i\partial y_2^j} \,.
\end{align}
\end{subequations}
Remarkably, the unique solutions of Eqs.~\eqref{KH} valid in the sense of distributions (and tending to zero when $r\to\infty$) can be written down under the explicit closed form
\begin{subequations}\label{KHexpl}
\begin{align}
K_1 &= -\frac{1}{r_2^3} + \frac{1}{r_2r_{12}^2}-\frac{1}{r_1^2r_2}+
\frac{r_2}{2r_1^2r_{12}^2}+\frac{r_{12}^2}{2r_1^2r_2^3}
+\frac{r_1^2}{2r_2^3r_{12}^2} \,, \\
H_1 &= -\frac{1}{2r_1^3}-\frac{1}{4r_{12}^3} - \frac{1}{4r_1^2r_{12}}-
\frac{r_2}{2 r_1^2r_{12}^2}+\frac{r_2}{2r_1^3r_{12}}
+\frac{3r_2^2}{4r_1^2r_{12}^3}+\frac{r_2^2}{2r_1^3r_{12}^2}-
\frac{r_2^3}{2r_1^3r_{12}^3} \,.
\end{align}\end{subequations}
With these solutions in hands, we control the non-compact potential $\hat{X}$ at the Newtonian order. Note that since $K_1$ and $H_1$ decrease at least like $1/r$ when $r\to +\infty$, there is no need to perform a matching to the far zone for these, as the matching equation~\eqref{eq:kernels_matching} [or the Poisson type version~\eqref{multP}] is automatically satisfied.

Finally, it remains to proceed with the extremely long computational task\footnote{This task is systematically done using a series of routines mainly developed by one of us (G.F.) with \textit{Mathematica} and the \textit{xAct} library~\cite{xtensor}.} of computing each of the non-compact support terms as a volume integral [see Appendix~\ref{app:MQAsPot}], using, in a first stage, the Finite Part regularization for the IR divergences and the Hadamard partie finie regularization for the UV ones. Unfortunately the result of this computation is too long to be presented here.

\subsection{The quadratic potential $\hat{W}_{ij}$ at 2PN order}
\label{sec:Wij2PN}

We now present more in details the calculation of the $\partial V \partial V$ pieces of the potentials relevant for the 4PN mass quadrupole, focusing on the case of $\hat{W}_{ij}$. We start with the derivation of the elementary kernels that are used to compute the potentials $\hat{W}_{ij}$ at 2PN order, as well as $\hat{Z}_{ij}$ and $\hat{R}_i$ at 1PN order, \textit{cf.} Sec.~\ref{sec:pot}. We treat as example the potential $\hat{W}_{ij}$ at 2PN order as it is the more cumbersome to compute, and as it involves all the elementary kernels that are needed to derive the other potentials $\hat{Z}_{ij}$ and $\hat{R}_i$. Our calculations are valid in 3 dimensions where this potential obeys the wave equation~\eqref{W3}. Since the first term has a compact support it is easily computed \emph{via} iterated Poisson integrals [see Sec.~\ref{sec:compact}] and we ignore it here. The second term is non-linear and with non-compact (NC) support; it reads
\begin{subequations}\label{WijNC}
	\begin{align}
	&\Box \hat{W}^\text{(NC)}_{ij} = -\partial_i V \partial_j V\,,\\
	\text{with}\quad &\Box V = -4 \pi G \tilde{\mu}_1 \delta_1 + 1 \leftrightarrow 2\,,
	\end{align}
\end{subequations}
where we recall our notation $\tilde{\mu}_1=\tilde{\mu}_1(t)$ for the effective time-dependent mass defined by Eq.~\eqref{mutilde}, see also~\eqref{mutilde1pot}. Up to 2PN order we have
\begin{equation}
V = \frac{G \tilde{\mu}_1}{r_1} + \frac{1}{c^2} \partial_t^2\Bigl(G \tilde{\mu}_1 \frac{r_1}{2}\Bigr) + \frac{1}{c^4} \partial_t^4\Bigl(G \tilde{\mu}_1 \frac{r_1^3}{24}\Bigr) + 1 \leftrightarrow 2 + \mathcal{O}(c^{-6})\,.
\end{equation}
By plugging this expression into the source term of $\hat{W}^\text{(NC)}_{ij}$, introducing the partial derivatives with respect to the source points $y^i_{1,2}$, and using the fact that $\tilde{\mu}_1$ and $\tilde{\mu}_2$ are just functions of time, we are able to express the solution up to 2PN order by means of a set of elementary kernel functions. It is convenient to split $\hat{W}^\text{(NC)}_{ij}$ up to 2PN order into ``self'' terms, which are essentially proportional to the mass squared of each particles $\tilde{\mu}_1^2$ or $\tilde{\mu}_2^2$, and ``interaction'' terms proportional to $\tilde{\mu}_1 \tilde{\mu}_2$ (or their time derivatives):
\begin{equation}\label{eq:Wijsplit}
\hat{W}^\text{(NC)}_{ij} = \hat{W}^\text{self}_{ij} + \hat{W}^\text{inter}_{ij} + \mathcal{O}(c^{-6})\,.
\end{equation}
The interaction part is expressed in terms of a series of elementary kernel functions $g$, $f$, $f^{12}$, $f^{21}$, $h$, $h^{12}$, $h^{21}$ and $k$ as
\begin{align}\label{eq:Wij2PNinter}
\mathop{\hat{W}}_{1}{\!}^\text{inter}_{ij} &= - G^2 \tilde{\mu}_1\tilde{\mu}_2\,\underset{1}{\partial_{(i}} \underset{2}{\partial_{j)}}g 
-\frac{G^2}{c^2}\left\{\partial_t^2\Bigl[\tilde{\mu}_1\tilde{\mu}_2\,\underset{1}{\partial_{(i}} \underset{2}{\partial_{j)}}f\Bigr] 
+ 2 \ddot{\tilde{\mu}}_1\tilde{\mu}_2\,\underset{1}{\partial_{(i}} \underset{2}{\partial_{j)}}f^{12} \right.\nonumber\\ &\quad\quad \left. + 4 \dot{\tilde{\mu}}_1\tilde{\mu}_2 v_1^k\,\underset{1}{\partial_{k(i}} \underset{2}{\partial_{j)}}f^{12} + 2 \tilde{\mu}_1\tilde{\mu}_2 a_1^k\,\underset{1}{\partial_{k(i}} \underset{2}{\partial_{j)}}f^{12} 
+ 2 \tilde{\mu}_1\tilde{\mu}_2 v_1^k v_1^l \,\underset{1}{\partial_{kl(i}} \underset{2}{\partial_{j)}}f^{12}\right\} 
\nonumber\\ &\quad - \frac{G^2}{c^4}\left\{\partial_t^4\Bigl[\tilde{\mu}_1\tilde{\mu}_2\,\underset{1}{\partial_{(i}} \underset{2}{\partial_{j)}}h\Bigr] 
+ \partial_t^2\Bigl[2 \ddot{\tilde{\mu}}_1\tilde{\mu}_2\,\underset{1}{\partial_{(i}} \underset{2}{\partial_{j)}}k^{21} + 4 \dot{\tilde{\mu}}_1\tilde{\mu}_2 v_1^k\,\underset{1}{\partial_{k(i}} \underset{2}{\partial_{j)}}k^{21} \right.\nonumber\\ &\quad\quad \left. + 2 \tilde{\mu}_1\tilde{\mu}_2 a_1^k\,\underset{1}{\partial_{k(i}} \underset{2}{\partial_{j)}}k^{21} + 2 \tilde{\mu}_1\tilde{\mu}_2 v_1^k v_1^l \,\underset{1}{\partial_{kl(i}} \underset{2}{\partial_{j)}}k^{21}\Bigr] 
\right.\nonumber\\ &\quad\quad \left. + \tilde{\mu}_1\tilde{\mu}_2 a_1^m a_2^k \,\underset{1}{\partial_{m(i}} \underset{2}{\partial_{j)k}}k + \tilde{\mu}_1\tilde{\mu}_2 a_1^m v_2^k v_2^l \,\underset{1}{\partial_{m(i}} \underset{2}{\partial_{j)kl}}k + \tilde{\mu}_1\tilde{\mu}_2 v_1^m v_1^n a_2^k\,\underset{1}{\partial_{mn(i}} \underset{2}{\partial_{j)k}}k \right.\nonumber\\ &\quad\quad \left. + \tilde{\mu}_1\tilde{\mu}_2 v_1^m v_1^n v_2^k v_2^l\,\underset{1}{\partial_{mn(i}} \underset{2}{\partial_{j)kl}}k 
+ 2 \tilde{\mu}_1\tilde{\mu}_2 c_1^k\,\underset{1}{\partial_{k(i}} \underset{2}{\partial_{j)}}h^{12} + 8\tilde{\mu}_1\tilde{\mu}_2 b_1^k v_1^l\,\underset{1}{\partial_{kl(i}} \underset{2}{\partial_{j)}}h^{12} \right.\nonumber\\ &\quad\quad \left. + 6\tilde{\mu}_1\tilde{\mu}_2 a_1^k a_1^l\,\underset{1}{\partial_{kl(i}} \underset{2}{\partial_{j)}}h^{12} + 12\tilde{\mu}_1\tilde{\mu}_2 a_1^k v_1^l v_1^m\,\underset{1}{\partial_{klm(i}} \underset{2}{\partial_{j)}}h^{12} \right.\nonumber\\ &\quad\quad \left. + 2\tilde{\mu}_1\tilde{\mu}_2 v_1^k v_1^l v_1^m v_1^n \,\underset{1}{\partial_{klmn(i}} \underset{2}{\partial_{j)}}h^{12}\right\}\,,
\end{align}
where we denote ${}_{1}{\partial_i} = \partial/\partial y_1^i$ and ${}_{2}{\partial_i} = \partial/\partial y_2^i$, with overdots meaning the time derivatives of $\tilde{\mu}_{1,2}$, and where $\bm{v}_{1,2}$ are the velocities, $\bm{a}_{1,2}$ the accelerations, $\bm{b}_{1,2}$ the time-derivatives of accelerations and $\bm{c}_{1,2}$ their second-time derivatives. To the terms given here, which can be called ``interaction''-terms, one must add the corresponding ``self''-terms, proportional to $G^2 \tilde{\mu}_1^2$ or, for instance, $G^2\dot{\tilde{\mu}}_1\tilde{\mu}_1$. The self terms are obtained from the interaction terms by performing the limit of source points $y_2^i\rightarrow y_1^i$ and replacing $\tilde{\mu}_2$ by $\tilde{\mu}_1$. However the terms become more divergent when performing the limit $y_2^i\rightarrow y_1^i$ and we have to carefully use the self-field regularization. With this caveat in mind we have
\begin{equation}\label{eq:Wij2PNself}
\mathop{\hat{W}}_{1}{\!}^\text{self}_{ij} = \lim_{\bm{y}_2\rightarrow \bm{y}_1\atop\tilde{\mu}_2\rightarrow\tilde{\mu}_1} \left[\mathop{\hat{W}}_{1}{\!}^\text{inter}_{ij}\right]\,.
\end{equation}
And, of course we have to add to~\eqref{eq:Wij2PNinter} and~\eqref{eq:Wij2PNself} the terms corresponding to $1 \leftrightarrow 2$.

The expression~\eqref{eq:Wij2PNinter} has been parametrized by means of the hierarchy of elementary kernel functions obeying the following Poisson equations:
\begin{subequations}\label{eq:kernels_def}
	\begin{align}
	\Delta g &= \frac{1}{r_1r_2}\,, \\ \Delta f &= g\,, \qquad\quad \Delta f^{12} = \frac{r_1}{2r_2}\,, \qquad\quad \Delta f^{21} = \frac{r_2}{2r_1}\,, \\ \Delta h &= f\,, \qquad\quad \Delta h^{12} = \frac{r_1^3}{24r_2}\,, \qquad\quad \Delta h^{21} = \frac{r_2^3}{24r_1}\,,\\\Delta k &= \frac{r_1 r_2}{4}\,, \qquad \Delta k^{12} = f^{21}\,, \qquad\qquad \Delta k^{21} = f^{12}\,,
	\end{align}
\end{subequations}
where the numerical coefficients have been introduced for later convenience, \textit{cf.} Ref.~\cite{BFeom}.

\subsubsection{Computing the particular solutions}

Particular solutions for the latter equations, for instance the Fock function given by~\eqref{Fock}, are known from a long time, see \textit{e.g.}~\cite{BDI95, JaraS98, BFeom}. The general structure of these solutions is constituted of two parts: an homogeneous regular solution of the Laplace or iterated Laplace operator multiplied by $\ln S$ (with $S=r_1 + r_2 + r_{12}$) and a specific polynomial of $r_1$, $r_2$ and $r_{12}$. As we seek for particular solutions, we are free to add a global homogeneous solution, for example by adding a numerical constant to the Fock function. This will not affect the true final solution as the proper homogeneous function will be selected by the matching procedure described hereafter. Thus we have chosen to start with the following particular solutions, that differ from those in Appendix A of~\cite{JaraS98} by homogeneous solutions and will be denoted with a hat:
\begin{subequations}\label{eq:kernels_part}
	\begin{align}
	\hat{g} &= \ln S + \frac{197}{810}\,, \\
	\hat{f} &= \frac{1}{12}\biggl[\bigl(r_1^2+r_2^2-r_{12}^2\bigr)\biggl(\ln S - \frac{73}{810}\biggr) + r_1r_{12}+r_2r_{12}-r_1r_2\biggr]\,,\\
	\hat{h} &= \frac{1}{320}\biggl[\Bigl(r_1^4+r_2^4-r_{12}^4-2r_{12}^2(r_1^2+r_2^2)+\frac{2}{3}r_1^2r_2^2\Bigr)\biggl(\ln S - \frac{37}{81}\biggr)\nonumber\\
	&\qquad\quad  + r_1r_2(r_{12}^2-r_1^2-r_2^2)+r_{12}(r_1+r_2)(r_1r_2-r_{12}) + \frac{4}{9}r_1^2r_2^2 + \frac{5}{3}r_{12}(r_1^3+r_2^3)\biggr]\,,\\
	\hat{k} &=\frac{1}{120}\biggl[ \Bigl(r_{12}^4-3r_1^4-3r_2^4+6r_1^2r_2^2+2r_{12}^2(r_1^2+r_2^2)\Bigr)\ln S  \nonumber\\ &\qquad\quad + \frac{21}{10}(r_1^4+r_2^4) -\frac{r_{12}^4}{30} + 3r_{12}(r_1^3+r_2^3) +  (r_1^2+r_2^2)\Bigl(3r_1r_2-\frac{31}{15}r_{12}^2\Bigr) \nonumber\\ &\qquad\quad +r_1r_2r_{12}^2 - \frac{21}{5}r_1^2r_2^2-r_{12}(r_1+r_2)\bigl(r_{12}^2-3r_1r_2\bigr)\biggr]\,,
	\end{align}
\end{subequations}
together with the functions $\hat{f}^{12}$, $\hat{h}^{12}$ and $\hat{k}^{12}$ obtained by exchanging the field point ${\bf x}$ with the source point ${\bf y_1}$:
\begin{equation}\label{eq:f12part}
\hat{f}^{12} = \hat{f}\Big|_{{\bf x}\longleftrightarrow{\bf y_1}}\,,\qquad\hat{h}^{12} = \hat{h}\Big|_{{\bf x}\longleftrightarrow{\bf y_1}}\,,\qquad\hat{k}^{12} = \hat{k}\Big|_{{\bf x}\longleftrightarrow{\bf y_1}}\,,
\end{equation}
and similarly $\hat{f}^{21}$, $\hat{h}^{21}$ and $\hat{k}^{21}$ obtained by exchanging ${\bf x}$ and ${\bf y_2}$.

It is straightforward to check that the kernel functions ~\eqref{eq:kernels_part} and~\eqref{eq:f12part} satisfy the constitutive relations~\eqref{eq:kernels_def} and also, in addition, the relations
\begin{equation}\label{eq:extra}
\Delta_1\hat{f}^{12} = \hat{g}\,,\qquad  \Delta_1\hat{h}^{12} = \hat{f}^{12}\,,\qquad  \Delta_1\hat{k} = \hat{f}^{21}\,,\qquad  \Delta_1\hat{k}^{21} = \hat{f}\,,
\end{equation}
together with the relations obtained from $1\leftrightarrow 2$. The extra relations~\eqref{eq:extra} are specifically true for the homogeneous solutions chosen in Eqs.~\eqref{eq:kernels_part}. All those relations suggest that there is an underlying algebra relating those particular kernel functions to higher order, and thus that we should be able to compute a particular solution to the general Poisson equation $\Delta \varphi_{nm} = r_1^{2n-1}r_2^{2m-1}$, with $(n,m) \in \mathbb{N}^2$. This would lead to the knowledge at all even PN orders of the quadratic potentials $\hat{W}_{ij}$, $\hat{Z}_{ij}$ and $\hat{R}_i$, up to the possible odd-odd couplings.

\subsubsection{Matching procedure}

We shall now define from the previous particular solutions some ``matched'' solutions, in such a way that a certain matching equation is fulfilled. It is convenient to define first some associated functions that obey d'Alembertian rather than Poisson equations, namely
\begin{equation}\label{eq:kernels_def}
\Box \mathcal{G} = \frac{1}{r_1r_2}\,, \qquad
\Box \mathcal{F}^{12} = \frac{r_1}{2r_2}\,, \qquad
\Box \mathcal{K} = \frac{r_1r_2}{4}\,, \qquad
\Box \mathcal{H}^{12} = \frac{r_1^3}{24r_2}\,.
\end{equation}
Performing a PN expansion we transform these wave equations~\eqref{eq:kernels_def} into Poisson like equations, and recover the definitions~\eqref{eq:kernels_def} of the previous Poisson kernels, with
\begin{subequations}\label{eq:kernels_def_PN_exp}
	\begin{align}
	& \mathcal{G} = g + \frac{1}{c^2}\,\partial_t^2 f+ \frac{1}{c^4}\,\partial_t^4 h + \mathcal{O}(c^{-6})\,,\\
	& \mathcal{F}^{12} = f^{12} + \frac{1}{c^2}\,\partial_t^2 k^{21}+ \mathcal{O}(c^{-4})\,,\\
	& \mathcal{H}^{12} = h^{12} + \mathcal{O}(c^{-2})\,,\\
	& \mathcal{K} = k + \mathcal{O}(c^{-2})\,.
	\end{align}
\end{subequations}
In order to have the correct prescription for the inverse d'Alembertian, we have to match the particular solutions~\eqref{eq:kernels_part} to the far-zone. This is taken into account by adding some specific homogeneous solutions to Eqs.~\eqref{eq:kernels_part} (see~\cite{BFeom} for a more detailed discussion).

Let $\Box\Psi = S({\bf x},t)$ be one of the wave equations~\eqref{eq:kernels_def}, with some non-compact support source $S$. The matching equation states that the multipolar expansion of the solution $\Psi$, denoted $\mathcal{M}(\Psi)$, should satisfy
\begin{equation}\label{eq:kernels_matching}
\mathcal{M}(\Psi) = \underset{B=0}{\text{FP}} \,\Box^{-1}_R\biggl[\biggl(\frac{r}{r_0}\biggr)^B\mathcal{M}(S)\biggr] -  \frac{1}{4\pi}\sum_{\ell=0}^{+\infty}\frac{(-)^\ell}{\ell !}\,\partial_L\left(\frac{1}{r}\,\mathcal{S}_L(t-r/c)\right)\,,
\end{equation} 
where the first term is a solution of the multipole expanded wave equation $\Box \mathcal{M}(\Psi)=\mathcal{M}(S)$, defined by means of the finite part procedure at $B=0$ applied to the standard retarded integral operator $\Box^{-1}_R$, and the second term is a homogeneous solution constructed out of the multipole moments: 
\begin{equation}\label{eq:source_matching}
\mathcal{S}_L(u) = \underset{B=0}{\text{FP}} \int \dd^3 \mathbf{x} \left(\frac{r}{r_0}\right)^B\!x_L\,S({\bf x},u)\,,
\end{equation}
which themselves integrate over the source $S$. Suppose now that we know a particular solution of the wave equation, say $\hat{\Psi}$ such that $\Box\hat{\Psi} = S$. We look for a homogeneous solution $\Psi^\text{hom}$ such that $\Psi=\hat{\Psi}+\Psi^\text{hom}$ satisfies Eqs.~\eqref{eq:kernels_matching}--\eqref{eq:source_matching}. Since the homogeneous solution is directly in the form of a multipole expansion, $\Psi^\text{hom}=\mathcal{M}(\Psi^\text{hom})$, we obtain the following relation:
\begin{equation}\label{eq:recipe_matching}
\Psi^\text{hom} = \widetilde{\Box_R^{-1}}\mathcal{M}\left(S\right) 
- \mathcal{M}\bigl(\hat{\Psi}\bigr) -  \frac{1}{4\pi}\sum_{\ell=0}^{+\infty}\frac{(-)^\ell}{\ell !}\,\partial_L\left(\frac{1}{r}\,\mathcal{S}_L(t-r/c)\right)\,,
\end{equation} 
where $\widetilde{\Box_R^{-1}}$ denotes the Hadamard-regularized retarded integral in~\eqref{eq:kernels_matching}.\footnote{Actually, in all this construction we are interested only in the conservative dynamics and we can use instead the symmetric integral operator $\widetilde{\Box_\text{sym}^{-1}}$.} The previous recipe~\eqref{eq:recipe_matching} completely determines the homogeneous solution, since all the terms in the right-hand side are known.

We have applied this method to determine all the relevant homogeneous solutions in the kernel functions $g$, $f$, \textit{etc}. For example, expanding the last term of Eq.~\eqref{eq:recipe_matching}, and identifying the relevant PN orders, it comes 
\begin{subequations}\label{eq:gfhom}
	\begin{align}
	g^\text{hom} &= \widetilde{\Delta^{-1}} \mathcal{M}\Bigl(\frac{1}{r_1r_2}\Bigr) - \mathcal{M}\bigl(\hat{g}\bigr)\,,\\
	f^\text{hom} &= \widetilde{\Delta^{-2}} \mathcal{M}\Bigl(\frac{1}{r_1r_2}\Bigr) - \mathcal{M}\bigl(\hat{f}\bigr) +\frac{1}{4}\bigl(r \,Y-n^i Y_i\bigr)\,,
	\end{align}
\end{subequations}
where we have denoted (notice the STF multipole factor $\hat{x}_L$)
\begin{equation}
Y_L = - \frac{1}{2\pi}\,\underset{B=0}{\text{FP}} \int \dd^3 \mathbf{x}\left(\frac{r}{r_0}\right)^B \frac{\hat{x}_L}{r_1 r_2} = \frac{r_{12}}{\ell+1} \sum_{m=0}^\ell y_1^{\langle M}y_2^{L-M\rangle}\,.
\end{equation}

We emphasize that although we introduced the ``Poisson'' kernels $g$, $f$, $f^{12}$, \textit{etc.} for convenience, it is better to consider the ``d'Alembertian'' kernels ($\mathcal{G}$, $\mathcal{F}^{12}$, \textit{etc.}) as more fundamental quantities. Indeed, when working with the Poisson kernels, we have to take into account the fact that the FP operation at $B=0$ and the inverse Laplacian do not commute, thus for instance
\begin{equation}\label{eq:noncommute}
\widetilde{\Delta^{-1}}\mathcal{M}\left(f\right) \neq 
\widetilde{\Delta^{-3}}\mathcal{M}\Bigl(\frac{1}{r_1r_2}\Bigr)\,,
\end{equation}
and the matching procedure in more complicated. Nonetheless, after adding some correction terms accounting \textit{e.g.} for the fact~\eqref{eq:noncommute}, the result comes out the same. Note however that the effect~\eqref{eq:noncommute} emerges at 2PN order, and only affects the computation of $h$.

Having matched the particular solutions, one has constructed the good prescription for the elementary kernels. The end results for $g$ and $f$ are for example
\begin{subequations}
	\begin{align}
	g =& \ln\left(\frac{S}{2r_0}\right)-1 \,,\\
	f =& \frac{r_1r_2\,{\bf n}_1\cdot{\bf n}_2}{6} \left[\ln\left(\frac{S}{2r_0}\right)+\frac{1}{6}\right] 
	+\frac{r_{12}r_1+r_{12}r_2-r_1r_2+2r\,{\bf n}\cdot({\bf y_1}+{\bf y_2})-3r^2}{12}\,.
	\end{align}
\end{subequations}
We do not display the other kernels since their homogeneous solutions are rather complicated, and we have described the general procedure to obtain them.

In the end, by means of this technique, we have obtained in the whole space (and 3 dimensions) the potential $\hat{W}_{ij}$ at 2PN order following Eqs.~\eqref{eq:Wij2PNinter}--\eqref{eq:Wij2PNself}, as well as the potentials $\hat{Z}_{ij}$ and $\hat{R}_{i}$ at 1PN order. Notably, this permits to check the value of the trace $\hat{W}=\hat{W}_{ii}$ at 2PN order evaluated at point 1, with respect to the direct calculation reported in Sec.~\ref{sec:potpart}.

To end with, let us mention a delicate issue that arises in the latter comparison. It comes from the fact that the particular solution of Eq.~\eqref{eqWii},
\begin{equation}\label{eq:Wiipart}
\hat{W}^{\text{part}} = -V^2/2+ \widetilde{\Box^{-1}_R} \left[8\pi G\left(\sigma_{ii}-\frac{1}{2}\sigma\,V\right) - \frac{1}{c^2}(\partial_t V)^2\right]\,,
\end{equation}
is not \textit{a priori} the matched one. Indeed, applying the operator $\widetilde{\Box^{-1}_R}$ to the right-hand side of the latter equality leads to an almost identical expression for $\hat{W}$, but where $V^2$ is now replaced by $\widetilde{\Box^{-1}_R} \Box V^2$. Thus, the homogeneous solution to be added to $\hat{W}^{\text{part}}$ is given by the ``commutator''
\begin{equation}\label{eq:Wiihom}
\hat{W}^{\text{hom}}= \bigl[\widetilde{\Box^{-1}_R}, \Box\bigr]\Bigl(-\dfrac{V^2}{2}\Bigr)\,,
\end{equation}
which does not vanish in general. More precisely, by means of techniques similar to those used in the derivation of the formula~\eqref{JL}, one can show that, for any function admitting an asymptotic expansion with general terms of the form $f_{p,q}(\mathbf{n}) r^{p} (\ln r)^q$ (for $p\le p_\text{max}$) near infinity, one has
\begin{subequations}\label{eq:commutator}
\begin{align}
&\bigl[\widetilde{\Box^{-1}_R},\Box\bigr]F = \sum_{\ell=0}^{+\infty}
  \frac{(-)^\ell}{\ell!} \sum_{s=0}^{+\infty}
\Delta^{-s} \hat{x}_L  \left(\frac{\dd}{\dd t}\right)^{2s}
  \hat{f}_L\, , \\
& \text{with}\quad
 \hat{f}_L = \!\!\!\!\!\!\sum^{+\infty}_{k=\max(0,\,\ell-p_\text{max})}
  \frac{(-)^k}{k!!(k-2\ell-1)!!} \left( \frac{\dd}{\dd t}\right)^k
  \left[(2k-2\ell-1) \hat{f}^L_{k-\ell,0} - \hat{f}^L_{k-\ell,1} \right]\, .
\end{align}
\end{subequations}
Here, the coefficients $\hat{f}^L_{p,q}$ are those of the decomposition of ${f}^L_{p,q}(\mathbf{n})$ in the spherical harmonic functions $\hat{n}_L=\text{STF} (n_{i_1}\cdots n_{i_\ell})$ for $\ell\geqslant 0$, while $\Delta^{-s} \hat{x}^L$ represents the solution
\begin{align}
\Delta^{-s}\hat{x}_L =
  \frac{\Gamma(\ell+3/2)}{2^{2s}\Gamma(s+1)\Gamma(s+\ell+3/2)} \hat{x}^L r^{2s}\,,
\end{align}
of the iterated Poisson equation $\Delta^s P =\hat{x}_L$. Working with the symmetric Green function, we find by application of the formula~\eqref{eq:commutator}:
\begin{align}
\hat{W}^{\text{sym hom}} =
-\frac{G^2 m}{c^4} \left[\frac{1}{6} I^{(4)} + \ddot{\tilde{\mu}} c^2\right]\, ,
\end{align}
where we have introduced for convenience the total mass $m=m_1+m_2$, the effective mass $\tilde{\mu}=\tilde{\mu}_1+\tilde{\mu}_2$, and the Newtonian moment of inertia $I=m_1 \bm{y}_1^2 +1\leftrightarrow 2$ of the binary system.

\subsection{Potentials at infinity}
\label{sec:PotAtInf}

Finally, as we have seen in Sec.~\ref{sec:MQPot} many of the potentials are required at $r\to+\infty$ in order to compute the surface terms. For these we need only their contributions in 3 dimensions. For the potentials that are already known for any $\mathbf{x} \in \mathbb{R}^3$, we just perform their expansions when $r\to+\infty$. In particular, we have computed in the whole space the potential $\hat{W}_{ij}$ at 2PN order in Sec.~\ref{sec:Wij2PN}, which is a great help for the calculation of surface terms. However, other potentials are not known, namely $\hat{X}$ at 1PN order, and $\hat{T}$, $\hat{Y}_i$ as well as $\hat{M}_{ij}$ at Newtonian order. For those potentials, we proceed differently.

To obtain the expansion when $r\to+\infty$ of the potential $P$, we consider the equation that it satisfies, $\Box P = S$, where the source $S$ is known, for instance given by Eq.~\eqref{Mij3d}. For a potential at the 4PN order, the equation reduces to a Poisson equation $\Delta P = S$. Then, we compute the asymptotic or multipole expansion $\mathcal{M}(P)$ from the source $S$ and its multipole expansion $\mathcal{M}(S)$ by the Poisson-like version of the matching formula~\eqref{eq:kernels_matching}, namely
\begin{equation}\label{multP}  
\mathcal{M}(P) = \mathop{\mathrm{FP}}_{B=0} \Delta^{-1} \biggl[\left(\frac{r}{r_0}\right)^B \mathcal{M}(S)
\biggr]-\frac{1}{4 \pi} \sum_{\ell=0}^{+\infty}
\frac{(-)^\ell}{\ell!} \partial_L 
\left(\frac{1}{r}\right) \mathcal{P}_L(t) \,, 
\end{equation}
where
\begin{equation}\label{calPL} 
\mathcal{P}_L(u) = \mathop{\mathrm{FP}}_{B=0} \int \dd^3\mathbf{x}~  
\left(\frac{r}{r_0} \right)^B x_L\,S(\mathbf{x},u)\,.
\end{equation}
The first term in~\eqref{multP} corresponds to integrating the multipole expansion of the known source term by term, while the second term represents an homogeneous solution parametrized by computable multipole moments~\eqref{calPL}.
In practice, to compute the first term in~\eqref{multP}, we apply the following formulas:
\begin{subequations}\label{EqPotInf}
\begin{align}
\widetilde{\Delta^{-1}}\Bigl[ r^{\alpha} \hat{n}_{L}\Bigr] &= \frac{r^{\alpha+2} \hat{n}_{L}}{(\alpha-\ell+2)(\alpha+\ell+3)}\,, \quad\text{for $\alpha \in \mathbb{C}\setminus\bigl\{\ell-2, -\ell-3\bigr\}$} \,, \\
\widetilde{\Delta^{-1}}\Bigl[ r^{\ell-2} \hat{n}_{L}\Bigr] &= \frac{1}{2 \ell+1} \left[\ln\left(\frac{r}{r_0}\right) - \frac{1}{2 \ell +1} \right] r^\ell \hat{n}_{L} \,,\\
\widetilde{\Delta^{-1}}\Bigl[ \frac{\hat{n}_{L}}{r^{\ell+3}}\Bigr] &= -\frac{1}{2 \ell+1} \left[\ln\left(\frac{r}{r_0}\right) + \frac{1}{2 \ell +1} \right]  \frac{\hat{n}_{L}}{r^{\ell+1}} \,,
\end{align}
\end{subequations}
where we abbreviated $\widetilde{\Delta^{-1}}=\mathrm{FP}_{B=0}\Delta^{-1}(r/r_0)^B$. Besides, for $\hat{X}$ at 1PN, we face the integration of terms involving also some logarithm $\ln(r/r_0)$. For that, we have, in the generic case $\alpha \in \mathbb{C}\setminus\{\ell-2, -\ell-3\}$,
\begin{equation}\label{intlog} 
\widetilde{\Delta^{-1}}\biggl[\ln\left(\frac{r}{r_0} \right) r^{\alpha} \hat{n}_{L}\biggr] = \frac{r^{\alpha+2} \hat{n}_{L}}{(\alpha-\ell+2)(\alpha+\ell+3)}\biggl[\ln\left(\frac{r}{r_0} \right) - \frac{2\alpha+5}{(\alpha-\ell+2)(\alpha+\ell+3)} \biggr]\,.
\end{equation}

\section{The 4PN mass quadrupole for circular orbits}
\label{sec:resultMQ}

We have applied the method described in this paper to compute the source mass quadrupole moment at the 4PN order in the case of circular orbits. As for the Fokker Lagrangian computation of the equations of motion~\cite{BBBFMa}, we first used Hadamard's partie finie to cure the UV divergences, and obtain a first result depending on $\ln s_1$ and $\ln s_2$. Then we computed the difference between the DR and the Hadamard partie finie regularization for the UV divergences, and obtained a new result free of $\ln s_1$ and $\ln s_2$ but containing poles in $1/\varepsilon$, and also the DR scale $\ell_0$.

Let us recall that these poles should cancel out when expressing physical observables such as the energy flux or the orbital phase of the system, but can still be present in intermediate non gauge invariant results such as the equations of motion or the source multipole moments. However, in that case, it is extremely useful to remove the UV poles by applying a shift of the particle's trajectories. This provides an important test of the result and also a substantial simplification. Indeed, at the 3PN order, it was already shown that applying the same shift as used for the 3PN equations of motion to the 3PN source mass quadrupole moment, indeed consistently removes all the UV poles~\cite{BDE04}.

At the 4PN order, the situation is a bit more complicated. The shift that we applied to the Fokker Lagrangian in order to obtain the final result for the 4PN equations of motion as obtained in~\cite{BBBFMc, MBBF17}, and from which we derived all the conserved quantities in~\cite{BBFM17}, is composed of three terms:
\begin{enumerate}
\item The shift $\bm{\xi}_{1,2}$ given in the Appendix C of~\cite{BBBFMa}\footnote{There are some missing terms in the equations~(C3) of~\cite{BBBFMa}; the correct expression, also taking into account the final determination of the ambiguity parameters~\cite{BBBFMc, MBBF17}, is given in Eqs.~\eqref{shift4PNxidecomp}--\eqref{shift4PNxi} below.} and which removed all the UV-type $1/\varepsilon$ poles in the Fokker Lagrangian; 
\item The shift $\bm{\chi}_{1,2}$ that was applied in~\cite{BBBFMc} and removes all the IR-type $1/\varepsilon$ poles of the Fokker Lagrangian (this shift has not yet been published in full form); 
\item Finally, the shift $\bm{\eta}_{1,2}$ given in the Appendix A of~\cite{BBFM17} that does not contain any pole and was merely used for convenience.
\end{enumerate}

For completeness, we provide in Appendix~\ref{app:shift} the full expressions of the shifts $\bm{\xi}_{1,2}$ and $\bm{\eta}_{1,2}$. Note that the shift $\bm{\chi}_{1,2}$ will not be used in the present paper since we treat the IR divergences by means of the Finite Part regularization instead of DR. However we intend to consider the shift $\bm{\chi}_{1,2}$ in future work when we investigate the problem of IR divergences in the 4PN mass quadrupole moment.

We have applied the sum of the shifts $\bm{\xi}_{1,2}$ and $\bm{\eta}_{1,2}$ to the 4PN quadrupole moment and checked that all the UV-type poles $1/\varepsilon$ (as well as the usual concomitant constants such as Euler's constant $\gamma_\text{E}$) cancel out as they should. Recall that the shifts have been determined from the separate calculation of the Fokker Lagrangian and equations of motion. Furthermore we have seen in Sec.~\ref{sec:potpart} that at the 4PN order some of the potentials needed to control the compact support terms do contain poles. These poles combine with those coming from the DR of the volume integrals of non-compact support terms in Sec.~\ref{sec:dimregUV}. The proper cancellation of all the poles constitutes a robust check of our UV DR computations, and a major confirmation that we understand the connection between the conservative equations of motion and multipole moments within the framework of the MPM-PN approach.

The next steps are to reduce our result to the frame of the center of mass (CM) and then to the case of quasi circular orbits. We only need the 3PN expressions of the CM coordinates, and the 3PN equations of motion for circular orbits, in order to express the mass quadrupole moment at the 4PN order in the CM frame for circular orbits. Therefore, even if our result does not yet use DR for the IR divergences, we can still consistently express it in the CM frame for circular orbits --- as the 3PN dynamics can be derived using the Finite Part regularization for the IR, and as the IR shift $\bm{\chi}_{1,2}$ only starts at the 4PN order. The result is then much more compact and is given as follows.

Finally, the UV-shifted mass quadrupole moment for circular orbits at the 4PN order, where the applied shifts $\bm{\xi}_{1,2}$ and $\bm{\eta}_{1,2}$ are given in Appendix~\ref{app:shift}, reads
\begin{equation}\label{Iij}
I_{ij} = \mu \left(A \, x_{\langle i}x_{j \rangle}+B \,
  \frac{r^2}{c^2}v_{\langle i}v_{j \rangle} + \frac{G^2
    m^2\nu}{c^5r}\,C\,x_{\langle i}v_{j \rangle}\right) + \mathcal{O}\left(\frac{1}{c^{9}}\right)\,,
\end{equation}
where the terms up to the 4PN order are explicitly given by\footnote{The terms $C$ represent the time-odd 2.5PN and 3.5PN contributions and are given here for completeness.}

\begin{subequations}\label{IijABC}
  \begin{align}
A &= 1
 + \gamma \biggl(- \frac{1}{42}
 -  \frac{13}{14} \nu \biggr)
 + \gamma^2 \biggl(- \frac{461}{1512}
 -  \frac{18395}{1512} \nu
 -  \frac{241}{1512} \nu^2\biggr)
\nonumber\\
& \quad + \gamma^3 \biggl(\frac{395899}{13200}
 -  \frac{428}{105} \ln\biggl(\frac{r}{r_{0}{}} \biggr)
 + \biggl[\frac{3304319}{166320}
 -  \frac{44}{3} \ln\biggl(\frac{r}{r'_{0}}\biggr) \biggr]\nu
 + \frac{162539}{16632} \nu^2 + \frac{2351}{33264} \nu^3
\biggr)
\nonumber\\
 &  \quad + \gamma^4 \biggl (- \frac{1023844001989}{12713500800}
      + \frac{31886}{2205} \ln\biggl(\frac{r}{r_{0}{}} \biggr)
      + \biggl[- \frac{18862022737}{470870400}
 -  \frac{2783}{1792} \pi^2 \nonumber\\
 & \qquad \quad -  \frac{24326}{735} \ln\biggl(\frac{r}{r_{0}{}} \biggr)
 + \frac{8495}{63} \ln\biggl(\frac{r}{r'_{0}} \biggr)\biggr] \nu
      + \biggl[\frac{1549721627}{40360320} + \frac{44909}{2688} \pi^2-  \frac{4897}{21}
      \ln\biggl(\frac{r}{r'_{0}} \biggr)\biggr]\nu^2\nonumber\\
  & \qquad \quad - \frac{22063949}{5189184} \nu^3 + \frac{71131}{314496} \nu^4
\biggl)\,, \\
B &=\frac{11}{21}
 -  \frac{11}{7} \nu
 + \gamma \biggl(\frac{1607}{378}
 -  \frac{1681}{378} \nu
 + \frac{229}{378} \nu^2\biggr) \nonumber\\
  & \quad + \gamma^2 \biggl(- \frac{357761}{19800}
      + \frac{428}{105} \ln\biggl(\frac{r}{r_{0}{}} \biggr)
 -  \frac{92339}{5544} \nu
 + \frac{35759}{924} \nu^2
 + \frac{457}{5544} \nu^3 \biggr)  \nonumber\\
 & \quad + \gamma^3 \biggl(\frac{17607264287}{1589187600} -  \frac{4922}{2205}
 \ln\biggl(\frac{r}{r_{0}{}} \biggr) + \biggl[\frac{5456382809}{529729200}
 + \frac{143}{192} \pi^2
   -  \frac{1714}{49} \ln\biggl(\frac{r}{r_{0}{}} \biggr)
   - \frac{968}{63} \ln\biggl(\frac{r}{r'_{0}} \biggr)\biggr] \nu  \nonumber\\
 & \qquad \quad + 
 \biggl[\frac{117172607}{1681680} 
 -  \frac{41}{24} \pi^2
 + \frac{968}{21} \ln\biggl(\frac{r}{r'_{0}} \biggr)\biggr] \nu^2
 -  \frac{1774615}{81081} \nu^3
 -  \frac{3053}{432432} \nu^4 \biggl)\,, \\
C &= \frac{48}{7} + \gamma \left(-\frac{4096}{315} - \frac{24512}{945}\nu \right)\,.
  \end{align}
\end{subequations}
Let us remind that this result has been obtained using the FP prescription for the IR divergences, and the DR for the UV divergences. In future work, we shall switch to DR for the IR divergences as well, in the form of the regularization $B\varepsilon$ which has been successfully applied recently to the 4PN equations of motion~\cite{MBBF17}. 

In our notation $\gamma$ is a PN parameter defined as 
\begin{equation}\label{gamma}
\gamma = \frac{G m}{r c^2}\,,
\end{equation}
where $r=\vert\bm{y}_1-\bm{y}_2\vert$ is the radial separation in harmonic coordinates, $\bm{x}=\bm{y}_1-\bm{y}_2$ is the relative distance and $\bm{v}=\bm{v}_1-\bm{v}_2$ is the relative velocity. The total mass is $m=m_1+m_2$, and the reduced mass $\mu$ and symmetric mass ratio $\nu$ are given by 
\begin{equation}\label{nu}
\nu = \frac{\mu}{m} = \frac{m_1m_2}{(m_1+m_2)^2}\,.
\end{equation}
Two constants parametrize the logarithmic terms of~\eqref{IijABC}. We have the constant $r_0$ which was introduced in the Finite Part regularization for the IR, see Eq.~\eqref{ValueILGeneral}. Then there is the constant $r_0'$ associated with the UV regularization, and which has been introduced by definition through the shift $\bm{\xi}_{1,2}$ in Eqs.~\eqref{shift4PNxi}.\footnote{In previous works on the 3PN/4PN equations of motion in harmonic coordinates, two gauge constants $r'_1$ and $r'_2$ in the logarithms were considered for the UV divergences instead of one~\cite{BFeom, BI03CM, BBBFMa}. In the CM frame this yielded the two convenient combinations (with $X_{1,2}=m_{1,2}/m$)
\begin{align*}
\ln r'_0 = X_1 \ln r'_1 + X_2 \ln r'_2\,,\\
\ln r''_0 = \frac{X_1^2 \ln r'_1 - X_2^2 \ln r'_2}{X_1-X_2}\,,
\end{align*}
with $r''_0$ entering specifically the expression of the particle's positions in the CM frame at the 3PN order~\cite{BI03CM}. Because of the factor $(X_1-X_2)^{-1}$ in $\ln r''_0$ there is an apparent divergence when the two masses are equal, but of course it is compensated by a factor $X_1-X_2$ in the CM relations. In the present paper, we make the choice $r_1' = r_2'$, which avoids such spurious divergence and has the advantage that the particle's positions are exactly $y_1^i=-y_2^i$ when the masses are equal. Hence we have only one UV constant $r'_0 = r''_0 = r'_1 = r'_2$.} 

The result~\eqref{Iij}--\eqref{IijABC} extends to the 4PN order the expression of the mass quadrupole moment that was known at the 3.5PN order~\cite{BIJ02, BI04mult, BFIS08, FMBI12}. It constitutes an important step in our program of completing the waveform and phase evolution of compact binary systems at the 4PN order, but remind that a thorough investigation of the IR divergences at the level of the 4PN multipole moment is still required and postponed to future work.  

\acknowledgments

The authors would like to thank Laura Bernard for providing us the files of the different shifts used for the 4PN equations of motion. We also thank Alejandro Boh\'e for discussions at an early stage of this work. This research made use of the computer facility Horizon Cluster funded by the Institut d'Astrophysique de Paris. We thank St\'ephane Rouberol for running smoothly this cluster for us.

\appendix

\section{The 4PN metric as a function of potentials}
\label{app:PNpotentials}

For insertion into the mass quadrupole moment we need the metric components developed up to order $c^{-8}$, $c^{-7}$ and $c^{-8}$ in the $00$, $0i$ and $ij$ components of the metric. This means the full 4PN accuracy for the spatial $ij$ components, but only the 3PN accuracy for the other $00$ and $0i$ components. However, for completeness and future use, we present the complete 4PN metric with our choice of parametrization, in $d$ dimensions since the metric is also used to compute the corrections due to DR. For the usual covariant components $g_{\mu\nu}$:
\begin{subequations}\label{metric4PN}
\begin{align}
g_{00} &= -1
 + 2 \frac{V}{c^{2}}
 + \frac{1}{c^{4}} \biggl[- \frac{4 (d -3) K}{d - 2}
 - 2 V^2\biggr] \nonumber\\
 & + \frac{1}{c^{6}} \biggl[\frac{8 (d -3) K V}{d - 2}
 + \frac{4}{3} V^3
 + 8 V_{a} V^{a}
 + 8 \hat{X}\biggr]
 \nonumber\\
 & + \frac{1}{c^{8}} \biggl[- \frac{8 (d -3)^2 K^2}{(d -2)^2}
 + 32 \hat{T} -  \frac{8 (d -3) K V^2}{d - 2}
 -  \frac{2}{3} V^4 + 32 \hat{R}_{a} V_{a}
 + V (-16 V_{a} V_{a}
 - 16 \hat{X})\biggr]\nonumber\\
& + \frac{1}{c^{10}} \biggl(64 \hat{P}
 + 32 \hat{R}_{a} \hat{R}_{a} + \frac{16 (d -3)^2 K^2 V}{(d -2)^2}
 - 64 \hat{T} V
 + \frac{4}{15} V^5
 -  \frac{32 (d -3) \hat{R}_{a} V V_{a}}{d - 2}\nonumber\\
&\qquad + V^2 \biggl[\frac{8 (9 - 10 d + 3 d^2) V_{a} V_{a}}{(d -2)^2}
 + 16 \hat{X}\biggr]
 + K \biggl[\frac{16 (d -3) V^3}{3 (d -2)}\nonumber\\
&\qquad + \frac{16 (d -3)^2 V_{a} V_{a}}{(d -2)^2}
 + \frac{32 (d -3) \hat{X}}{d - 2}\biggr]
 + 64 V_{a} \hat{Y}_{a}\biggr) + \mathcal{O}\left(\frac{1}{c^{12}}\right)\, , \\[1ex]
g_{0i} &= -4 \frac{V_{i}}{c^3}
 + \frac{1}{c^{5}} \biggl[-8 \hat{R}_{i}
 + \frac{4 (d -3) V V_{i}}{d - 2}\biggr]
 \nonumber\\
 & + \frac{1}{c^{7}} \biggl[\frac{8 (d -3) \hat{R}_{i} V}{d - 2}
  -  \frac{8 (d -3)^2 K V_{i}}{(d -2)^2} -  \frac{4 (5 - 4 d + d^2) V^2 V_{i}}{(d -2)^2}
 - 8 V_{a} \hat{W}_{ia}
 - 16 \hat{Y}^{i}\biggr]
 \nonumber\\
 & + \frac{1}{c^{9}} \biggl(-32 \hat{N}_{i} + \hat{R}_{i} \biggl[- \frac{16 (d -3)^2 K}{(d -2)^2}
 -  \frac{4 (d -3)^2 V^2}{(d -2)^2}\biggr]
 + \frac{8 (d -3)^3 K V V^{i}}{(d -2)^3}\nonumber\\
&\qquad + \frac{8 (d -3) (3 - 3 d + d^2) V^3 V_{i}}{3 (d -2)^3}
 - 16 \hat{R}_{a} \hat{W}_{ia}
 + V_{i} \biggl[\frac{16 (-5 + 2 d) V_{a} V_{a}}{d - 2}\nonumber\\
&\qquad + \frac{16 (d -3) \hat{X}}{d - 2}\biggr]
 + V \biggl[\frac{8 (d -3) V_{a} \hat{W}_{ia}}{d - 2}
 + \frac{16 (d -3) \hat{Y}_{i}}{d - 2}\biggr]
 - 32 V_{a} \hat{Z}_{ia}\biggr) \nonumber\\
 & + \mathcal{O}\left(\frac{1}{c^{11}}\right) \, , \\[1ex]
g_{ij} &= \delta_{ij}
 + \frac{1}{c^{2}} \frac{2 V}{d - 2}\delta_{ij}
 + \frac{1}{c^{4}} \biggl(\delta_{ij} \biggl[- 
\frac{4 (d -3) K}{(d -2)^2}
 + \frac{2 V^2}{(d -2)^2}\biggr]
 + 4 \hat{W}_{ij}\biggr)
 \nonumber\\
 & + \frac{1}{c^{6}} \biggl(-16 V_{i} V_{j} + \frac{8 V \hat{W}_{ij}}{d - 2}
 + \delta_{ij} \biggl[- \frac{8 (d -3) K V}{(d -2)^3}
 + \frac{4 V^3}{3 (d -2)^3}
 + \frac{8 V_{a} V_{a}}{d - 2}
 + \frac{8 \hat{X}}{d - 2}\biggr] + 16 \hat{Z}_{ij}\biggr)
 \nonumber\\
 & + \frac{1}{c^{8}} \Biggl[32 \hat{M}_{ij}
 - 32 \hat{R}_{(i} V_{j)}
 -  \frac{16 (d -3) K \hat{W}_{ij}}{(d -2)^2}
 + \frac{8 V^2 \hat{W}_{ij}}{(d -2)^2}\nonumber\\
&\qquad + 8 \hat{W}_{ia} \hat{W}_{ja}
 + \delta_{ij} \biggl(- \frac{4 (d -3)^3 K^2}{(d -2)^4}
 -  \frac{32 \hat{M}}{d - 2}
 + \frac{32 \hat{T}}{d - 2}
 -  \frac{8 (d -3) K V^2}{(d -2)^4}\nonumber\\
&\qquad + \frac{2 V^4}{3 (d -2)^4}
 + \frac{32 \hat{R}_{a} V_{a}}{d - 2}
 -  \frac{4 \hat{W}_{ab} \hat{W}_{ab}}{d - 2}
 + V \biggl[\frac{16 V_{a} V_{a}}{(d -2)^2}
 -  \frac{8 (d -3) \hat{X}}{(d -2)^2}\biggr]\biggr)\nonumber\\
&\qquad + V (- \frac{32 V_{i} V_{j}}{d - 2}
 + \frac{32 \hat{Z}_{ij}}{d - 2})\Biggr] + \mathcal{O}\left(\frac{1}{c^{10}}\right)\,.
\end{align}
\end{subequations}
Up to 3PN order there are nine potentials $V$, $V_i$, $K$, $\hat W_{ij}$, $\hat R_i$, $\hat X$, $\hat Z_{ij}$, $\hat Y_i$, $\hat T$, which agree with our previous definitions in~\cite{BFeom, BDE04, BDEI05dr}. The new generation of potentials at the 4PN order are denoted $\hat P$, $\hat N_i$ and $\hat M_{ij}$. Since we need for the 4PN quadrupole moment the metric components in gothic form $h^{\mu\nu}$ up to order $c^{-8}$, $c^{-7}$ and $c^{-8}$, we also present them here:
\begin{subequations}\label{metricgothic}
	\begin{align}
	h^{00} &= - \frac{1}{c^{2}}\frac{2 (d - 1) V}{d - 2} + \frac{1}{c^{4}} \left[\frac{4 (d - 3) (d - 1) K}{(d - 2)^2} -  \frac{2 (d - 1)^2 V^2}{(d - 2)^2} - 2 \hat W\right] \nonumber\\
	&\qquad + \frac{1}{c^{6}} \left[- \frac{4 (d - 1)^3 V^3}{3 (d - 2)^3} + \frac{8 (d - 3) V_{a} V_{a}}{d - 2} + V (\frac{8 (d - 3) (d - 1)^2 K}{(d - 2)^3}\right.\nonumber\\
	&\qquad\left. -  \frac{4 (d - 1) \hat W}{d - 2}) -  \frac{8 (d - 1) \hat X}{d - 2} - 8 \hat Z\right]\nonumber \\
	&\qquad + \frac{1}{c^{8}} \left[\frac{2 d \hat W_{ab} \hat W_{ab}}{d - 2}  -  \frac{2 (d - 3)^2 (d - 1) (-4 + 3 d) K^2}{(d - 2)^4} + \frac{32 \hat M}{d - 2} -  \frac{32 (d - 1) \hat T}{d - 2} \right.\nonumber\\
	&\qquad -  \frac{2 (d - 1)^4 V^4}{3 (d - 2)^4} + \frac{16 (d - 4) \hat R_{a} V_{a}}{d - 2} + \frac{8 (d - 3) (d - 1) K \hat W}{(d - 2)^2} - 2 \hat W^2\nonumber\\
	&\qquad + \frac{8 (d - 3) (d - 1)^3 KV^2}{(d - 2)^4} -  \frac{4 (d - 1)^2 \hat W V^2}{(d - 2)^2} +  \frac{16 (d - 3) (d - 1) V V_{a} V_{a}}{(d - 2)^2}\nonumber\\
	&\qquad\left. -  \frac{4 (d - 1) (-4 + 3 d) \hat X V}{(d - 2)^2} -  \frac{16 (d - 1) \hat Z V}{d - 2}\right] + \mathcal{O}\left(\frac{1}{c^{10}}\right)\,, \\
	h^{0i} &=  -\frac{4}{c^{3}} V_{i}  + \frac{1}{c^{5}} \left[-8 \hat R_{i} - \frac{4 (d - 1) V_{i} V}{d - 2}\right] \nonumber\\
	& + \frac{1}{c^7}\left[-16 \hat Y_{i} +  \frac{8 (d - 3) (d - 1) V_{i} K}{(d - 2)^2} - \frac{8 (d - 1) \hat R_{i} V}{d - 2} - \frac{4 (d - 1)^2 V_{i} V^2}{(d - 2)^2} \right.\nonumber\\
	&\qquad + 8 \hat W_{ia} V_{a} - 8 V_{i} \hat W\bigg] + \mathcal{O}\left(\frac{1}{c^{9}}\right)\,,\\
	h^{ij} &= \frac{1}{c^{4}} \bigg[-4 \hat W_{ij}  + 2 \delta_{ij} \hat W\bigg] + \frac{1}{c^6}\bigg[-16 \hat Z_{ij} + 8 \delta_{ij} \hat Z\bigg] \nonumber\\
	&\qquad+\frac{1}{c^8}\bigg[-32 \hat M_{ij}  - 16 V_{i} \hat R_{j} - 16 \hat R_{i} V_{j} - 8 \hat W_{ij} \hat W + 8 \hat W_{ia} \hat W_{ja}\nonumber\\
	&\qquad + \delta_{ij} \bigg\{- \frac{2 (d - 3)^2 (d - 1) K^2}{(d - 2)^3} + 16 \hat R_{a} V_{a} + 2 \hat W^2 - 2 \hat W_{ab}\hat W_{ab} -  \frac{4 (d - 1) V \hat X}{d - 2}\bigg\}\bigg]\nonumber\\
	& + \mathcal{O}\left(\frac{1}{c^{10}}\right)\,.
	\end{align}
\end{subequations}
The compact-support parts of the potentials are generated by the matter stress energy-tensor $T^{\mu\nu}$ through the definitions
\begin{equation}\label{sigma2}
\sigma = 2\frac{(d-2)T^{00}+T^{ii}}{(d-1)c^2}\,,\qquad\sigma_i = \frac{T^{0i}}{c}\,, \qquad \sigma_{ij} = T^{ij}\,.
\end{equation}
All the potentials obey flat space-time wave equations in $d$ dimensions (where $\Box=\eta^{\mu\nu}\partial_{\mu}\partial_{\nu}$) given at 3PN order by:
\begin{subequations}\label{defpotentials}
\begin{align}
	\Box V &= - 4 \pi G\, \sigma \,,\label{dalV}\\
	\Box V_{i} &= - 4 \pi G\, \sigma_{i}\,,\label{dalVi}\\
	\Box K&=-4\pi G \sigma V\,,
	\label{dalK}\\
	\Box\hat W_{ij}&=-4\pi G\left(\sigma_{ij}
	-\delta_{ij}\,\frac{\sigma_{kk}}{d-2}\right)
	-\frac{1}{2}\left(\frac{d-1}{d-2}\right)\partial_i V \partial_j V\,,
	\label{dalWij}\\
	\Box\hat R_i&=-\frac{4\pi G}{d-2}\left(\frac{5-d}{2}\, V \sigma_i
	-\frac{d-1}{2}\, V_i\, \sigma\right)\nonumber\\
	& -\frac{d-1}{d-2}\,\partial_k V\partial_i V_k
	-\frac{d(d-1)}{4(d-2)^2}\,\partial_t V \partial_i V\,,
	\label{dalRi}\\
	\Box\hat X&=-4\pi G\left[\frac{V\sigma_{ii}}{d-2}
	+2\left(\frac{d-3}{d-1}\right)\sigma_i V_i
	+\left(\frac{d-3}{d-2}\right)^2
	\sigma\left(\frac{V^2}{2} +K\right)\right]\nonumber\\
	& +\hat W_{ij}\, \partial_{ij}V
	+2 V_i\,\partial_t\partial_i V
	+\frac{1}{2}\left(\frac{d-1}{d-2}\right) V \partial^2_t V
	\nonumber\\
	& +\frac{d(d-1)}{4(d-2)^2}\left(\partial_t V\right)^2
	-2 \partial_i V_j\, \partial_j V_i\,,
	\label{dalX}\\
	\Box\hat Z_{ij}&=-\frac{4\pi G}{d-2}\, V\left(\sigma_{ij}
	-\delta_{ij}\,\frac{\sigma_{kk}}{d-2}\right)
	-\frac{d-1}{d-2}\, \partial_t V_{(i}\, \partial_{j)}V
	+\partial_i V_k\, \partial_j V_k\nonumber\\
	& +\partial_k V_i\, \partial_k V_j
	-2 \partial_k V_{(i}\, \partial_{j)}V_k
	-\frac{\delta_{ij}}{d-2}\, \partial_k V_m
	\left(\partial_k V_m -\partial_m V_k\right)\nonumber\\
	& -\frac{d(d-1)}{8(d-2)^3}\, \delta_{ij}\left(\partial_t V\right)^2
	+\frac{(d-1)(d-3)}{2(d-2)^2}\, \partial_{(i} V\partial_{j)} K\,,
	\label{dalZij}\\
	\Box\hat Y_i&=-4\pi G
	\biggl[-\frac{1}{2}\left(\frac{d-1}{d-2}\right)\sigma\hat R_i
	-\frac{(5-d)(d-1)}{4(d-2)^2}\, \sigma V V_i
	+\frac{1}{2}\, \sigma_k\hat W_{ik}
	+\frac{1}{2}\sigma_{ik} V_k\nonumber\\
	& +\frac{1}{2(d-2)}\, \sigma_{kk}V_i
	-\frac{d-3}{(d-2)^2}\, \sigma_i \left(V^2 +\frac{5-d}{2}\,
	K\right)\biggr]
	\nonumber\\
	& +\hat W_{kl}\, \partial_{kl} V_i
	-\frac{1}{2}\left(\frac{d-1}{d-2}\right)
	\partial_t\hat W_{ik}\, \partial_k V
	+\partial_i\hat W_{kl}\, \partial_k V_l
	-\partial_k\hat W_{il}\, \partial_l V_k
	\nonumber\\
	& -\frac{d-1}{d-2}\, \partial_k V \partial_i \hat R_k
	-\frac{d(d-1)}{4 (d-2)^2}\, V_k\, \partial_i V \partial_k V
	-\frac{d(d-1)^2}{8 (d-2)^3}\, V\partial_t V\partial_i V
	\nonumber\\
	& -\frac{1}{2}\left(\frac{d-1}{d-2}\right)^2 V \partial_k V
	\partial_k V_i
	+\frac{1}{2}\left(\frac{d-1}{d-2}\right) V \partial_tV_i
	+2 V_k\, \partial_k\partial_t V_i
	\nonumber\\
	& +\frac{(d-1)(d-3)}{(d-2)^2}\, \partial_k K \partial_i V_k
	+\frac{d(d-1)(d-3)}{4(d-2)^3}
	\left(\partial_t V\partial_i K +\partial_i V\partial_t K\right)\,,
	\label{dalYi}\\
	\Box\hat T&=-4\pi G\biggl[\frac{1}{2(d-1)}\, \sigma_{ij} \hat W_{ij}
	+\frac{5-d}{4(d-2)^2}\, V^2\sigma_{ii}
	+\frac{1}{d-2}\, \sigma V_i V_i
	-\frac{1}{2}\left(\frac{d-3}{d-2}\right)\sigma\hat X\nonumber\\
	& -\frac{1}{12}\left(\frac{d-3}{d-2}\right)^3
	\sigma V^3 -\frac{1}{2}\left(\frac{d-3}{d-2}\right)^3 \sigma V K
	+\frac{(5-d)(d-3)}{2(d-1)(d-2)}\, \sigma_i V_i V\nonumber\\
	& +\frac{d-3}{d-1}\, \sigma_i\hat R_i
	-\frac{d-3}{2(d-2)^2}\, \sigma_{ii} K\biggr]
	+\hat Z_{ij}\, \partial_{ij}V
	+\hat R_i\, \partial_t\partial_i V\nonumber\\
	& -2 \partial_i V_j\, \partial_j\hat R_i
	-\partial_i V_j\, \partial_t\hat W_{ij}
	+\frac{1}{2}\left(\frac{d-1}{d-2}\right) V V_i\, \partial_t\partial_i V
	+\frac{d-1}{d-2}\, V_i\, \partial_j V_i\, \partial_j V\nonumber\\
	& +\frac{d(d-1)}{4(d-2)^2}\, V_i\, \partial_t V\, \partial_i V
	+\frac{1}{8}\left(\frac{d-1}{d-2}\right)^2 V^2\partial^2_t V
	+\frac{d(d-1)^2}{8(d-2)^3}\, V\left(\partial_t V\right)^2\nonumber\\
	& -\frac{1}{2}\left(\partial_t V_i\right)^2
	-\frac{(d-1)(d-3)}{4(d-2)^2}\, V \partial^2_t K
	-\frac{d(d-1)(d-3)}{4(d-2)^3}\, \partial_t V\, \partial_t K\nonumber\\
	& -\frac{(d-1)(d-3)}{4(d-2)^2}\, K \partial^2_t V
	-\frac{d-3}{d-2}\, V_i\, \partial_t\partial_i K
	-\frac{1}{2}\left(\frac{d-3}{d-2}\right)\hat W_{ij}\,
	\partial_{ij} K\,.
\end{align}
\end{subequations}
At the 4PN order, only one new potential is required to control the mass quadrupole moment: namely the potential $\hat{M}_{ij}$ and we have already given the equation it obeys in 3 dimensions, see Eq.~\eqref{Mij3d}. The full equations obeyed by the 4PN potentials in $d$ dimensions are:
\begin{subequations}\label{defpotentials4PN}
\begin{align}
  \Box \hat{P}={}
  &4 \pi G \Biggl[\biggl(- \frac{(d -3)^2 
    (19 - 13 d + 2 d^2) K^2}{8 (d -2)^4}
    + \frac{\hat{M}^{a}{}_{a}}{d - 2}
    + \frac{(d -3) \hat{T}}{d - 2}\nonumber\\
  & -  \frac{(d -3)^4 K V^2}{4 (d -2)^4}
    -  \frac{(d -3)^4 V^4}{48 (d -2)^4}
    + \frac{(d - 5) \hat{R}^{a} V_{a}}{2 (d -2)}
    + \frac{\hat{W}_{ab} \hat{W}^{ab}}{8 (d -2)}\nonumber\\
  & + V \biggl[- \frac{(8 - 5 d + d^2) V_{a} V^{a}}{2 (d -2)^2}
    -  \frac{(19 - 13 d + 2 d^2) \hat{X}}
{4 (d -2)^2}\biggr]\biggr) \sigma
    + \biggl[\frac{(8 - 5 d + d^2) \hat{R}^{a} V}
{(d -2) (d - 1)}\nonumber\\
  & -  \frac{(d -3) (8 - 5 d + d^2) K V^{a}}
{(d -2)^2 (d - 1)}
    -  \frac{(-22 + 21 d - 8 d^2 + d^3) V^2 V^{a}}
{2 (d -2)^2 (d - 1)}\nonumber\\
  & -  \frac{(d -3) \hat{Y}^{a}}{d - 1}\biggr] \sigma_{a}
    -  \frac{(d -3) V^{a} \hat{W}_{ab} \sigma^{b}}{2 (d - 1)}
    + \biggl[- \frac{(d -3) V^{a} V^{b}}{2 (d - 1)}\nonumber\\
  & -  \frac{V \hat{W}^{ab}}{(d -2) (d - 1)}
    -  \frac{\hat{Z}^{ab}}{d - 1}\biggr] \sigma_{ab}
    + \biggl[- \frac{(d - 5) (d -3) K V}
{2 (d -2)^3}\nonumber\\
  & -  \frac{(19 - 8 d + d^2) V^3}{12 (d -2)^3}
    -  \frac{\hat{X}}{2 (d -2)}\biggr] \sigma^{a}{}_{a}\Biggr]
    + \frac{(d -3)^2 
(d - 1) d (\partial_t K)^2}{8 (d -2)^4}\nonumber\\
  & -  \partial_t \hat{R}^{a} \partial_t V_{a}
    + \frac{(d - 1) d \partial_t V \partial_t \hat{X}}{4 (d -2)^2}
    + \hat{Y}^{a} \partial_t \partial_{a}V
    + \frac{(d - 1)^3 V^3 \partial_t^{2} V}{24 (d -2)^3}\nonumber\\
  & -  \frac{(d -3) V_{a} V^{a} \partial_t^{2} V}{4 (d -2)}
    + \frac{(d - 1) \hat{X} \partial_t^{2} V}{4 (d -2)}
    -  \frac{(d -3) \hat{Z}^{ab} \partial_{b}\partial_{a}K}{d - 2}
    + \hat{M}^{ab} \partial_{b}\partial_{a}V\nonumber\\
  & + \hat{R}^{a} \biggl[- \frac{(d -3) \partial_t \partial_{a}K}{d - 2}
    + \frac{(d - 1) d \partial_t V \partial_{a}V}{4 (d -2)^2}
    + V^{b} \partial_{b}\partial_{a}V\biggr]
    + \partial_t \hat{W}_{ab} (- \frac{1}{8} \partial_t \hat{W}^{ab}\nonumber\\
  & -  \partial^{b}\hat{R}^{a})
    -  \partial_{a}\hat{R}_{b} \partial^{b}\hat{R}^{a}
    - 2 \partial_t \hat{Z}_{ab} \partial^{b}V^{a}
    - 2 \partial_{a}\hat{Y}_{b} \partial^{b}V^{a}\nonumber\\
  & + V \biggl(- \frac{(d -3) (d - 1)^2 
d \partial_t K \partial_t V}{4 (d -2)^4}
    -  \frac{(d - 1) \partial_t V_{a} \partial_t V^{a}}
{2 (d -2)}\nonumber\\
  & + \frac{(d - 1) \hat{R}^{a} \partial_t \partial_{a}V}{2 (d -2)}
    + \frac{(d - 1) \partial_t^{2} \hat{X}}{4 (d -2)}
    + V^{a} \biggl[- \frac{(d -3) 
(d - 1) \partial_t \partial_{a}K}{2 (d -2)^2}\nonumber\\
  & + \frac{(d - 1)^2 d \partial_t V \partial_{a}V}
{4 (d -2)^3}\biggr]
    -  \frac{(d - 1) \partial_{b}V_{a} \partial^{b}\hat{R}^{a}}{d - 2}
    -  \frac{(d - 1) \partial_t \hat{W}_{ab} \partial^{b}V^{a}}
{2 (d -2)}\biggr)\nonumber\\
  & + V^2 \biggl[\frac{(d - 1)^3 d (\partial_t V)^2}
{16 (d -2)^4}
    + \frac{(d - 1)^2 V^{a} \partial_t \partial_{a}V}{4 (d -2)^2}
    -  \frac{(d -3) (d - 1)^2 \partial_t^{2} K}
{8 (d -2)^3}\nonumber\\
  & -  \frac{(d - 1)^2 \partial_{a}V_{b} \partial^{b}V^{a}}
{4 (d -2)^2}\biggr]
    + K \biggl[- \frac{(d -3) 
(d - 1)^2 d (\partial_t V)^2}{8 (d -2)^4}\nonumber\\
  & -  \frac{(d -3) (d - 1) V^{a} 
\partial_t \partial_{a}V}{2 (d -2)^2}
    + \frac{(d -3)^2 (d - 1) \partial_t^{2} K}
{4 (d -2)^3}\nonumber\\
  & -  \frac{(d -3) (d - 1)^2 V \partial_t^{2} V}
{4 (d -2)^3}
    + \frac{(d -3) (d - 1) 
\partial_{b}V_{a} \partial^{b}V^{a}}{2 (d -2)^2}\biggr]
    + \hat{W}^{ab} (\frac{1}{2} \partial_{b}\partial_{a}\hat{X}\nonumber\\
  & -  \frac{1}{4} \hat{W}_{a}{}^{i} \partial_{i}\partial_{b}V)
    + \hat{W}_{ai} \partial^{b}V^{a} \partial^{i}V_{b}
    + V^{a} \biggl(\partial_t \partial_{a}\hat{X}
    -  \frac{1}{2} \hat{W}_{ab} \partial_t \partial^{b}V\nonumber\\
  & + \partial_t V \biggl[- \frac{(d - 4) (d - 1) 
\partial_t V_{a}}{2 (d -2)^2}
    -  \frac{(d -3) (d - 1) d \partial_{a}K}
{4 (d -2)^3}\biggr]\nonumber\\
  & -  \frac{(d -3) (d - 1) d 
\partial_t K \partial_{a}V}{4 (d -2)^3}
    + \frac{(d - 1) V^{b} \partial_{a}V \partial_{b}V}{4 (d -2)^2}
    + 2 \partial_t V_{b} \partial^{b}V_{a}
    -  \partial_{a}\hat{W}_{bi} \partial^{i}V^{b}\nonumber\\
  & + \partial_{i}\hat{W}_{ab} \partial^{i}V^{b}\biggr) \, , \\[1ex]
  \Box \hat{N}_i={}
  & 4 \pi G \biggl(\biggl[- \frac{(d -3) 
(d - 1) \hat{R}^{i} V}{2 (d -2)^2}
    + \frac{(d -3)^2 (d - 1) K V^{i}}
{2 (-2 +d)^3}\nonumber\\
  & + \frac{(d - 1) (37 - 26 d + 5 d^2) V^2 V^{i}}
{16 (d -2)^3}
    + \frac{(d - 1) \hat{Y}^{i}}{2 (d -2)}\biggr] \sigma
    + V^{a} V^{i} \sigma_{a}\nonumber\\
  & + \biggl[\frac{(d - 5) V \hat{W}^{i}{}_{a}}{4 (d -2)}
    -  \hat{Z}^{i}{}_{a}\biggr] \sigma^{a}
    + \biggl[\frac{(d - 5)^2 (d -3) K V}{4 (d -2)^3}
    + \frac{(d - 5)^3 V^3}{48 (d -2)^3}\nonumber\\
  & -  \frac{3 V_{a} V^{a}}{2 (d -2)}
    + \frac{(d - 5) \hat{X}}{2 (d -2)}\biggr] \sigma^{i}
    + \biggl[- \frac{\hat{R}^{i}}{2 (d -2)}
    + \frac{(d -3) V V^{i}}{2 (d -2)^2}\biggr] 
\sigma^{a}{}_{a}\nonumber\\
  & + (- \frac{1}{2} \hat{R}^{a}
    -  \frac{V V^{a}}{d - 2}) \sigma^{i}{}_{a}\biggr)
    -  \frac{1}{2} \partial_t V^{a} \partial_t \hat{W}^{i}{}_{a}
    + \frac{(d -3) (d - 1) 
\partial_t \hat{W}^{i}{}_{a} \partial^{a}K}{2 (d -2)^2}\nonumber\\
  & + \frac{(d - 1) d \hat{W}^{i}{}_{a} 
\partial_t V \partial^{a}V}{8 (d -2)^2}
    -  \frac{(d - 1) \partial_t \hat{Z}^{i}{}_{a} \partial^{a}V}{d - 2}
    + \frac{(d - 1)^2 \hat{R}^{i} 
\partial_{a}V \partial^{a}V}{4 (d -2)^2}\nonumber\\
  & + \hat{W}^{ab} \partial_{b}\partial_{a}\hat{R}^{i}
    + 2 \hat{Z}^{ab} \partial_{b}\partial_{a}V^{i}
    -  \partial_{a}\hat{W}^{i}{}_{b} \partial^{b}\hat{R}^{a}
    + \frac{(d - 1) \hat{W}^{i}{}_{b} 
\partial^{a}V \partial^{b}V_{a}}{2 (d -2)}\nonumber\\
  & + V^{i} \biggl[\frac{(d - 1) d 
(\partial_t V)^2}{8 (d -2)^2}
    -  \frac{(d -3) (d - 1)^2 \partial_{a}V 
\partial^{a}K}{2 (d -2)^3}
    + \frac{(d - 1) \partial_t V_{a} \partial^{a}V}{d - 2}\nonumber\\
  & + \partial_{a}V_{b} \partial^{b}V^{a}\biggr]
    -  \frac{(d -3)^2 (d - 1) 
d \partial_t K \partial^{i}K}{4 (d -2)^4}
    + \frac{(d -3) (d - 1) \partial^{a}K 
\partial^{i}\hat{R}_{a}}{(d -2)^2}\nonumber\\
  & -  \frac{(d - 1) d \partial_t \hat{X} \partial^{i}V}
{4 (d -2)^2}
    + K \biggl[- \frac{(d -3) (d - 1) \partial_t^{2} V^{i}}
{2 (d -2)^2}\nonumber\\
  & + \frac{(d -3) (d - 1)^2 d 
\partial_t V \partial^{i}V}{8 (d -2)^4}\biggr]
    + \hat{R}^{a} \biggl[2 \partial_t \partial_{a}V^{i}
    -  \frac{(d - 1) d \partial_{a}V \partial^{i}V}
{4 (d -2)^2}\biggr]\nonumber\\
  & + V \biggl(\frac{(d - 1) \partial_t^{2} \hat{R}^{i}}{2 (d -2)}
    -  \frac{(d - 1)^2 \partial_t \hat{W}^{i}{}_{a} 
\partial^{a}V}{4 (d -2)^2}
    + \partial_t V \biggl[\frac{(d - 1)^2 \partial_t V^{i}}
{4 (d -2)^2}\nonumber\\
  & + \frac{(d -3) (d - 1)^2 d \partial^{i}K}
{8 (d -2)^4}\biggr]
    + \frac{(d -3) (d - 1)^2 d 
\partial_t K \partial^{i}V}{8 (d -2)^4}\nonumber\\
  & + V^{a} \biggl[\frac{(d - 1) \partial_t \partial_{a}V^{i}}{d - 2}
    -  \frac{(d - 1)^3 \partial_{a}V \partial^{i}V}
{4 (d -2)^3}\biggr]\biggr)
    + V^2 \biggl[\frac{(d - 1)^2 \partial_t^{2} V^{i}}
{4 (d -2)^2}\nonumber\\
  & -  \frac{(d - 1)^3 d \partial_t V \partial^{i}V}{32 (d -2)^4}
    -  \frac{(d - 1)^3 \partial^{a}V \partial^{i}V_{a}}
{8 (d -2)^3}\biggr]
    -  \frac{(d - 1) \partial_{a}\hat{X} \partial^{i}V^{a}}{d - 2}\nonumber\\
  & + \frac{(d - 1) \hat{W}_{ab} \partial^{a}V \partial^{i}V^{b}}
{2 (d -2)}
    + V^{a} \biggl[2 \partial_t \partial_{a}\hat{R}^{i}
    + \frac{(d - 1) \partial_t V^{i} \partial_{a}V}
{2 (d -2)}\nonumber\\
  & -  \frac{(d - 1) \partial_{a}\hat{W}^{i}{}_{b} \partial^{b}V}
{2 (d -2)}
    + \frac{(d - 1) \partial_{b}\hat{W}^{i}{}_{a} \partial^{b}V}
{2 (d -2)}
    + \partial_{b}V_{a} \partial^{b}V^{i}\nonumber\\
  & + \frac{(d -3) (d - 1) d \partial_{a}V \partial^{i}K}
{4 (d -2)^3}
    -  \frac{(d - 1) \partial_t V_{a} \partial^{i}V}
{(d -2)^2}\nonumber\\
  & + \frac{(d -3) (d - 1) d 
\partial_{a}K \partial^{i}V}{4 (d -2)^3}
    + \frac{(d - 4) (d - 1) 
\partial_t V \partial^{i}V_{a}}{2 (d -2)^2}
    + \partial_{a}V_{b} (\partial^{b}V^{i}\nonumber\\
  & -  \partial^{i}V^{b})
    -  \frac{d \partial_{b}V_{a} \partial^{i}V^{b}}{d - 2}\biggr]
    + \partial^{b}\hat{R}^{a} \partial^{i}\hat{W}_{ab}
    + \partial_t \hat{W}_{ab} (- \frac{1}{2} \partial^{b}\hat{W}^{ia}
    + \frac{1}{4} \partial^{i}\hat{W}^{ab})\nonumber\\
  & -  \frac{(d - 1) d \partial_t V \partial^{i}\hat{X}}
       {4 (d -2)^2}
    -  \frac{(d - 1) \partial^{a}V \partial^{i}\hat{Y}_{a}}{d - 2}
    + \partial^{b}V^{a} (-2 \partial_{a}\hat{Z}^{i}{}_{b}
    + 2 \partial^{i}\hat{Z}_{ab}) \, , \\[1ex]
\Box \hat{M}_{ij}={}&4 \pi G \Biggl[\biggl(- \frac{(d - 1) V^{i} V^{j}}{2 (d - 2)}
 + \delta^{ij} \biggl[\frac{(d - 3)^2 (d - 1) K V}{4 (d - 2)^3}
 + \frac{(d - 3)^2 (d - 1) V^3}{16 (d - 2)^3}\nonumber\\
& + \frac{(d - 1) V_{a} V^{a}}{4 (d - 2)}
 + \frac{(d - 1) \hat{X}}{8 (d - 2)}\biggr]\biggr) \sigma
 + \delta^{ij} \biggl[- \frac{1}{2} \hat{R}^{a}
 + \frac{(d - 4) V V^{a}}{2 (d - 2)}\biggr] \sigma_{a}
 + \biggl[\hat{R}^{(i}\nonumber\\
& -  \frac{(d - 5) V V^{(i}}{2 (d - 2)}\biggr] \sigma^{j)}
 + \frac{1}{8} \delta^{ij} \hat{W}^{ab} \sigma_{ab}
 + \frac{(d - 1) \delta^{ij} V^2 \sigma^{a}{}_{a}}{8 (d - 2)^2}
 -  \frac{\delta^{ij} \hat{W}^{a}{}_{a} \sigma^{b}{}_{b}}{8 (d - 2)}\nonumber\\
& + \biggl[\frac{(d - 3)K}{(d - 2)^2}
 -  \frac{V^2}{(d - 2)^2}\biggr] \sigma^{ij}
 -  \frac{1}{2} \hat{W}^{a(i} \sigma^{j)}{}_{a}\Biggr]
 + \frac{1}{2} \hat{W}^{ab} \partial_{b}\partial_{a}\hat{W}^{ij}\nonumber\\
& -  \frac{1}{2} \partial_{a}\hat{W}^{i}{}_{b} \partial^{b}\hat{W}^{ja}
 -  \frac{(d - 3)^2 (d - 1) \partial^{i}K \partial^{j}K}{4 (d - 2)^3}
 + \partial_t V^{(i} \biggl[- \partial_t V^{j)}\nonumber\\
& + \frac{(d - 3) (d - 1) \partial^{j)}K}{(d - 2)^2}\biggr]
 - 2 \partial_{a}V^{(i} \partial^{j)}\hat{R}^{a}
 + 2\partial^{(i}V_{a} \partial^{j)}\hat{R}^{a}
 -  \frac{(d - 1) \partial_t \hat{R}^{(i} \partial^{j)}V}{d - 2}\nonumber\\
& + \frac{(d - 1) \hat{W}^{(i}{}_{a} \partial^{j)}V \partial^{a}V}{4 (d - 2)} 
 -  \frac{(d - 1) \partial^{(i}\hat{X} \partial^{j)}V}{2 (d - 2)}
 + V^{(i} \biggl[\frac{(d - 1) \partial_{a}V^{j)} \partial^{a}V}{d - 2}\nonumber\\
& + \frac{(d - 1) \partial_t V \partial^{j)}V}{4 (d - 2)}\biggr]
 + V \biggl[\frac{(d - 1) \partial_t^{2} \hat{W}^{ij}}{4 (d - 2)}
 -  \frac{(d - 1)^2 \partial_t V^{(i} \partial^{j)}V}{2 (d - 2)^2}\biggr]
 + V^{a} \biggl[\partial_t \partial_{a}\hat{W}^{ij}\nonumber\\
& -  \frac{(d - 1) \partial_{a}V^{(i} \partial^{j)}V}{d - 2}\biggr]
 - 2 \partial^{a}\hat{R}^{(i} \partial^{j)}V_{a}
  + \partial_t \hat{W}^{(i}{}_{a} (- \partial^{a}V^{j)}
 + \partial^{j)}V^{a})\nonumber\\
& + \partial^{b}\hat{W}^{a(i} \partial^{j)}\hat{W}_{ab}
 -  \frac{1}{4} \partial^{i}\hat{W}^{ab} \partial^{j}\hat{W}_{ab}
 \nonumber\\
& + \delta^{ij} \biggl(- \frac{(d - 3) (d - 1) d \partial_t K \partial_t V}{8 (d - 2)^3}
 -  \frac{(d - 1)^2 V^2 \partial_t^{2} V}{16 (d - 2)^2}
 + V^{a} \biggl[- \frac{1}{2} \partial_t \partial_{a}\hat{W}^{b}{}_{b}\nonumber\\
& -  \frac{(d - 1) \partial_t V \partial_{a}V}{8 (d - 2)}\biggr]
 -  \frac{(d - 3) (d - 1) \partial_t V_{a} \partial^{a}K}{2 (d - 2)^2}
 + \frac{(d - 1) \partial_t \hat{R}_{a} \partial^{a}V}{2 (d - 2)}\nonumber\\
& -  \frac{1}{4} \hat{W}^{ab} \partial_{b}\partial_{a}\hat{W}^{k}{}_{k}
 + \partial_{a}V_{b} \partial^{b}\hat{R}^{a}
 -  \frac{(d - 1) \hat{W}_{ab} \partial^{a}V \partial^{b}V}{16 (d - 2)}
 -  \frac{1}{2} \partial_t \hat{W}_{ab} \partial^{b}V^{a}\nonumber\\
& + V \biggl[\frac{(d - 1)^2 d (\partial_t V)^2}{32 (d - 2)^3}
 -  \frac{(d - 1) V^{a} \partial_t \partial_{a}V}{4 (d - 2)}
 -  \frac{(d - 1) \partial_t^{2} \hat{W}^{a}{}_{a}}{8 (d - 2)}\nonumber\\
& + \frac{(d - 1)^2 \partial_t V_{a} \partial^{a}V}{4 (d - 2)^2}
 -  \frac{(d - 1) \hat{W}^{ab} \partial_{b}\partial_{a}V}{8 (d - 2)}
 + \frac{(d - 1) \partial_{a}V_{b} \partial^{b}V^{a}}{4 (d - 2)}\biggr]\nonumber\\
& -  \frac{1}{4} \partial_{b}\hat{W}_{ak} \partial^{k}\hat{W}^{ab}\biggr) \, .
\end{align}
\end{subequations}

Those potentials obey the usual harmonicity conditions $\partial_\mu\bar{h}^{\mu\nu} = 0$, which explicitly read up to the 4PN order
\begin{subequations}\label{harmonicityconds4PN}
\begin{align}
0={}& \partial_t \bigg\{V_{i} + \frac{1}{c^{2}} \biggl(2 \hat{R}_{i} + \frac{d -  1}{d - 2} V V_{i}\biggl) \nonumber \\ 
& \quad + \frac{1}{c^{4}} \biggl(4 \hat{Y}_{i} + \frac{2 (d -  1)}{d -  2}\hat{R}_{i} V  + \biggl [- \frac{2 (d -  1) (d -  3)}{(d -  2)^2} \hat{K} + 2
\hat{W}\biggl] V_{i} -  2 V_{k} \hat{W}_{ik} + \frac{(d - 1)^2}{(d -  2)^2} V^2 V_{i}\biggl)\bigg\}\nonumber\\
& + \partial_{j}\bigg\{\hat{W}_{ij} -  \frac{1}{2} \delta_{ij} \hat{W} 
+\frac{1}{c^{2}} (- 2 \delta_{ij} \hat{Z} + 4 \hat{Z}_{ij})\nonumber \\
 & \quad +
\frac{1}{c^{4}} \biggl(8 \hat{M}_{ij}+ 4 \hat{R}_{j} V_{i} + 4 \hat{R}_{i} V_{j} + 2 \hat{W} \hat{W}_{ij} -  2 \hat{W}_{ik} \hat{W}_{jk}
\nonumber \\
 & \qquad \quad
  + \delta_{ij} \biggl [\frac{(d -  1) (d - 3)^2}{2 (d -  2)^3} \hat{K}^2 -  4 \hat{R}_{k} V_{k} -  \frac{1}{2} \hat{W}^2 +\frac{1}{2} \hat{W}_{kl} \hat{W}_{kl} + \frac{d -  1}{d -  2} V \hat{X}\biggl]\biggl) \bigg\}\,,\\
0={}& \partial_t \bigg\{\frac{d -  1}{2 (d -  2)} V 
+ \frac{1}{c^{2}} \biggl[-\frac{(d -  1) (d -  3)}{(d-  2)^2} \hat{K} + \frac{1}{2} \hat{W} + \frac{(d - 1)^2}{2 (d -  2)^2} V^2\biggl] \nonumber \\
& \quad + \frac{1}{c^{4}} \biggl(\frac{2(d -  1)}{d -  2} \hat{X} + 2 \hat{Z} -  \frac{2 (d -  3)}{d -  2} V_{k} V_{k}  + \frac{(d -  1)^3}{3 (d -  2)^3} V^3 \nonumber \\ 
& \quad \qquad +\biggl [- \frac{2 (d -  1)^2 (d -  3)}{(d -  2)^3} \hat{K} + \frac{d -  1}{d -  2}
\hat{W}\biggl] V \biggl) \nonumber \\ 
& \quad +
\frac{1}{c^{6}} \biggl(- \frac{8}{d -  2} \hat{M} + \frac{8 (d -  1)}{d -  2}\hat{T} 
+ \biggl [\frac{(d -  1) (3 d -  4)}{(d -  2)^2} \hat{X} + \frac{4 (d -  1)}{d -  2} \hat{Z}\biggl] V\nonumber\\
& \qquad \qquad  + \frac{1}{2} \hat{W}^2  -  \frac{d}{2 (d-  2)} \hat{W}_{kl} \hat{W}_{kl}-  \frac{2 (d -  1) (d -  3)}{(d -  2)^2}\hat{K} \hat{W} \nonumber \\ 
& \qquad \qquad + \frac{(d -  1) (d -  3)^2 (3 d -  4)}{2 (d -  2)^4} \hat{K}^2 -  \frac{4(d -  4)}{d -  2} \hat{R}_{k} V_{k}-  \frac{4 (d -  1) (d -3)}{(d -  2)^2} V V_{k} V_{k}  \nonumber \\ 
& \qquad \qquad  + \biggl[- \frac{2 (d - 1)^3 (d -  3)}{(d -  2)^4} \hat{K} + \frac{(d -  1)^2}{(d -  2)^2} \hat{W}\biggl] V^2 + \frac{(d -  1)^4}{6 (d -  2)^4} V^4\biggl)\bigg\}\nonumber\\
&
 + \partial_{i}\bigg\{V_{i} + \frac{1}{c^{2}} \biggl(2 \hat{R}_{i} + \frac{d -  1}{d -2} V V_{i}\biggl) \nonumber\\
& \quad + \frac{1}{c^{4}} \biggl(4 \hat{Y}_{i} + \frac{2 (d -  1)}{d -  2}\hat{R}_{i} V  -  2 V_{k} \hat{W}_{ik}+ \biggl [- \frac{2 (d -  1) (d -  3)}{(d -  2)^2} \hat{K} + 2\hat{W}\biggl] V_{i}  + \frac{(d - 1)^2}{(d -  2)^2} V^2 V_{i}\biggl) \nonumber \\
 & \quad + \frac{1}{c^{6}} \Biggl[8\hat{N}_{i} -  \frac{4 (d -  1) (d -  3)}{(d -  2)^2} \hat{K} \hat{R}_{i} + 4\hat{R}_{i} \hat{W} -  4 \hat{R}_{k} \hat{W}_{ik} + \frac{4 (d -  1)}{d -  2} V\hat{Y}_{i} -  8 V_{k} \hat{Z}_{ik}\nonumber\\
& \qquad \qquad  + \biggl [\frac{4 (d -  1)}{d -  2} \hat{X} + 8\hat{Z}\biggl] V_{i}  + \frac{(d -  1)^2}{(d -  2)^2}\hat{R}_{i} V^2 + \frac{4}{d -  2} V_{i} V_{k} V_{k} + \frac{2 (d -  1)^3}{3 (d -  2)^3} V^3 V_{i} \nonumber \\ 
 & \qquad \qquad + \biggl(\biggl [- \frac{2 (d - 1)^2 (d -  3)}{(d -  2)^3} \hat{K} + \frac{2 (d -  1)}{d -  2} \hat{W}\biggl] V_{i}-  \frac{2 (d -  1)}{d -  2} V_{k} \hat{W}_{ik}\biggl) V 
\Biggl]\bigg\}\,.
\end{align}
\end{subequations}

Finally we give the expressions of the effective masses $\mu_1$ and $\tilde{\mu}_1$ in $d$ dimensions which parametrize the source densities~\eqref{sigma} for point particle sources. They are defined by Eqs.~\eqref{mutilde} and~\eqref{mu}, namely
\begin{subequations}\label{defrecallmu}
\begin{align}
\mu_1 &= \frac{1}{\sqrt{-(g)_1}}\frac{m_1}{\sqrt{-(g_{\mu\nu})_1 \frac{v_1^\mu v_1^\nu}{c^2}}} \,,\\ \tilde{\mu}_1  &= \frac{2}{d-1}\left(d-2 + \frac{v_1^2}{c^2}\right) \mu_1\,,
\end{align}
\end{subequations}
where $g$ is the determinant of the metric, and $(g)_1$ and $(g_{\mu\nu})_1$ are evaluated at $\mathbf{x} = \bm{y}_1$ following the DR. Most importantly the expression of $\tilde{\mu}_1$ with full 4PN accuracy is required for the 4PN mass quadrupole moment. It is given in terms of the potentials by
\begin{align}\label{mutilde1pot}
\tilde{\mu}_1 =& \frac{2 m_1}{d-1}\Biggl( d-2 + \frac{1}{c^2}\left[\frac{d}{2}  v_1^2  + (d - 4) V\right] \nonumber\\
&\quad+ \frac{1}{c^4} \left[ -4 (d - 2) v_{1}^{i} V_{i}  -  \frac{2 (d - 4) (d - 3) K}{d - 2}  + \frac{1}{8} (-2  + 3 d) v_1^4\right.\nonumber\\
&\qquad\quad\left. + \frac{(4 - 10 d + 3 d^2) v_1^2 V}{2 (d - 2)}  + \frac{(d - 4)^2 V^2}{2 (d - 2)}  - 2 (d - 2) \hat W \right] \nonumber\\
&\quad+ \frac{1}{c^6} \left[-8 (d - 2) \hat R^{i} v_{1i} + 4 (d - 4) V_{i} V^{i} -  \frac{(d - 3) (4 - 10 d + 3 d^2) v_1^2 K}{(d - 2)^2} \right.\nonumber\\
&\qquad\quad + 2 (-2+ d) \hat W_{ij} v_{1}^{i} v_{1}^{j} + \frac{1}{16} (-4 + 5 d) v_1^6 - 4 (-5 + 2 d) v_{1}^{i} V_{i} V\nonumber\\
&\qquad\quad -  \frac{2 (d - 4)^2 (d - 3) K V}{(d - 2)^2} + \frac{3}{8} (-4 + 5 d) v_1^4 V\nonumber\\
&\qquad\quad + \frac{(-40 + 92 d - 52 d^2 + 9 d^3) v_1^2 V^2}{4 (d - 2)^2} + \frac{(d - 4)^3 V^3}{6 (d - 2)^2} - 2 (-4 + 3 d) v_1^2 v_{1}^{j} V_{j}\nonumber\\
&\qquad\quad -  d v_1^2 \hat W - 2 (d - 4) V \hat W + 4 (d - 4) \hat X - 8 (d - 2) \hat Z \bigg]\nonumber\\
&\quad+ \frac{1}{c^8} \left[ -32 \hat R^{i} V_{i} + 2 d \hat W_{ij}^2 - 16 (d - 2) v_{1}^{i} \hat Y_{i} + \frac{8 (d - 3) (-5 + 2 d) v_{1}^{i} V_{i} K}{d - 2}\right.\nonumber\\
&\qquad\quad + \frac{2 (d - 3)^2 (16 - 9 d + 2 d^2) K^2}{(d - 2)^3} + 32 \hat M + 16 (d - 4) \hat T + 8 (d - 2)  \hat Z_{ij} v_{1}^{i} v_{1}^{j}\nonumber\\
&\qquad\quad - 4 (-4 + 3 d) \hat R^{i} v_{1i} v_1^2 -  \frac{3 (d - 3) (-4 + 5 d) K v_1^4}{4 (d - 2)} + \frac{5}{128} (-6+ 7 d)  v_1^4\nonumber\\
&\qquad\quad  - 8 (-5 + 2 d) \hat R^{i} v_{1i} V + \frac{4 (d - 4)^2 V_{i} V^{i} V}{d - 2}\nonumber\\
&\qquad\quad -  \frac{(d - 3) (-40 + 92 d - 52 d^2 + 9 d^3) v_1^2 K V}{(d - 2)^3} + 6 (d - 2) \hat W_{ij} v_{1}^{i} v_{1}^{j} V\nonumber\\
&\qquad\quad + \frac{(52 - 90 d + 35 d^2) v_1^6 V}{16 (d - 2)} -  \frac{2 (26 - 22 d + 5 d^2) v_{1}^{i} V_{i} V^2}{d - 2}\nonumber\\
&\qquad\quad -  \frac{(d - 4)^3 (d - 3) K V^2}{(d - 2)^3} + \frac{3 (40 - 68 d + 25 d^2) v_1^4 V^2}{16 (d - 2)}\nonumber\\
&\qquad\quad + \frac{(304 - 768 d + 624 d^2 - 214 d^3 + 27 d^4) v_1^2 V^3}{12 (d - 2)^3} + \frac{(d - 4)^4 V^4}{24 (d - 2)^3} -  \frac{3}{2} (-6+ 5 d) v_1^4 v_{1}^{a} V_{a}\nonumber\\
&\qquad\quad  + 16 (d - 2) v_{1}^{i} v_{1}^{j}  V_{i} V_{j} -  \frac{2 (32 - 41 d + 12 d^2) v_1^2 v_{1}^{j} V V_{j}}{d - 2}\nonumber\\
&\qquad\quad - 8 (d - 2) v_{1}^{i} \hat W_{ij} V^{j} + \frac{2 (4 - 10 d + 3 d^2) v_1^2 V_{j} V^{j}}{d - 2} + 8 (d - 2) v_{1}^{i} V_{i} \hat W\nonumber\\
&\qquad\quad + \frac{4 (d - 4) (d - 3) K \hat W}{d - 2} -  \frac{1}{4} (-2 + 3 d) v_1^4 \hat W -  \frac{(4 - 10 d + 3 d^2) v_1^2 V \hat W}{d - 2}\nonumber\\
&\qquad\quad -  \frac{(d - 4)^2 V^2 \hat W}{d - 2} + 2 (d - 2) \hat W^2 + (-4 + 3 d) v_1^2 v_{1}^{i} v_{1}^{j} \hat W_{ij} - 4 d v_1^2 \hat Z\nonumber\\
&\qquad\quad\left. + \frac{2 (4 - 10 d + 3 d^2) v_1^2 \hat X}{d - 2} + \frac{4 (16 - 9 d + 2 d^2) V \hat X}{d - 2} - 8 (d - 4) V \hat Z \right]\Biggr)_1\,,
\end{align}
where it is understood that all the potentials are evaluated at the point $\mathbf{x} = \bm{y}_1$ following DR. Note that we must apply the rule of ``distributivity'', \textit{e.g.} $(V \hat Z)_1 = (V)_1 (\hat Z)_1$, see~\cite{BDE04} for discussions. Besides the 4PN expression of $\tilde{\mu}_1$  we require also the 3PN expression of $\mu_1$ which we provide for completeness:
\begin{align}\label{mu1pot}
\mu_1 &= m_1\Biggl(1 + \frac{1}{c^2}\left[ \frac{1}{2} v_1^2  + \frac{(d - 4) V}{d - 2}\right] \nonumber\\
&\quad+ \frac{1}{c^4}\left[-4 v_{1}^{i} V_{i} -  \frac{2 (d - 4) (d - 3) K}{(d - 2)^2} + \frac{3}{8} v_1^4 + \frac{3}{2} v_1^2 V + \frac{(d - 4)^2 V^2}{2 (d - 2)^2} - 2 \hat W\right]\nonumber\\
&\quad+ \frac{1}{c^6}\left[-8 \hat R^{i} v_{1i} + \frac{4 (d - 4) V_{i} V^{i}}{d - 2} -  \frac{3 (d - 3) v_1^2 K}{d - 2} + 2  \hat W_{ij} v_{1}^{i} v_{1}^{j} + \frac{5}{16}  v_1^6\right.\nonumber\\
&\qquad\quad -  \frac{4 (-5 + 2 d) v_{1}^{i} V_{i} V}{d - 2} -  \frac{2 (d - 4)^2 (d - 3) K V}{(d - 2)^3} + \frac{3 (-8 + 5 d) v_1^4 V}{8 (d - 2)}\nonumber\\
&\qquad\quad + \frac{9}{4} v_1^2 V^2 + \frac{(d - 4)^3 V^3}{6 (d - 2)^3} - 6 v_1^2 v_{1}^{j} V_{j} -  v_1^2 \hat W -  \frac{2 (d - 4) V \hat W}{d - 2}\nonumber\\
&\qquad\quad\left. + \frac{4 (d - 4) \hat X}{d - 2} - 8 \hat Z \right]\Biggr)_1\,, 
\end{align}

\section{The shifts applied in the 4PN equations of motion}
\label{app:shift}

In the recent work on the 4PN equations of motion~\cite{BBBFMa, BBBFMb, BBBFMc, MBBF17, BBFM17}, which succeeded in computing from first principles all the ``ambiguities'' in the problem, we applied to our ``brute'' calculation a series of shifts of the trajectories so as to remove UV- and IR-types poles and simplify the end result. Now, in order to be consistent we have to apply after our ``brute'' derivation of the 4PN quadrupole moment the \textit{same} series of shifts. As recalled in Sec.~\ref{sec:resultMQ} the total shift is composed of three pieces. A first shift $\bm{\xi}_{1,2}$ removed the poles $1/\varepsilon$ corresponding to UV divergences and led to the Lagrangian provided in~\cite{BBBFMb,BBBFMc}; a second shift $\bm{\chi}_{1,2}$, used in~\cite{BBBFMc}, dealt with the $1/\varepsilon$ poles due to IR divergences; and a third shift $\bm{\eta}_\text{1,2}$ presented in~\cite{BBFM17} was mainly used for convenience. We provide here the shifts $\bm{\xi}_{1,2}$ and $\bm{\eta}_\text{1,2}$ \textit{in extenso}. Note that the second shift $\bm{\chi}$ has not been used in the present paper but will be part of our next program to investigate the IR divergences occuring in the 4PN quadrupole moment.

The shift $\bm{\xi}_1$ is composed of 3PN and 4PN contributions and is given by
\begin{equation}\label{xi1}
\xi_1^i = \frac{11}{3}\frac{G^2\,m_1^2}{c^6} \left[
\frac{1}{\varepsilon}-2\ln\left(
\frac{\overline{q}^{1/2}r'_0}{\ell_0}\right)
-\frac{327}{1540}\right] a^i_1 +
\frac{1}{c^8}\xi^i_{1,\,\mathrm{4PN}}\,.
\end{equation}
The first term represents the 3PN contribution determined in~\cite{BDE04}. Here $\overline{q}=4\pi \de^{\gamma_\text{E}}$ where $\gamma_\text{E}$ is the Euler constant, and $\bm{a}_{1}$ represents the Newtonian acceleration of 1 in $d$ dimensions. The 4PN contributions have already been shown in the Appendix C of~\cite{BBBFMa}, however the terms proportional to $y_1^i$ [see Eq.~\eqref{termy1} below] were inadvertently missing there. Furthermore we have adjusted one coefficient in the expression of the shift to take into account the subsequent determination of the ambiguity parameters in~\cite{BBBFMc, MBBF17}. The 4PN terms in the shift are conveniently written in the form
\begin{equation}\label{shift4PNxidecomp}
\xi^i_{1,\, \text{4PN}} = \frac{1}{\varepsilon}\xi^{i\, (-1)}_{1,\,
  \text{4PN}} + \xi^{(0,y_{1})}_{1,\, \text{4PN}} y^i_{1}
+ \xi^{(0,n_{12})}_{1,\, \text{4PN}} n^i_{12} +
\xi^{(0,v_{1})}_{1,\,\text{4PN}} v^i_1 +
\xi^{(0,v_{12})}_{1,\,\text{4PN}} v^i_{12} \,,
\end{equation}
with $1/\varepsilon$ being the UV pole and $v^i_{12}=v^i_{1}-v^i_{2}$. We have
\begin{subequations}\label{shift4PNxi}
\begin{align}
 \xi^{i\, (-1)}_{1,\, \text{4PN}}={}
 &\frac{G^3 m_{1}^2 m_{2} v_{12}^{i}}{r_{12}^2} \Bigl(11 (n_{12}{} v_{12}{})
 + \frac{11}{3} (n_{12}{} v_{1}{})\Bigr)
 + n_{12}^{i} \biggl [\frac{G^4}{r_{12}^3}\Bigl(\frac{55}{3} m_{1}^3 m_{2}
 + \frac{22}{3} m_{1}^2 m_{2}^2
 + 4 m_{1} m_{2}^3\Bigr) \nonumber\\
& + \frac{G^3 m_{1}^2 m_{2}}{r_{12}^2} \Bigl(\frac{11}{2} (n_{12}{} v_{12}{})^2
 - 11 (n_{12}{} v_{12}{}) (n_{12}{} v_{1}{})
 + \frac{11}{2} (n_{12}{} v_{1}{})^2
 -  \frac{22}{3} v_{12}^{2}\Bigr)\biggr]\,, \\[1ex]
\xi^{(0,y_{1})}_{1,\, \text{4PN}}={}& G^3 (m_{1}
 + m_{2}) \frac{m_{1} m_{2}}{15 r_{12}^3} \biggl(\biggl [-67
 + 48 \ln\Bigl(\frac{r_{12}}{r_{0}}\Bigr)\biggr] (n_{12}{} v_{12}{})^2
 + \biggl [17
 - 16 \ln\Bigl(\frac{r_{12}}{r_{0}}\Bigr)\biggr] v_{12}{}^{2}\biggr)\nonumber\\
& +G^4(m_{1}
 + m_{2})^2 \frac{m_{1} m_{2}}{15r_{12}^4} \biggl [-17
 + 16 \ln\Bigl(\frac{r_{12}}{r_{0}}\Bigr)\biggr]\,, \label{termy1}\\[1ex]
  \xi^{(0,n_{12})}_{1,\, \text{4PN}}={}&
G^3 m_{1} m_{2}^2 \Biggl[\frac{1}{r_{12}^{2}} \biggl(\biggl [- \frac{1753}{80}
 + \frac{72}{5} \ln\Bigl(\frac{r_{12}}{r_{0}}\Bigr)\biggr] (n_{12}{} v_{12}{})^2
 + \biggl [\frac{1753}{40}
 + \frac{96}{5} \ln\Bigl(\frac{r_{12}}{r_{0}}\Bigr)\biggr] (n_{12}{} v_{12}{})
  (n_{12}{} v_{1}{}) \nonumber \\ & \quad + \biggl [- \frac{1753}{60}
  -  \frac{64}{5} \ln\Bigl(\frac{r_{12}}{r_{0}}\Bigr)\biggr] (v_{12}{} v_{1}{})
 + \biggl [\frac{1753}{120}
 -  \frac{32}{5} \ln\Bigl(\frac{r_{12}}{r_{0}}\Bigr)\biggr] v_{12}^{2}\biggr)
 \nonumber \\ &+ \frac{1}{r_{12}^{3}} \biggl(\biggl [- \frac{431}{10}+ 48
 \ln\Bigl(\frac{r_{12}}{r_{0}}\Bigr)\biggr] (n_{12}{} v_{12}{})^2 (n_{12}{} y_{1}{}) 
 + \biggl [\frac{53}{2}
 -  \frac{144}{5} \ln\Bigl(\frac{r_{12}}{r_{0}}\Bigr)\biggr]
                (n_{12}{} v_{12}{}) (v_{12}{} y_{1}{}) \nonumber \\
  & \quad + \biggl [\frac{67}{10}
-  \frac{48}{5} \ln\Bigl(\frac{r_{12}}{r_{0}}\Bigr)\biggr] (n_{12}{}
  y_{1}{}) v_{12}^{2}\biggr)\Biggr] \nonumber\\ &
 + G^3 m_{1}^2 m_{2} \Biggl[\frac{1}{r_{12}^{2}} \biggl(\biggl [- \frac{2761}{168}
 -  \frac{33}{2} \ln\Bigl(\frac{\bar{q}^{1/2} r'_{0}}{\ell_{0}}\Bigr)
 -  \frac{11}{2} \ln\Bigl(\frac{r_{12}}{r'_{0}}\Bigr)\biggr]
(n_{12}{} v_{12}{})^2 \nonumber\\ & \quad + \biggl [\frac{4367}{168}
 + 33 \ln\Bigl(\frac{\bar{q}^{1/2} r'_{0}}{\ell_{0}}\Bigr)
 + \frac{96}{5} \ln\Bigl(\frac{r_{12}}{r_{0}}\Bigr)
 + 11 \ln\Bigl(\frac{r_{12}}{r'_{0}}\Bigr)\biggr] (n_{12}{} v_{12}{}) (n_{12}{} v_{1}{})
 \nonumber\\ & \quad + \biggl [\frac{7489}{840}
 -  \frac{33}{2} \ln\Bigl(\frac{\bar{q}^{1/2} r'_{0}}{\ell_{0}}\Bigr)
 -  \frac{11}{2} \ln\Bigl(\frac{r_{12}}{r'_{0}}\Bigr)\biggr] (n_{12}{} v_{1}{})^2
 + \biggl [- \frac{1753}{60}
 -  \frac{64}{5} \ln\Bigl(\frac{r_{12}}{r_{0}}\Bigr)\biggr] (v_{12}{} v_{1}{})
 \nonumber\\ & \quad + \biggl [\frac{1753}{120}
 + 22 \ln\Bigl(\frac{\bar{q}^{1/2} r'_{0}}{\ell_{0}}\Bigr)
 + \frac{16}{5} \ln\Bigl(\frac{r_{12}}{r_{0}}\Bigr)\biggr] v_{12}^{2}\biggr)
\nonumber\\ & + \frac{1}{r_{12}^{3}} \biggl(\biggl [- \frac{431}{10}
 + 48 \ln\Bigl(\frac{r_{12}}{r_{0}}\Bigr)\biggr] (n_{12}{} v_{12}{})^2 (n_{12}{} y_{1}{})
 + \biggl [\frac{53}{2}
 -  \frac{144}{5} \ln\Bigl(\frac{r_{12}}{r_{0}}\Bigr)\biggr] (n_{12}{}
 v_{12}{}) (v_{12}{} y_{1}{}) \nonumber\\ & \quad + \biggl [\frac{67}{10}
 -  \frac{48}{5} \ln\Bigl(\frac{r_{12}}{r_{0}}\Bigr)\biggr] (n_{12}{} y_{1}{})
  v_{12}^{2}\biggr)\Biggr]\nonumber\\ 
  & + G^4 m_{1}^2 m_{2}^2\biggl( \frac{1}{r_{12}^{3}}
 \biggl [- \frac{88}{3} \ln\Bigl(\frac{\bar{q}^{1/2} r'_{0}}{\ell_{0}}\Bigr)
 -  \frac{16}{5} \ln\Bigl(\frac{r_{12}}{r_{0}}\Bigr)\biggr]
 + \frac{1}{r_{12}^{4}} \biggl [\frac{67}{15}-  \frac{32}{5}
   \ln\Bigl(\frac{r_{12}}{r_{0}}\Bigr)\biggr] (n_{12}{} y_{1}{})\biggr)\nonumber\\
  & + G^4 m_{1}^3 m_{2} \biggl(\frac{1}{r_{12}^{3}}
  \biggl [- \frac{220}{3} \ln\Bigl(\frac{\bar{q}^{1/2} r'_{0}}{\ell_{0}}\Bigr)
 -  \frac{32}{5} \ln\Bigl(\frac{r_{12}}{r_{0}}\Bigr)\biggr]
 + \frac{1}{r_{12}^{4}} \biggl [\frac{67}{30}
    -  \frac{16}{5} \ln\Bigl(\frac{r_{12}}{r_{0}}\Bigr)\biggr]
    (n_{12}{} y_{1}{})\biggr)\nonumber\\
 & + G^4 m_{1} m_{2}^3 \biggl(\frac{1}{r_{12}^{3}}
  \biggl [-16 \ln\Bigl(\frac{\bar{q}^{1/2} r'_{0}}{\ell_{0}}\Bigr)
 + \frac{16}{5} \ln\Bigl(\frac{r_{12}}{r_{0}}\Bigr)\biggr]
 + \frac{1}{r_{12}^{4}} \biggl [\frac{67}{30}
 -  \frac{16}{5} \ln\Bigl(\frac{r_{12}}{r_{0}}\Bigr)\biggr] (n_{12}{}
   y_{1}{})\biggr)\,, \\[1ex]
  \xi^{(0,v_{1})}_{1,\, \text{4PN}}={}
  &G^3 \biggl(\frac{m_{1}^2 m_{2}}{r_{12}^2} \biggl [- \frac{269}{120}
 + \frac{32}{15} \ln\Bigl(\frac{r_{12}}{r_{0}}\Bigr)\biggr] (n_{12}{} v_{12}{})
 + \frac{m_{1} m_{2}^2}{r_{12}^2} \biggl [- \frac{269}{120}
 + \frac{32}{15} \ln\Bigl(\frac{r_{12}}{r_{0}}\Bigr)\biggr] (n_{12}{}
    v_{12}{})\biggr)\,, \\[1ex]
  \xi^{(0,v_{12})}_{1,\, \text{4PN}}={}
  & G^3 m_{1} m_{2}^2 \Biggl[\frac{1}{r_{12}^{2}} \biggl(\biggl [\frac{755}{48}
 - 8 \ln\Bigl(\frac{r_{12}}{r_{0}}\Bigr)\biggr] (n_{12}{} v_{12}{})
 + \biggl [- \frac{1753}{60}
 -  \frac{64}{5} \ln\Bigl(\frac{r_{12}}{r_{0}}\Bigr)\biggr] (n_{12}{} v_{1}{})\biggr)
 \nonumber\\ &+ \frac{1}{r_{12}^{3}} \biggl(\biggl [\frac{53}{2}
  -  \frac{144}{5} \ln\Bigl(\frac{r_{12}}{r_{0}}\Bigr)\biggr]
    (n_{12}{} v_{12}{}) (n_{12}{} y_{1}{})
 + \biggl [- \frac{134}{15}
 + \frac{64}{5} \ln\Bigl(\frac{r_{12}}{r_{0}}\Bigr)\biggr] (v_{12}{} y_{1}{})\biggr)\Biggr]
 \nonumber\\ & + G^3 m_{1}^2 m_{2} \Biggl[\frac{1}{r_{12}^{2}}
 \biggl(\biggl [\frac{23113}{1680}
- 33 \ln\Bigl(\frac{\bar{q}^{1/2} r'_{0}}{\ell_{0}}\Bigr)
 + \frac{11}{3} \ln\Bigl(\frac{r_{12}}{r'_{0}}\Bigr)\biggr] (n_{12}{} v_{12}{})
 \nonumber\\ & + \biggl [- \frac{2572}{105}
 - 11 \ln\Bigl(\frac{\bar{q}^{1/2} r'_{0}}{\ell_{0}}\Bigr)
 -  \frac{64}{5} \ln\Bigl(\frac{r_{12}}{r_{0}}\Bigr)
 -  \frac{11}{3} \ln\Bigl(\frac{r_{12}}{r'_{0}}\Bigr)\biggr] (n_{12}{} v_{1}{})\biggr)
 \nonumber\\ & + \frac{1}{r_{12}^{3}} \biggl(\biggl [\frac{53}{2}
   -  \frac{144}{5} \ln\Bigl(\frac{r_{12}}{r_{0}}\Bigr)\biggr]
   (n_{12}{} v_{12}{}) (n_{12}{} y_{1}{})
 + \biggl [- \frac{134}{15}
 + \frac{64}{5} \ln \Bigl(\frac{r_{12}}{r_{0}}\Bigr)\biggr]
    (v_{12}{} y_{1}{})\biggr)\Biggr]\, .
\end{align}
\end{subequations}

On the other hand the shift $\bm{\eta}_1$ is quite simple. It starts only at the 4PN order and it is made only of $G^3$ and $G^4$ terms,
\begin{equation}\label{shifteta}
\bm{\eta}_1 = \frac{G^3}{c^8}\bm{\eta}^{(3)}_{1,\,\text{4PN}}+ \frac{G^4}{c^8}\bm{\eta}^{(4)}_{1,\,\text{4PN}}\,.
\end{equation}
where
\begin{subequations}\label{detailshifteta}
	\begin{align}
	\bm{\eta}^{(3)}_{1,\,\text{4PN}}={}
	&\frac{\bm{v}_{12}}{r_{12}^2} \Bigl(\frac{769}{24} m_{1}^2 m_{2} (n_{12} v_{12})
	+ \frac{561}{35} m_{1} m_{2}^2 (n_{12} v_{12})\Bigr)\\
	& + \frac{\bm{n}_{12}}{r_{12}^2} \biggl[m_{1} m_{2}^2 \Bigl(\frac{21719}{1400} (n_{12} v_{12})^2
	-  \frac{2096}{175} v_{12}^{2}\Bigr) + m_{1}^2 m_{2} \Bigl(- \frac{2119}{50} (n_{12} v_{12})^2
	+ \frac{58769}{2100} v_{12}^{2}\Bigr)\biggr]\nonumber\,,
	\\
	\bm{\eta}^{(4)}_{1,\,\text{4PN}}={}
	&\Bigl(\frac{8861}{2100} m_{1}^3 m_{2}
	+ \frac{613}{350} m_{1}^2 m_{2}^2
	-  \frac{5183}{2100} m_{1} m_{2}^3\Bigr) \frac{\bm{n}_{12}}{r_{12}^3}
	\,.
	\end{align}\end{subequations}

\section{The 4PN mass quadrupole as a function of the potentials}
\label{app:MQAsPot}

\subsection{Nomenclature of terms and conventions}
\label{nomencl}

In $d$ dimensions, the mass-type quadrupole moment is given by Eq.~\eqref{ILexpr2} for $\ell=2$. The first three terms of~\eqref{ILexpr2} will be denoted respectively by $S_{ij}$, $V_{ij}$ and $T_{ij}$ with $S$, $V$ and $T$ standing for scalar, vector and tensor. We recall that the fourth term in~\eqref{ILexpr2} does not contribute. Thus we write
\begin{equation}\label{nomencl1}
I_{ij} = S_{ij} + V_{ij} + T_{ij}\,.
\end{equation}
Furthermore, we know from the series expansion~\eqref{series} that each term $S$, $V$ and $T$ corresponds to a finite sum at 4PN order indexed by the integer $k$ in~\eqref{series}. At the 4PN order this sum has 5 elements ($k=0,1,2,3,4$) for $S$, 4 elements for $V$ and 3 for $T$. We denote these terms using capital roman numerals hence
\begin{subequations}\label{nomencl2}
	\begin{align}
	S &= S\mathrm{I} + S\mathrm{II} + S\mathrm{III} + S\mathrm{IV} + S\mathrm{V} \,,\\
	V &= V\mathrm{I} + V\mathrm{II} + V\mathrm{III} + V\mathrm{IV} \,,\\
	T &= T\mathrm{I} + T\mathrm{II} + T\mathrm{III} \,.
	\end{align}
\end{subequations}
In our notation we forget about the always understood quadrupole indices $ij$, and in practical calculations we let these two indices to be free, being understood that at the end of the calculation the STP projection must be applied. Since there are poles $\propto 1/\varepsilon$ it is crucial to apply the STF projection in $d$ dimensions, \textit{i.e.} $\hat{T}^{ij}=T^{ij}-\frac{1}{d}\delta^{ij}T^{kk}$. However, alternatively, one can wait till the end of the calculation when the poles have been removed by the appropriate shifts [see Appendix~\ref{app:shift}] and apply finally the usual STF projection in 3 dimensions.

In addition, each of these terms contains non-compact support terms which we denote with the suffix NC (such as $S$INC), compact support terms (suffix C) and surface terms of two sorts, either surface Laplacian terms (SL) or surface divergence terms (SD). Hence we shall write for the term $S$I:
\begin{equation}\label{nomencl3}
S\mathrm{I} = S\mathrm{IC} + S\mathrm{INC} + S\mathrm{ISL} + S\mathrm{ISD} \,,
\end{equation}
and similarly for the other terms. Finally we shall indicate the different PN pieces with the index $n$PN, it being understood that the term is to be multiplied by the appropriate factor $1/c^{2n}$. For simplicity we omit writing the finite part integral $\text{FP}_{B=0}\int \dd^d\mathbf{x}(r/r_0)^B$ in front of each terms, and we also do not write the necessary time derivatives $(\dd/\dd t)^{2k}$ present in the series expansion, see Eq.~\eqref{series}. Also we do not write the terms that are equal to zero. All the non-compact and surface terms shown below have to be divided by a factor $G \pi$. Remind also our notation~\eqref{PsiL} for the terms involving super-potentials, for instance $\Psi^{\partial_{ab}V}_{ij}$ or $\Psi^{\partial_a V}_{ijk}$.

\subsection{Exhaustive list of terms}
\label{exhaustive}

We exhaustively provide all the terms composing the 4PN quadrupole moment following this nomenclature and conventions. We start with the list of compact (C) terms.
{\footnotesize
\begin{align*}
S\text{IC}_{0\text{PN}}={}&\frac{(d - 1)^{}}{2 (d - 2)} \hat{x}_{ij} \sigma ,\\
S\text{IC}_{1\text{PN}}={}&- \frac{(d - 3)^{} (d - 1)^{}}{(d - 2)^2} V \hat{x}_{ij} \sigma,\\
S\text{IC}_{2\text{PN}}={}&\Bigl(\frac{(d - 3)^2 (d - 1)^{}}{(d - 2)^3} V^2 \sigma
 + \frac{4 (d - 3)^{}}{d - 2} V_{a} \sigma_{a}
 + \frac{2 (d - 1)^{}}{(d - 2)^2} V \sigma_{a}{}_{a}
 + \frac{2 (d - 3)^2 (d - 1)^{}}{(d - 2)^3} K \sigma \Bigr) \hat{x}_{ij}\nonumber\\
& + \frac{2 (d - 1)^{}}{d - 2} \sigma_{ab} \Psi_{ij}^{\partial_{ab} V}
 -  \frac{2 (d - 1)^{}}{(d - 2)^2} \sigma_{b}{}_{b} \Psi_{ij}^{\partial_{aa} V},\\
S\text{IC}_{3\text{PN}}={}&\Bigl(- \frac{2 (d - 3)^3 (d - 1)^{}}{3 (d - 2)^4} V^3 \sigma
 + \frac{8 (d - 1)^{}}{(d - 2)^2} V_{a} V_{a} \sigma
 -  \frac{4 (d - 3)^{} (d - 1)^{}}{(d - 2)^2} \hat{X} \sigma
 -  \frac{4 (d - 5)^{} (d - 3)^{}}{(d - 2)^2} V V_{a} \sigma_{a}\nonumber\\
& + \frac{8 (d - 3)^{}}{d - 2} \hat{R}_{a} \sigma_{a}
 -  \frac{2 (d - 5)^{} (d - 1)^{}}{(d - 2)^3} V^2 \sigma_{a}{}_{a}
 + \frac{4}{d - 2} \hat{W}_{ab} \sigma_{ab}
 -  \frac{4 (d - 3)^3 (d - 1)^{}}{(d - 2)^4} K V \sigma\nonumber\\
& -  \frac{4 (d - 3)^{} (d - 1)^{}}{(d - 2)^3} K \sigma_{a}{}_{a}\Bigr) \hat{x}_{ij}
 + \frac{8 (d - 1)^2}{(d - 2)^2} \sigma \Psi_{ij}^{\partial_{a}V_{b}} \partial_{b}V_{a}
 + \frac{8 (d - 1)^2}{(d - 2)^2} V_{a} \Psi_{ij}^{\partial_{a}V_{b}} \partial_{b}\sigma\nonumber\\
& -  \frac{4 (d - 1)^2}{(d - 2)^2} V_{a} \sigma \Psi_{ij}^{\partial_{ta} V}
 + \frac{8 (d - 5)^{} (d - 1)^{}}{(d - 2)^2} V \Psi_{ij}^{\partial_{a}V_{b}} \partial_{b}\sigma_{a}
 + \frac{8 (d - 5)^{} (d - 1)^{}}{(d - 2)^2} \sigma_{a} \Psi_{ij}^{\partial_{a}V_{b}} \partial_{b}V\nonumber\\
& -  \frac{4 (d - 5)^{} (d - 1)^{}}{(d - 2)^2} V \sigma_{a} \Psi_{ij}^{\partial_{ta} V}
 + \frac{8 (d - 1)^{}}{(d - 2)^2} V \sigma_{ab} \Psi_{ij}^{\partial_{ab} V}
 -  \frac{8 (d - 1)^{}}{(d - 2)^3} V \sigma_{b}{}_{b} \Psi_{ij}^{\partial_{aa} V},\\
S\text{IC}_{4\text{PN}}={}&\Bigl(\frac{(d - 3)^{} (d - 1)^{} (3 d^3 - 23 d^2 + 55 d - 43)^{}}{3 (d - 2)^5} V^4 \sigma
 + \frac{4 (d - 1)^{} (2 d^2 - 13 d + 19)^{}}{(d - 2)^3} V V_{a} V_{a} \sigma\nonumber\\
& -  \frac{4 (d - 3)^{} (d - 1)^2}{(d - 2)^3} V^2 \hat{W} \sigma
 -  \frac{(d - 1)^{} (d + 2)^{}}{(d - 2)^2} \hat{W}_{ab} \hat{W}_{ab} \sigma
 -  \frac{8 (d - 8)^{} (d - 1)^{}}{(d - 2)^2} \hat{R}_{a} V_{a} \sigma\nonumber\\
& + \frac{4 (d - 3)^{} (d - 1)^{} (3 d - 7)^{}}{(d - 2)^3} V \hat{X} \sigma
 + \frac{8 (d - 1)^2}{(d - 2)^2} V \hat{Z} \sigma
 -  \frac{16 (d - 3)^{} (d - 1)^{}}{(d - 2)^2} \hat{T} \sigma
 -  \frac{16 (d - 1)^{}}{(d - 2)^2} \hat{M} \sigma\nonumber\\
& + \frac{4 (3 d^3 - 25 d^2 + 71 d - 65)^{}}{(d - 2)^3} V^2 V_{a} \sigma_{a}
 -  \frac{8 (d - 3)^{}}{d - 2} V_{a} \hat{W} \sigma_{a}
 + \frac{8 (d - 3)^{}}{d - 2} V_{a} \hat{W}_{ab} \sigma_{b}\nonumber\\
& -  \frac{8 (2 d^2 - 13 d + 19)^{}}{(d - 2)^2} \hat{R}_{a} V \sigma_{a}
 + \frac{16 (d - 3)^{}}{d - 2} \hat{Y}_{a} \sigma_{a}
 + \frac{2 (d - 1)^{} (5 d^2 - 37 d + 56)^{}}{3 (d - 2)^4} V^3 \sigma_{a}{}_{a}
 + \frac{8 (d - 3)^{}}{d - 2} V_{a} V_{b} \sigma_{ab}\nonumber\\
& + \frac{8 (d - 3)^{}}{(d - 2)^2} V_{a} V_{a} \sigma_{b}{}_{b}
 -  \frac{2 (d^2 -  d - 8)^{}}{(d - 2)^2} V \hat{W}_{ab} \sigma_{ab}
 + \frac{2 d (d - 1)^{}}{(d - 2)^3} V \hat{W} \sigma_{a}{}_{a}
 + \frac{8 (d - 1)^{}}{(d - 2)^2} \hat{X} \sigma_{a}{}_{a}
 + \frac{16}{d - 2} \hat{Z}_{ab} \sigma_{ab}\nonumber\\
& + \frac{(d - 3)^{} (d - 1)^{} (11 d^3 - 76 d^2 + 175 d - 142)^{}}{(d - 2)^5} K V^2 \sigma
 + \frac{(d - 3)^2 (d - 1)^{} (7 d^2 - 33 d + 42)^{}}{(d - 2)^5} K^2 \sigma\nonumber\\
& -  \frac{4 (d - 3)^{} (d - 1)^2}{(d - 2)^3} K \hat{W} \sigma
 + \frac{16 (d - 3)^{} (d^2 - 5 d + 8)^{}}{(d - 2)^3} K V_{a} \sigma_{a}
 + \frac{16 (d - 3)^2 (d - 1)^{}}{(d - 2)^4} K V \sigma_{a}{}_{a}\Bigr) \hat{x}_{ij}\nonumber\\
& -  \frac{8 (d - 5)^{} (d - 1)^2}{(d - 2)^3} V \sigma \Psi_{ij}^{\partial_{a}V_{b}} \partial_{b}V_{a}
 -  \frac{8 (d - 5)^{} (d - 1)^2}{(d - 2)^3} V_{a} \sigma \Psi_{ij}^{\partial_{a}V_{b}} \partial_{b}V
 -  \frac{8 (d - 5)^{} (d - 1)^2}{(d - 2)^3} V V_{a} \Psi_{ij}^{\partial_{a}V_{b}} \partial_{b}\sigma\nonumber\\
& + \frac{16 (d - 1)^2}{(d - 2)^2} \sigma \Psi_{ij}^{\partial_{a}V_{b}} \partial_{b}\hat{R}_{a}
 + \frac{16 (d - 1)^2}{(d - 2)^2} \hat{R}_{a} \Psi_{ij}^{\partial_{a}V_{b}} \partial_{b}\sigma
 + \frac{4 (d - 5)^{} (d - 1)^2}{(d - 2)^3} V V_{a} \sigma \Psi_{ij}^{\partial_{ta} V}\nonumber\\
& -  \frac{8 (d - 1)^2}{(d - 2)^2} \hat{R}_{a} \sigma \Psi_{ij}^{\partial_{ta} V}
 -  \frac{(d - 3)^2 (d - 1)^2}{(d - 2)^4} V^3 \sigma \Psi_{ij}^{\partial_{aa} V}
 + \frac{8 (d - 1)^2}{(d - 2)^2} V_{a} V_{b} \sigma \Psi_{ij}^{\partial_{ab} V}\nonumber\\
& -  \frac{4 (d - 1)^2}{(d - 2)^2} V_{b} V_{b} \sigma \Psi_{ij}^{\partial_{aa} V}
 -  \frac{2 (d - 1)^2}{(d - 2)^2} \hat{X} \sigma \Psi_{ij}^{\partial_{aa} V}
 + \frac{32 (d - 3)^{} (d - 1)^{}}{(d - 2)^3} V^2 \Psi_{ij}^{\partial_{a}V_{b}} \partial_{b}\sigma_{a}\nonumber\\
& + \frac{64 (d - 3)^{} (d - 1)^{}}{(d - 2)^3} V \sigma_{a} \Psi_{ij}^{\partial_{a}V_{b}} \partial_{b}V
 -  \frac{16 (d - 1)^{}}{d - 2} \hat{W}_{a}{}_{k} \Psi_{ij}^{\partial_{k}V_{b}} \partial_{b}\sigma_{a}
 -  \frac{16 (d - 1)^{}}{d - 2} \sigma_{a} \Psi_{ij}^{\partial_{b}V_{k}} \partial_{k}\hat{W}_{ab}\nonumber\\
& -  \frac{16 (d - 3)^{} (d - 1)^{}}{(d - 2)^3} V^2 \sigma_{a} \Psi_{ij}^{\partial_{ta} V}
 + \frac{8 (d - 1)^{}}{d - 2} \hat{W}_{a}{}_{b} \sigma_{a} \Psi_{ij}^{\partial_{tb} V}
 + \frac{8 (d - 5)^{} (d - 1)^{}}{(d - 2)^2} V V_{a} \sigma_{b} \Psi_{ij}^{\partial_{ab} V}\nonumber\\
& -  \frac{8 (d - 4)^{} (d - 1)^{}}{(d - 2)^2} V V_{b} \sigma_{b} \Psi_{ij}^{\partial_{aa} V}
 -  \frac{16 (d - 1)^{}}{d - 2} \hat{R}_{a} \sigma_{b} \Psi_{ij}^{\partial_{ab} V}
 + \frac{8 (d - 1)^{}}{d - 2} \hat{R}_{b} \sigma_{b} \Psi_{ij}^{\partial_{aa} V}\nonumber\\
& -  \frac{16 (d - 1)^{}}{(d - 2)^2} \sigma_{k}{}_{k} \Psi_{ij}^{\partial_{a}V_{b}} \partial_{b}V_{a}
 -  \frac{16 (d - 1)^{}}{d - 2} \sigma_{a}{}_{k} \Psi_{ij}^{\partial_{k}V_{b}} \partial_{b}V_{a}
 -  \frac{16 (d - 1)^{}}{d - 2} V_{a} \Psi_{ij}^{\partial_{b}V_{k}} \partial_{k}\sigma_{ab}\nonumber\\
& -  \frac{16 (d - 1)^{}}{(d - 2)^2} V_{a} \Psi_{ij}^{\partial_{a}V_{b}} \partial_{b}\sigma_{k}{}_{k}
 + \frac{8 (d - 1)^{}}{(d - 2)^2} V_{a} \sigma_{b}{}_{b} \Psi_{ij}^{\partial_{ta} V}
 + \frac{8 (d - 1)^{}}{d - 2} V_{a} \sigma_{a}{}_{b} \Psi_{ij}^{\partial_{tb} V}
 + \frac{16 (d - 1)^{}}{(d - 2)^3} V^2 \sigma_{ab} \Psi_{ij}^{\partial_{ab} V}\nonumber\\
& -  \frac{2 (d - 1)^2}{(d - 2)^3} V^2 \sigma_{b}{}_{b} \Psi_{ij}^{\partial_{aa} V}
 + \frac{8 (d - 1)^{}}{d - 2} \hat{W}_{ab} \sigma_{a}{}_{k} \Psi_{ij}^{\partial_{bk} V}
 -  \frac{2 (d - 1)^{}}{d - 2} \hat{W}_{ka} \sigma_{ka} \Psi_{ij}^{\partial_{bb} V}
 + \frac{2 (d - 1)^{}}{(d - 2)^2} \hat{W} \sigma_{k}{}_{k} \Psi_{ij}^{\partial_{aa} V}\nonumber\\
& -  \frac{4 (d - 3)^2 (d - 1)^2}{(d - 2)^4} K V \sigma \Psi_{ij}^{\partial_{aa} V}
 -  \frac{16 (d - 5)^{} (d - 3)^{} (d - 1)^{}}{(d - 2)^3} K \Psi_{ij}^{\partial_{a}V_{b}} \partial_{b}\sigma_{a}\nonumber\\
& -  \frac{16 (d - 5)^{} (d - 3)^{} (d - 1)^{}}{(d - 2)^3} \sigma_{a} \Psi_{ij}^{\partial_{a}V_{b}} \partial_{b}K
 + \frac{8 (d - 5)^{} (d - 3)^{} (d - 1)^{}}{(d - 2)^3} K \sigma_{a} \Psi_{ij}^{\partial_{ta} V}\nonumber\\
& -  \frac{16 (d - 3)^{} (d - 1)^{}}{(d - 2)^3} K \sigma_{ab} \Psi_{ij}^{\partial_{ab} V},
\nonumber\\
S\text{IIC}_{1\text{PN}}={}&\frac{(d - 1)^{}}{4 (d - 2) (d + 4)} r^2 \hat{x}_{ij} \sigma, \\
S\text{IIC}_{2\text{PN}}={}&\frac{(d - 1)^{}}{(d - 2)^2 (d + 4)} r^2 V \hat{x}_{ij} \sigma, \nonumber \\
S\text{IIC}_{3\text{PN}}={}&\Bigl(\frac{2 (d - 1)^{}}{(d - 2)^3 (d + 4)} V^2
 + \frac{(d - 1)^{}}{2 (d - 2) (d + 4)} \hat{W}
 -  \frac{2 (d - 3)^{} (d - 1)^{}}{(d - 2)^3 (d + 4)} K\Bigr) r^2 \hat{x}_{ij} \sigma,\\
S\text{IIC}_{4\text{PN}}={}&\Bigl(\frac{8 (d - 1)^{}}{3 (d - 2)^4 (d + 4)} V^3
 + \frac{4 (d - 1)^{}}{(d - 2)^2 (d + 4)} V_{a} V_{a}
 + \frac{2 (d - 1)^{}}{(d - 2)^2 (d + 4)} V \hat{W}
 + \frac{4 (d - 1)^{}}{(d - 2)^2 (d + 4)} \hat{X}\nonumber\\
& + \frac{2 (d - 1)^{}}{(d - 2) (d + 4)} \hat{Z}
 -  \frac{8 (d - 3)^{} (d - 1)^{}}{(d - 2)^4 (d + 4)} K V\Bigr) r^2 \hat{x}_{ij} \sigma,
\nonumber\\
S\text{IIIC}_{2\text{PN}}={}&\frac{(d - 1)^{}}{16 (d - 2) (d + 4) (d + 6)} r^4 \hat{x}_{ij} \sigma,\\
S\text{IIIC}_{3\text{PN}}={}&\frac{(d - 1)^{}}{4 (d - 2)^2 (d + 4) (d + 6)} r^4 V \hat{x}_{ij} \sigma,\\
S\text{IIIC}_{4\text{PN}}={}&\Bigl(\frac{(d - 1)^{}}{2 (d - 2)^3 (d + 4) (d + 6)} V^2
 + \frac{(d - 1)^{}}{8 (d - 2) (d + 4) (d + 6)} \hat{W}
 -  \frac{(d - 3)^{} (d - 1)^{}}{2 (d - 2)^3 (d + 4) (d + 6)} K\Bigr) r^4 \hat{x}_{ij} \sigma, \nonumber\\
S\text{IVNC}_{4\text{PN}}={}&- \frac{(d - 1)^2}{192 (d - 2)^2 (d + 4) (d + 6) (d + 8)} r^6 \hat{x}_{ij} \partial_{a}V \partial_{a}V \nonumber\\
S\text{VC}_{4\text{PN}}={}&\frac{(d - 1)^{}}{768 (d - 2) (d + 4) (d + 6) (d + 8) (d + 10)} r^8 \hat{x}_{ij} \sigma, \nonumber\\
V\text{IC}_{1\text{PN}}={}&- \frac{2 (d - 1)^{} (d + 2)^{}}{d (d - 2) (d + 4)} \hat{x}_{ija} \sigma_{a},\\
V\text{IC}_{2\text{PN}}={}&\Bigl(\frac{2 (d - 1)^2 (d + 2)^{}}{d (d - 2)^2 (d + 4)} V_{a} \sigma
 + \frac{2 (d - 5)^{} (d - 1)^{} (d + 2)^{}}{d (d - 2)^2 (d + 4)} V \sigma_{a}\Bigr) \hat{x}_{ija},\\
V\text{IC}_{3\text{PN}}={}&\Bigl(- \frac{4 (d - 3)^{} (d - 1)^2 (d + 2)^{}}{d (d - 2)^3 (d + 4)} V V_{a} \sigma
 + \frac{4 (d - 1)^2 (d + 2)^{}}{d (d - 2)^2 (d + 4)} \hat{R}_{a} \sigma
 -  \frac{(d - 5)^2 (d - 1)^{} (d + 2)^{}}{d (d - 2)^3 (d + 4)} V^2 \sigma_{a}\nonumber\\
& -  \frac{4 (d - 1)^{} (d + 2)^{}}{d (d - 2) (d + 4)} \hat{W}_{ab} \sigma_{b}
 -  \frac{4 (d - 1)^{} (d + 2)^{}}{d (d - 2) (d + 4)} V_{b} \sigma_{ab}
 -  \frac{4 (d - 1)^{} (d + 2)^{}}{d (d - 2)^2 (d + 4)} V_{a} \sigma_{b}{}_{b}\nonumber\\
& -  \frac{4 (d - 5)^{} (d - 3)^{} (d - 1)^{} (d + 2)^{}}{d (d - 2)^3 (d + 4)} K \sigma_{a}\Bigr) \hat{x}_{ija}
 -  \Bigl(\frac{4 (d - 1)^3 (d + 2)^{}}{d (d - 2)^3 (d + 4)} \sigma \partial_{b}V_{a}
 + \frac{4 (d - 1)^3 (d + 2)^{}}{d (d - 2)^3 (d + 4)} V_{a} \partial_{b}\sigma\nonumber\\
& + \frac{4 (d - 5)^{} (d - 1)^2 (d + 2)^{}}{d (d - 2)^3 (d + 4)} \sigma_{a} \partial_{b}V
 + \frac{4 (d - 5)^{} (d - 1)^2 (d + 2)^{}}{d (d - 2)^3 (d + 4)} V \partial_{b}\sigma_{a}\Bigr) \Psi_{ijb}^{\partial_{a} V},\\
V\text{IC}_{4\text{PN}}={}&\Bigl(- \frac{16 (d - 1)^2 (d + 2)^{}}{d (d - 2)^3 (d + 4)} V^2 V_{a} \sigma
 -  \frac{4 (d - 5)^{} (d - 1)^2 (d + 2)^{}}{d (d - 2)^3 (d + 4)} \hat{R}_{a} V \sigma
 + \frac{8 (d - 1)^2 (d + 2)^{}}{d (d - 2)^2 (d + 4)} \hat{Y}_{a} \sigma\nonumber\\
& -  \frac{2 (d - 1)^{} (d + 2)^{} (3 d^3 - 9 d^2 - 15 d + 53)^{}}{3 d (d - 2)^4 (d + 4)} V^3 \sigma_{a}
 + \frac{24 (d - 1)^{} (d + 2)^{}}{d (d - 2) (d + 4)} V_{a} V_{b} \sigma_{b}\nonumber\\
& + \frac{4 (d - 8)^{} (d - 1)^{} (d + 2)^{}}{d (d - 2)^2 (d + 4)} V_{b} V_{b} \sigma_{a}
 + \frac{4 (d - 5)^{} (d - 1)^{} (d + 2)^{}}{d (d - 2)^2 (d + 4)} V \hat{W}_{ab} \sigma_{b}
 + \frac{8 (d - 5)^{} (d - 1)^{} (d + 2)^{}}{d (d - 2)^2 (d + 4)} \hat{X} \sigma_{a}\nonumber\\
& -  \frac{16 (d - 1)^{} (d + 2)^{}}{d (d - 2) (d + 4)} \hat{Z}_{ab} \sigma_{b}
 -  \frac{16 (d - 1)^{} (d + 2)^{}}{d (d - 2)^2 (d + 4)} V V_{b} \sigma_{ab}
 + \frac{8 (d - 3)^{} (d - 1)^{} (d + 2)^{}}{d (d - 2)^3 (d + 4)} V V_{a} \sigma_{b}{}_{b}\nonumber\\
& -  \frac{8 (d - 1)^{} (d + 2)^{}}{d (d - 2) (d + 4)} \hat{R}_{b} \sigma_{ab}
 -  \frac{8 (d - 1)^{} (d + 2)^{}}{d (d - 2)^2 (d + 4)} \hat{R}_{a} \sigma_{b}{}_{b}
 + \frac{8 (d - 3)^2 (d - 1)^2 (d + 2)^{}}{d (d - 2)^4 (d + 4)} K V_{a} \sigma\nonumber\\
& + \frac{4 (d - 5)^2 (d - 3)^{} (d - 1)^{} (d + 2)^{}}{d (d - 2)^4 (d + 4)} K V \sigma_{a}\Bigr) \hat{x}_{ija}
 + \frac{4 (d - 5)^{} (d - 1)^3 (d + 2)^{}}{d (d - 2)^4 (d + 4)} V_{a} \sigma \Psi_{ijb}^{\partial_{a} V} \partial_{b}V\nonumber\\
& + \frac{4 (d - 5)^{} (d - 1)^3 (d + 2)^{}}{d (d - 2)^4 (d + 4)} V \sigma \Psi_{ijb}^{\partial_{a} V} \partial_{b}V_{a}
 + \frac{4 (d - 5)^{} (d - 1)^3 (d + 2)^{}}{d (d - 2)^4 (d + 4)} V V_{a} \Psi_{ijb}^{\partial_{a} V} \partial_{b}\sigma\nonumber\\
& -  \frac{8 (d - 1)^3 (d + 2)^{}}{d (d - 2)^3 (d + 4)} \sigma \Psi_{ijb}^{\partial_{a} V} \partial_{b}\hat{R}_{a}
 -  \frac{8 (d - 1)^3 (d + 2)^{}}{d (d - 2)^3 (d + 4)} \hat{R}_{a} \Psi_{ijb}^{\partial_{a} V} \partial_{b}\sigma\nonumber\\
& -  \frac{32 (d - 3)^{} (d - 1)^2 (d + 2)^{}}{d (d - 2)^4 (d + 4)} V \sigma_{a} \Psi_{ijb}^{\partial_{a} V} \partial_{b}V
 -  \frac{16 (d - 3)^{} (d - 1)^2 (d + 2)^{}}{d (d - 2)^4 (d + 4)} V^2 \Psi_{ijb}^{\partial_{a} V} \partial_{b}\sigma_{a}\nonumber\\
& + \frac{8 (d - 1)^2 (d + 2)^{}}{d (d - 2)^2 (d + 4)} \hat{W}_{a}{}_{k} \Psi_{ijb}^{\partial_{k} V} \partial_{b}\sigma_{a}
 + \frac{8 (d - 1)^2 (d + 2)^{}}{d (d - 2)^2 (d + 4)} \sigma_{a} \Psi_{ijk}^{\partial_{b} V} \partial_{k}\hat{W}_{a}{}_{b}\nonumber\\
& + \frac{8 (d - 1)^2 (d + 2)^{}}{d (d - 2)^3 (d + 4)} \sigma_{k}{}_{k} \Psi_{ijb}^{\partial_{a} V} \partial_{b}V_{a}
 + \frac{8 (d - 1)^2 (d + 2)^{}}{d (d - 2)^2 (d + 4)} \sigma_{a}{}_{k} \Psi_{ijb}^{\partial_{k} V} \partial_{b}V_{a}
 + \frac{8 (d - 1)^2 (d + 2)^{}}{d (d - 2)^3 (d + 4)} V_{a} \Psi_{ijk}^{\partial_{a} V} \partial_{k}\sigma_{b}{}_{b}\nonumber\\
& + \frac{8 (d - 1)^2 (d + 2)^{}}{d (d - 2)^2 (d + 4)} V_{a} \Psi_{ijk}^{\partial_{b} V} \partial_{k}\sigma_{a}{}_{b}
 + \frac{8 (d - 5)^{} (d - 3)^{} (d - 1)^2 (d + 2)^{}}{d (d - 2)^4 (d + 4)} \sigma_{a} \Psi_{ijb}^{\partial_{a} V} \partial_{b}K\nonumber\\
& + \frac{8 (d - 5)^{} (d - 3)^{} (d - 1)^2 (d + 2)^{}}{d (d - 2)^4 (d + 4)} K \Psi_{ijb}^{\partial_{a} V} \partial_{b}\sigma_{a}, \nonumber\\
V\text{IIC}_{2\text{PN}}={}&- \frac{(d - 1)^{} (d + 2)^{}}{d (d - 2) (d + 4) (d + 6)} r^2 \hat{x}_{ija} \sigma_{a},\\
V\text{IIC}_{3\text{PN}}={}&- \frac{4 (d - 1)^{} (d + 2)^{}}{d (d - 2)^2 (d + 4) (d + 6)} r^2 V \hat{x}_{ija} \sigma_{a},\\
V\text{IIC}_{4\text{PN}}={}&\Bigl(- \frac{8 (d - 1)^{} (d + 2)^{}}{d (d - 2)^3 (d + 4) (d + 6)} V^2
 -  \frac{2 (d - 1)^{} (d + 2)^{}}{d (d - 2) (d + 4) (d + 6)} \hat{W}
 + \frac{8 (d - 3)^{} (d - 1)^{} (d + 2)^{}}{d (d - 2)^3 (d + 4) (d + 6)} K\Bigr) r^2 \hat{x}_{ija} \sigma_{a}, \nonumber\\
V\text{IIIC}_{3\text{PN}}={}&- \frac{(d - 1)^{} (d + 2)^{}}{4 d (d - 2) (d + 4) (d + 6) (d + 8)} r^4 \hat{x}_{ija} \sigma_{a},\\
V\text{IIIC}_{4\text{PN}}={}&- \frac{(d - 1)^{} (d + 2)^{}}{d (d - 2)^2 (d + 4) (d + 6) (d + 8)} r^4 V \hat{x}_{ija} \sigma_{a},\nonumber\\
V\text{IVC}_{4\text{PN}}={}&- \frac{(d - 1)^{} (d + 2)^{}}{24 d (d - 2) (d + 4) (d + 6) (d + 8) (d + 10)} r^6 \hat{x}_{ija} \sigma_{a}, \nonumber\\
T\text{IC}_{2\text{PN}}={}&\frac{(d - 1)^{} (d + 2)^{}}{d (d - 2) (d + 1) (d + 6)} \hat{x}_{ijab} \sigma_{ab},\\
T\text{IC}_{3\text{PN}}={}&\frac{4 (d - 1)^{} (d + 2)^{}}{d (d - 2)^2 (d + 1) (d + 6)} V \hat{x}_{ijab} \sigma_{ab},\\
T\text{IC}_{4\text{PN}}={}&\Bigl(\frac{8 (d - 1)^{} (d + 2)^{}}{d (d - 2)^3 (d + 1) (d + 6)} V^2
 -  \frac{8 (d - 3)^{} (d - 1)^{} (d + 2)^{}}{d (d - 2)^3 (d + 1) (d + 6)} K
 + \frac{2 (d - 1)^{} (d + 2)^{}}{d (d - 2) (d + 1) (d + 6)} \hat{W}\nonumber\\
& -  \frac{8 (d - 3)^{} (d - 1)^{} (d + 2)^{}}{d (d - 2)^3 (d + 1) (d + 6)} K\Bigr) \hat{x}_{ijab} \sigma_{ab}, \nonumber\\
T\text{IIC}_{3\text{PN}}={}&\frac{(d - 1)^{} (d + 2)^{}}{2 d (d - 2) (d + 1) (d + 6) (d + 8)} r^2 \hat{x}_{ijab} \sigma_{ab},\\
T\text{IIC}_{4\text{PN}}={}&\frac{2 (d - 1)^{} (d + 2)^{}}{d (d - 2)^2 (d + 1) (d + 6) (d + 8)} r^2 V \hat{x}_{ijab} \sigma_{ab},\nonumber\\
T\text{IIIC}_{4\text{PN}}={}&\frac{(d - 1)^{} (d + 2)^{}}{8 d (d - 2) (d + 1) (d + 6) (d + 8) (d + 10)} r^4 \hat{x}_{ijab} \sigma_{ab}.
\end{align*}}
\noindent
Next we present the long list of the non-compact (NC) support terms. Remind that all the terms have to be divided by a factor $G \pi$. 
{\footnotesize \begin{align*}
S\text{INC}_{2\text{PN}}={}&\Bigl(\frac{(d - 4)^{} (d - 1)^2}{8 (d - 2)^3} (\partial_t V)^2
 -  \frac{(d - 1)^{}}{d - 2} V_{a} \partial_t \partial_{a}V
 + \frac{(d - 1)^{}}{d - 2} \partial_{a}V_{b} \partial_{b}V_{a}\Bigr) \hat{x}_{ij}
 + \frac{(d - 1)^2}{4 (d - 2)^2} \Psi_{ij}^{\partial_{ab} V} \partial_{a}V \partial_{b}V,\\
S\text{INC}_{3\text{PN}}={}&\Bigl(\frac{(d - 4)^{} (d - 1)^3}{4 (d - 2)^4} (\partial_t V)^2 V
 -  \frac{(d - 1)^2}{(d - 2)^2} V V_{a} \partial_t \partial_{a}V
 -  \frac{d (d - 1)^2}{2 (d - 2)^3} V_{a} \partial_t V \partial_{a}V
 + \frac{2}{d - 2} \partial_t V_{a} \partial_t V_{a}\nonumber\\
& -  \frac{(d - 3)^{}}{d - 2} V_{a} \partial_t^{2} V_{a}
 -  \frac{2 (d - 1)^2}{(d - 2)^2} V_{a} \partial_{b}V_{a} \partial_{b}V
 + \frac{(d - 1)^{}}{2 (d - 2)} \partial_t V \partial_t \hat{W}
 + \frac{(d - 1)^{}}{4 (d - 2)} \hat{W} \partial_t^{2} V
 + \frac{(d - 1)^{}}{4 (d - 2)} V \partial_t^{2} \hat{W}\nonumber\\
& + \frac{2 (d - 1)^{}}{d - 2} \partial_t \hat{W}_{ab} \partial_{b}V_{a}
 -  \frac{(d - 4)^{} (d - 3)^{} (d - 1)^2}{2 (d - 2)^4} \partial_t K \partial_t V
 + \frac{2 (d - 3)^{} (d - 1)^{}}{(d - 2)^2} V_{a} \partial_t \partial_{a}K\nonumber\\
& + \frac{(d - 3)^{} (d - 1)^{}}{(d - 2)^2} \hat{W}_{ab} \partial_{ba}K\Bigr) \hat{x}_{ij}
 -  \frac{d (d - 1)^2}{(d - 2)^3} \Psi_{ij}^{\partial_{a}V_{b}} \partial_t \partial_{b}V \partial_{a}V
 -  \frac{d (d - 1)^2}{(d - 2)^3} \Psi_{ij}^{\partial_{a}V_{b}} \partial_t V \partial_{ba}V\nonumber\\
& -  \frac{4 (d - 1)^2}{(d - 2)^2} \Psi_{ij}^{\partial_{b}V_{k}} \partial_{b}V_{a} \partial_{ka}V
 -  \frac{4 (d - 1)^2}{(d - 2)^2} \Psi_{ij}^{\partial_{b}V_{k}} \partial_{a}V \partial_{kb}V_{a}
 + \frac{d (d - 1)^2}{2 (d - 2)^3} \Psi_{ij}^{\partial_{ta} V} \partial_t V \partial_{a}V\nonumber\\
& + \frac{2 (d - 1)^2}{(d - 2)^2} \Psi_{ij}^{\partial_{tb} V} \partial_{a}V \partial_{b}V_{a}
 + \frac{d (d - 1)^2}{4 (d - 2)^4} \Psi_{ij}^{\partial_{aa} V} (\partial_t V)^2
 + \frac{2 (d - 1)^2}{(d - 2)^2} \Psi_{ij}^{\partial_{ab} V} \partial_t V_{b} \partial_{a}V\nonumber\\
& -  \frac{2 (d - 1)^{}}{(d - 2)^2} \Psi_{ij}^{\partial_{bb} V} \partial_{a}V_{k} \partial_{k}V_{a}
 + \frac{2 (d - 1)^{}}{(d - 2)^2} \Psi_{ij}^{\partial_{bb} V} \partial_{a}V_{k} \partial_{a}V_{k}
 -  \frac{2 (d - 1)^{}}{d - 2} \Psi_{ij}^{\partial_{ak} V} \partial_{b}V_{k} \partial_{b}V_{a}\nonumber\\
& -  \frac{2 (d - 1)^{}}{d - 2} \Psi_{ij}^{\partial_{bk} V} \partial_{b}V_{a} \partial_{k}V_{a}
 + \frac{4 (d - 1)^{}}{d - 2} \Psi_{ij}^{\partial_{ak} V} \partial_{b}V_{a} \partial_{k}V_{b}
 -  \frac{(d - 1)^{}}{2 (d - 2)} \Psi_{ij}^{\partial_{ab} V} \partial_t^{2} \hat{W}_{ab}\nonumber\\
& -  \frac{(d - 3)^{} (d - 1)^2}{(d - 2)^3} \Psi_{ij}^{\partial_{ab} V} \partial_{a}K \partial_{b}V,\\
S\text{INC}_{4\text{PN}}={}&\Bigl(\frac{(d - 1)^3}{4 (d - 2)^3} (\partial_t V)^2 V^2
 -  \frac{(d - 1)^3}{4 (d - 2)^3} V^3 \partial_t^{2} V
 -  \frac{(d - 1)^2 (2 d - 3)^{}}{(d - 2)^3} V^2 V_{a} \partial_t \partial_{a}V\nonumber\\
& -  \frac{d (d - 1)^2 (2 d - 5)^{}}{2 (d - 2)^4} V V_{a} \partial_t V \partial_{a}V
 + \frac{(d - 1)^3}{(d - 2)^3} V^2 \partial_t V_{a} \partial_{a}V
 -  \frac{2 (d - 1)^{} (d^2 - 5 d + 8)^{}}{(d - 2)^3} V_{a} \partial_t V \partial_t V_{a}\nonumber\\
& + \frac{4 (d - 1)^{}}{(d - 2)^2} V \partial_t V_{a} \partial_t V_{a}
 -  \frac{2 (d - 3)^{} (d - 1)^{}}{(d - 2)^2} V V_{a} \partial_t^{2} V_{a}
 -  \frac{(d - 1)^2}{(d - 2)^3} V_{a} V_{b} \partial_{a}V \partial_{b}V\nonumber\\
& + \frac{6 (d - 1)^2}{(d - 2)^3} V V_{a} \partial_{a}V_{b} \partial_{b}V
 -  \frac{(d - 3)^{} (d - 1)^{}}{2 (d - 2)^2} V_{a} V_{a} \partial_{b}V \partial_{b}V
 + \frac{(d - 1)^2 (3 d - 7)^{}}{(d - 2)^3} V^2 \partial_{a}V_{b} \partial_{b}V_{a}\nonumber\\
& -  \frac{(d - 4)^{} (d - 1)^2}{(d - 2)^3} V^2 \partial_{b}V_{a} \partial_{b}V_{a}
 -  \frac{8 (d - 1)^{}}{d - 2} V_{a} \partial_t V_{b} \partial_{b}V_{a}
 + \frac{(d - 1)^2}{2 (d - 2)^2} (\partial_t V)^2 \hat{W}
 + \frac{(d - 1)^2}{(d - 2)^2} V \partial_t V \partial_t \hat{W}\nonumber\\
& + \frac{(d - 1)^2}{2 (d - 2)^2} V \hat{W} \partial_t^{2} V
 + \frac{(d - 1)^2}{4 (d - 2)^2} V^2 \partial_t^{2} \hat{W}
 -  \frac{(d - 1)^2}{2 (d - 2)^2} V^2 \hat{W}_{ab} \partial_{ba}V
 -  \frac{d (d - 1)^2}{4 (d - 2)^3} V \hat{W}_{ab} \partial_{a}V \partial_{b}V\nonumber\\
& + \frac{2 (d - 1)^{}}{d - 2} V_{a} \hat{W}_{ab} \partial_t \partial_{b}V
 + \frac{2 (d - 1)^2}{(d - 2)^2} V \partial_t \hat{W}_{ab} \partial_{b}V_{a}
 + \frac{4 (d - 3)^{}}{d - 2} V_{a} \partial_{b}\hat{W} \partial_{b}V_{a}
 + \frac{2 (d - 3)^{}}{d - 2} \hat{W} \partial_{b}V_{a} \partial_{b}V_{a}\nonumber\\
& -  \frac{4 (d - 1)^{}}{d - 2} \hat{W}_{ai} \partial_{b}V_{a} \partial_{i}V_{b}
 + \frac{4 (d - 1)^{}}{d - 2} V_{a} \partial_{a}\hat{W}_{bi} \partial_{i}V_{b}
 -  \frac{4 (d - 1)^{}}{d - 2} V_{a} \partial_{i}\hat{W}_{ab} \partial_{i}V_{b}
 + \frac{(d - 3)^{}}{2 (d - 2)} \partial_t \hat{W}_{ab} \partial_t \hat{W}_{ab}\nonumber\\
& + \frac{1}{2 (d - 2)} (\partial_t \hat{W})^2
 -  \frac{1}{d - 2} \hat{W}_{ab} \partial_t^{2} \hat{W}_{ab}
 + \frac{1}{2 (d - 2)} \hat{W} \partial_t^{2} \hat{W}
 + \frac{d (d - 1)^{}}{(d - 2)^2} \hat{W}_{bi} \partial_{a}\hat{W}_{bi} \partial_{a}V\nonumber\\
& + \frac{(d - 1)^{}}{d - 2} \hat{W}_{a}{}_{i} \hat{W}_{ab} \partial_{ib}V
 + \frac{d (d - 1)^{}}{2 (d - 2)^2} V \partial_{i}\hat{W}_{ab} \partial_{i}\hat{W}_{ab}
 -  \frac{2 (d - 1)^2}{(d - 2)^2} \hat{R}_{a} V \partial_t \partial_{a}V
 -  \frac{d (d - 1)^2}{(d - 2)^3} \hat{R}_{a} \partial_t V \partial_{a}V\nonumber\\
& + \frac{8}{d - 2} \partial_t \hat{R}_{a} \partial_t V_{a}
 -  \frac{2 (d - 3)^{}}{d - 2} V_{a} \partial_t^{2} \hat{R}_{a}
 -  \frac{2 (d - 3)^{}}{d - 2} \hat{R}_{a} \partial_t^{2} V_{a}
 -  \frac{4 (d - 1)^{}}{d - 2} \hat{R}_{a} V_{b} \partial_{ba}V\nonumber\\
& -  \frac{12 (d - 1)^{}}{(d - 2)^2} V_{a} \partial_{b}V \partial_{b}\hat{R}_{a}
 + \frac{4 (d - 4)^{} (d - 1)^{}}{(d - 2)^2} V \partial_{b}V_{a} \partial_{b}\hat{R}_{a}
 -  \frac{12 (d - 1)^{}}{(d - 2)^2} \hat{R}_{a} \partial_{b}V_{a} \partial_{b}V\nonumber\\
& + \frac{4 (d - 1)^{}}{d - 2} \partial_t \hat{W}_{ab} \partial_{b}\hat{R}_{a}
 + \frac{4 (d - 1)^{}}{d - 2} \partial_{a}\hat{R}_{b} \partial_{b}\hat{R}_{a}
 + \frac{(d - 4)^{} (d - 1)^2}{(d - 2)^3} \partial_t V \partial_t \hat{X}
 -  \frac{(d - 1)^2}{(d - 2)^2} \hat{X} \partial_{a}V \partial_{a}V\nonumber\\
& -  \frac{2 (d - 1)^2}{(d - 2)^2} V \partial_{a}\hat{X} \partial_{a}V
 -  \frac{4 (d - 1)^{}}{d - 2} V_{a} \partial_t \partial_{a}\hat{X}
 -  \frac{2 (d - 1)^{}}{d - 2} \hat{W}_{ab} \partial_{ba}\hat{X}
 + \frac{2 (d - 1)^{}}{d - 2} \partial_t V \partial_t \hat{Z}\nonumber\\
& + \frac{(d - 1)^{}}{d - 2} \hat{Z} \partial_t^{2} V
 + \frac{(d - 1)^{}}{d - 2} V \partial_t^{2} \hat{Z}
 -  \frac{2 (d - 1)^2}{(d - 2)^2} \hat{Z} \partial_{a}V \partial_{a}V
 -  \frac{4 (d - 1)^2}{(d - 2)^2} V \partial_{a}\hat{Z} \partial_{a}V\nonumber\\
& + \frac{8 (d - 1)^{}}{d - 2} \partial_t \hat{Z}_{ab} \partial_{b}V_{a}
 -  \frac{(d - 3)^2 (d - 1)^2 (3 d - 4)^{}}{2 (d - 2)^5} V \partial_{a}K \partial_{a}K
 -  \frac{(d - 3)^2 (d - 1)^2 (3 d - 4)^{}}{(d - 2)^5} K \partial_{a}V \partial_{a}K\nonumber\\
& -  \frac{(d - 4)^{} (d - 3)^{} (d - 1)^3}{2 (d - 2)^5} K (\partial_t V)^2
 -  \frac{(d - 4)^{} (d - 3)^{} (d - 1)^3}{(d - 2)^5} V \partial_t K \partial_t V\nonumber\\
& + \frac{(d - 4)^{} (d - 3)^2 (d - 1)^2}{2 (d - 2)^5} (\partial_t K)^2
 + \frac{2 (d - 3)^{} (d - 1)^2}{(d - 2)^3} V V_{a} \partial_t \partial_{a}K
 + \frac{2 (d - 3)^{} (d - 1)^2}{(d - 2)^3} K V_{a} \partial_t \partial_{a}V\nonumber\\
& + \frac{d (d - 3)^{} (d - 1)^2}{(d - 2)^4} V_{a} \partial_t V \partial_{a}K
 + \frac{d (d - 3)^{} (d - 1)^2}{(d - 2)^4} V_{a} \partial_t K \partial_{a}V
 -  \frac{2 (d - 3)^{} (d - 1)^2}{(d - 2)^3} K \partial_{b}V_{a} \partial_{b}V_{a}\nonumber\\
& -  \frac{(d - 3)^{} (d - 1)^{}}{(d - 2)^2} \partial_t K \partial_t \hat{W}
 -  \frac{(d - 3)^{} (d - 1)^{}}{2 (d - 2)^2} \hat{W} \partial_t^{2} K
 -  \frac{(d - 3)^{} (d - 1)^{}}{2 (d - 2)^2} K \partial_t^{2} \hat{W}\nonumber\\
& + \frac{2 (d - 3)^{} (d - 1)^2}{(d - 2)^3} \hat{W} \partial_{a}V \partial_{a}K
 + \frac{2 (d - 3)^{} (d - 1)^2}{(d - 2)^3} V \partial_{a}\hat{W} \partial_{a}K
 + \frac{2 (d - 3)^{} (d - 1)^2}{(d - 2)^3} K \partial_{a}\hat{W} \partial_{a}V\nonumber\\
& + \frac{4 (d - 3)^{} (d - 1)^{}}{(d - 2)^2} \hat{R}_{a} \partial_t \partial_{a}K
 + \frac{4 (d - 3)^{} (d - 1)^{}}{(d - 2)^2} \hat{Z}_{ab} \partial_{ba}K\Bigr) \hat{x}_{ij}
 -  \frac{d (d - 1)^3}{(d - 2)^4} V \Psi_{ij}^{\partial_{a}V_{b}} \partial_t \partial_{b}V \partial_{a}V\nonumber\\
& -  \frac{d (d - 1)^3}{(d - 2)^4} \Psi_{ij}^{\partial_{a}V_{b}} \partial_t V \partial_{a}V \partial_{b}V
 -  \frac{d (d - 1)^3}{(d - 2)^4} V \Psi_{ij}^{\partial_{a}V_{b}} \partial_t V \partial_{ba}V
 + \frac{4 (d - 1)^2}{(d - 2)^2} V \Psi_{ij}^{\partial_{a}V_{b}} \partial_t^{2} \partial_{b}V_{a}\nonumber\\
& + \frac{4 (d - 1)^2}{(d - 2)^2} \Psi_{ij}^{\partial_{b}V_{a}} \partial_t^{2} V_{b} \partial_{a}V
 -  \frac{2 d (d - 1)^2}{(d - 2)^3} V_{a} \Psi_{ij}^{\partial_{b}V_{k}} \partial_{b}V \partial_{ka}V
 -  \frac{2 d (d - 1)^2}{(d - 2)^3} V_{b} \Psi_{ij}^{\partial_{k}V_{a}} \partial_{b}V \partial_{ka}V\nonumber\\
& -  \frac{2 d (d - 1)^2}{(d - 2)^3} \Psi_{ij}^{\partial_{b}V_{k}} \partial_{a}V \partial_{b}V \partial_{k}V_{a}
 -  \frac{4 (d - 1)^3}{(d - 2)^3} \Psi_{ij}^{\partial_{k}V_{b}} \partial_{a}V_{k} \partial_{a}V \partial_{b}V
 -  \frac{4 (d - 1)^3}{(d - 2)^3} V \Psi_{ij}^{\partial_{b}V_{k}} \partial_{a}V \partial_{ka}V_{b}\nonumber\\
& -  \frac{4 (d - 1)^3}{(d - 2)^3} V \Psi_{ij}^{\partial_{a}V_{k}} \partial_{b}V_{a} \partial_{kb}V
 + \frac{16 (d - 1)^{}}{d - 2} V_{a} \Psi_{ij}^{\partial_{b}V_{k}} \partial_t \partial_{ka}V_{b}
 + \frac{16 (d - 1)^{}}{d - 2} \Psi_{ij}^{\partial_{k}V_{b}} \partial_t \partial_{a}V_{k} \partial_{b}V_{a}\nonumber\\
& -  \frac{4 (d - 1)^2}{(d - 2)^2} \Psi_{ij}^{\partial_{b}V_{k}} \partial_t \partial_{k}\hat{W}_{ab} \partial_{a}V
 -  \frac{4 (d - 1)^2}{(d - 2)^2} \Psi_{ij}^{\partial_{a}V_{b}} \partial_t \hat{W}_{ak} \partial_{kb}V
 + \frac{8 (d - 1)^{}}{d - 2} \hat{W}_{ab} \Psi_{ij}^{\partial_{k}V_{l}} \partial_{lba}V_{k}\nonumber\\
& + \frac{8 (d - 1)^{}}{d - 2} \Psi_{ij}^{\partial_{k}V_{l}} \partial_{b}V_{a} \partial_{lk}\hat{W}_{ab}
 -  \frac{8 (d - 1)^{}}{d - 2} \Psi_{ij}^{\partial_{k}V_{l}} \partial_{b}V_{a} \partial_{la}\hat{W}_{bk}
 + \frac{8 (d - 1)^{}}{d - 2} \Psi_{ij}^{\partial_{a}V_{b}} \partial_{a}\hat{W}_{kl} \partial_{lb}V_{k}\nonumber\\
& -  \frac{8 (d - 1)^{}}{d - 2} \Psi_{ij}^{\partial_{a}V_{b}} \partial_{k}\hat{W}_{al} \partial_{lb}V_{k}
 + \frac{8 (d - 1)^{}}{d - 2} \Psi_{ij}^{\partial_{a}V_{b}} \partial_{b}\hat{W}_{kl} \partial_{lk}V_{a}
 + \frac{4 (d - 1)^{}}{d - 2} \Psi_{ij}^{\partial_{a}V_{b}} \partial_t^{2} \partial_{b}\hat{R}_{a}\nonumber\\
& -  \frac{8 (d - 1)^2}{(d - 2)^2} \Psi_{ij}^{\partial_{b}V_{k}} \partial_{b}\hat{R}_{a} \partial_{ka}V
 -  \frac{8 (d - 1)^2}{(d - 2)^2} \Psi_{ij}^{\partial_{b}V_{k}} \partial_{a}V \partial_{kb}\hat{R}_{a}
 + \frac{d (d - 1)^3}{2 (d - 2)^4} V \Psi_{ij}^{\partial_{ta} V} \partial_t V \partial_{a}V\nonumber\\
& -  \frac{2 (d - 1)^2}{(d - 2)^2} V \Psi_{ij}^{\partial_{ta} V} \partial_t^{2} V_{a}
 + \frac{d (d - 1)^2}{(d - 2)^3} V_{b} \Psi_{ij}^{\partial_{ta} V} \partial_{a}V \partial_{b}V
 + \frac{2 (d - 1)^3}{(d - 2)^3} V \Psi_{ij}^{\partial_{tb} V} \partial_{a}V_{b} \partial_{a}V\nonumber\\
& -  \frac{8 (d - 1)^{}}{d - 2} V_{a} \Psi_{ij}^{\partial_{tb} V} \partial_t \partial_{a}V_{b}
 + \frac{2 (d - 1)^2}{(d - 2)^2} \Psi_{ij}^{\partial_{tb} V} \partial_t \hat{W}_{ab} \partial_{a}V
 -  \frac{4 (d - 1)^{}}{d - 2} \hat{W}_{ab} \Psi_{ij}^{\partial_{tk} V} \partial_{ba}V_{k}\nonumber\\
& -  \frac{4 (d - 1)^{}}{d - 2} \Psi_{ij}^{\partial_{tk} V} \partial_{b}V_{a} \partial_{k}\hat{W}_{ab}
 + \frac{4 (d - 1)^{}}{d - 2} \Psi_{ij}^{\partial_{tk} V} \partial_{a}\hat{W}_{bk} \partial_{b}V_{a}
 -  \frac{2 (d - 1)^{}}{d - 2} \Psi_{ij}^{\partial_{ta} V} \partial_t^{2} \hat{R}_{a}\nonumber\\
& + \frac{4 (d - 1)^2}{(d - 2)^2} \Psi_{ij}^{\partial_{tb} V} \partial_{a}V \partial_{b}\hat{R}_{a}
 -  \frac{d (d - 1)^3}{8 (d - 2)^4} V \Psi_{ij}^{\partial_{aa} V} (\partial_t V)^2
 + \frac{(d - 1)^3}{4 (d - 2)^3} V^2 \Psi_{ij}^{\partial_{aa} V} \partial_t^{2} V\nonumber\\
& + \frac{(d - 1)^2}{(d - 2)^2} V V_{b} \Psi_{ij}^{\partial_{aa} V} \partial_t \partial_{b}V
 + \frac{(d - 1)^2}{2 (d - 2)^2} V_{b} \Psi_{ij}^{\partial_{aa} V} \partial_t V \partial_{b}V
 -  \frac{(d - 1)^3}{(d - 2)^3} V \Psi_{ij}^{\partial_{aa} V} \partial_t V_{b} \partial_{b}V\nonumber\\
& + \frac{2 (d - 1)^3}{(d - 2)^3} V \Psi_{ij}^{\partial_{ab} V} \partial_t V_{b} \partial_{a}V
 -  \frac{(d - 1)^2}{(d - 2)^2} V_{a} \Psi_{ij}^{\partial_{ab} V} \partial_t V \partial_{b}V
 + \frac{4 (d - 1)^{}}{d - 2} \Psi_{ij}^{\partial_{ab} V} \partial_t V_{a} \partial_t V_{b}\nonumber\\
& + \frac{4 (d - 1)^2}{(d - 2)^2} V_{a} \Psi_{ij}^{\partial_{bk} V} \partial_{a}V_{k} \partial_{b}V
 -  \frac{4 (d - 1)^2}{(d - 2)^2} V_{a} \Psi_{ij}^{\partial_{ak} V} \partial_{b}V_{k} \partial_{b}V
 -  \frac{(d - 1)^2}{(d - 2)^2} V \Psi_{ij}^{\partial_{bb} V} \partial_{a}V_{k} \partial_{k}V_{a}\nonumber\\
& -  \frac{(d - 1)^2}{(d - 2)^2} V \Psi_{ij}^{\partial_{ab} V} \partial_t^{2} \hat{W}_{ab}
 + \frac{(d - 1)^2}{2 (d - 2)^2} V \Psi_{ij}^{\partial_{aa} V} \partial_t^{2} \hat{W}
 + \frac{(d - 1)^2}{2 (d - 2)^2} V \hat{W}_{ka} \Psi_{ij}^{\partial_{bb} V} \partial_{ka}V\nonumber\\
& -  \frac{(d - 1)^2}{(d - 2)^2} \hat{W}_{a}{}_{k} \Psi_{ij}^{\partial_{bk} V} \partial_{a}V \partial_{b}V
 + \frac{(d - 1)^2}{4 (d - 2)^2} \hat{W}_{ka} \Psi_{ij}^{\partial_{bb} V} \partial_{a}V \partial_{k}V
 -  \frac{4 (d - 1)^{}}{d - 2} V_{a} \Psi_{ij}^{\partial_{bk} V} \partial_t \partial_{a}\hat{W}_{bk}\nonumber\\
& + \frac{2 (d - 1)^{}}{d - 2} V_{b} \Psi_{ij}^{\partial_{aa} V} \partial_t \partial_{b}\hat{W}
 + \frac{2 (d - 1)^{}}{d - 2} \Psi_{ij}^{\partial_{bb} V} \partial_t \hat{W}_{ka} \partial_{a}V_{k}
 -  \frac{4 (d - 1)^{}}{d - 2} \Psi_{ij}^{\partial_{bk} V} \partial_t \hat{W}_{a}{}_{k} \partial_{b}V_{a}\nonumber\\
& + \frac{4 (d - 1)^{}}{d - 2} \Psi_{ij}^{\partial_{ak} V} \partial_t \hat{W}_{b}{}_{k} \partial_{b}V_{a}
 + \frac{(d - 1)^{}}{d - 2} \Psi_{ij}^{\partial_{kl} V} \partial_{k}\hat{W}_{ab} \partial_{l}\hat{W}_{ab}
 -  \frac{2 (d - 1)^{}}{d - 2} \hat{W}_{ab} \Psi_{ij}^{\partial_{kl} V} \partial_{ba}\hat{W}_{kl}\nonumber\\
& + \frac{(d - 1)^{}}{d - 2} \hat{W}_{ka} \Psi_{ij}^{\partial_{bb} V} \partial_{ka}\hat{W}
 -  \frac{4 (d - 1)^{}}{d - 2} \Psi_{ij}^{\partial_{bl} V} \partial_{k}\hat{W}_{ab} \partial_{l}\hat{W}_{ak}
 + \frac{2 (d - 1)^{}}{d - 2} \Psi_{ij}^{\partial_{bl} V} \partial_{a}\hat{W}_{k}{}_{l} \partial_{k}\hat{W}_{ab}\nonumber\\
& + \frac{(d - 1)^{}}{d - 2} \Psi_{ij}^{\partial_{ll} V} \partial_{b}\hat{W}_{ak} \partial_{k}\hat{W}_{ab}
 -  \frac{2 (d - 1)^2}{(d - 2)^2} \Psi_{ij}^{\partial_{aa} V} \partial_t \hat{R}_{b} \partial_{b}V
 + \frac{4 (d - 1)^2}{(d - 2)^2} \Psi_{ij}^{\partial_{ab} V} \partial_t \hat{R}_{b} \partial_{a}V\nonumber\\
& -  \frac{4 (d - 1)^{}}{d - 2} \Psi_{ij}^{\partial_{bb} V} \partial_{a}\hat{R}_{k} \partial_{k}V_{a}
 + \frac{8 (d - 1)^{}}{d - 2} \Psi_{ij}^{\partial_{bk} V} \partial_{a}V_{k} \partial_{b}\hat{R}_{a}
 -  \frac{8 (d - 1)^{}}{d - 2} \Psi_{ij}^{\partial_{bk} V} \partial_{b}\hat{R}_{a} \partial_{k}V_{a}\nonumber\\
& + \frac{8 (d - 1)^{}}{d - 2} \Psi_{ij}^{\partial_{ak} V} \partial_{b}\hat{R}_{a} \partial_{k}V_{b}
 + \frac{2 (d - 1)^2}{(d - 2)^2} \Psi_{ij}^{\partial_{ab} V} \partial_{a}V \partial_{b}\hat{X}
 -  \frac{2 (d - 1)^{}}{d - 2} \Psi_{ij}^{\partial_{ab} V} \partial_t^{2} \hat{Z}_{ab}\nonumber\\
& + \frac{2 d (d - 3)^{} (d - 1)^2}{(d - 2)^4} \Psi_{ij}^{\partial_{a}V_{b}} \partial_t \partial_{b}V \partial_{a}K
 + \frac{2 d (d - 3)^{} (d - 1)^2}{(d - 2)^4} \Psi_{ij}^{\partial_{a}V_{b}} \partial_t \partial_{b}K \partial_{a}V\nonumber\\
& + \frac{2 d (d - 3)^{} (d - 1)^2}{(d - 2)^4} \Psi_{ij}^{\partial_{a}V_{b}} \partial_t V \partial_{ba}K
 + \frac{2 d (d - 3)^{} (d - 1)^2}{(d - 2)^4} \Psi_{ij}^{\partial_{a}V_{b}} \partial_t K \partial_{ba}V\nonumber\\
& + \frac{8 (d - 3)^{} (d - 1)^2}{(d - 2)^3} \Psi_{ij}^{\partial_{b}V_{k}} \partial_{b}V_{a} \partial_{ka}K
 + \frac{8 (d - 3)^{} (d - 1)^2}{(d - 2)^3} \Psi_{ij}^{\partial_{b}V_{k}} \partial_{a}K \partial_{kb}V_{a}\nonumber\\
& -  \frac{d (d - 3)^{} (d - 1)^2}{(d - 2)^4} \Psi_{ij}^{\partial_{ta} V} \partial_t V \partial_{a}K
 -  \frac{d (d - 3)^{} (d - 1)^2}{(d - 2)^4} \Psi_{ij}^{\partial_{ta} V} \partial_t K \partial_{a}V\nonumber\\
& -  \frac{4 (d - 3)^{} (d - 1)^2}{(d - 2)^3} \Psi_{ij}^{\partial_{tb} V} \partial_{a}K \partial_{b}V_{a}
 + \frac{(d - 3)^2 (d - 1)^2}{(d - 2)^4} \Psi_{ij}^{\partial_{ab} V} \partial_{a}K \partial_{b}K\nonumber\\
& + \frac{d (d - 3)^{} (d - 1)^2}{2 (d - 2)^4} \Psi_{ij}^{\partial_{aa} V} \partial_t K \partial_t V
 + \frac{2 (d - 3)^{} (d - 1)^2}{(d - 2)^3} \Psi_{ij}^{\partial_{aa} V} \partial_t V_{b} \partial_{b}K\nonumber\\
& -  \frac{4 (d - 3)^{} (d - 1)^2}{(d - 2)^3} \Psi_{ij}^{\partial_{ab} V} \partial_t V_{b} \partial_{a}K, \nonumber\\
S\text{IINC}_{2\text{PN}}={}&- \frac{(d - 1)^2}{8 (d - 2)^2 (d + 4)} r^2 \hat{x}_{ij} \partial_{a}V \partial_{a}V,\\
S\text{IINC}_{3\text{PN}}={}&\Bigl(\frac{(d - 4)^{} (d - 1)^2}{16 (d - 2)^3 (d + 4)} (\partial_t V)^2
 -  \frac{(d - 1)^2}{8 (d - 2)^2 (d + 4)} V \partial_t^{2} V
 -  \frac{(d - 1)^2 (3 d - 2)^{}}{16 (d - 2)^3 (d + 4)} V \partial_{a}V \partial_{a}V\nonumber\\
& -  \frac{(d - 1)^{}}{2 (d - 2) (d + 4)} V_{a} \partial_t \partial_{a}V
 + \frac{(d - 1)^{}}{2 (d - 2) (d + 4)} \partial_{a}V_{b} \partial_{b}V_{a}
 + \frac{(d - 3)^{}}{2 (d - 2) (d + 4)} \partial_{b}V_{a} \partial_{b}V_{a}\nonumber\\
& -  \frac{(d - 1)^{}}{4 (d - 2) (d + 4)} \partial_{a}\hat{W} \partial_{a}V
 -  \frac{(d - 1)^{}}{4 (d - 2) (d + 4)} \hat{W}_{ab} \partial_{ba}V
 + \frac{(d - 3)^{} (d - 1)^2}{2 (d - 2)^3 (d + 4)} \partial_{a}V \partial_{a}K\Bigr) r^2 \hat{x}_{ij},\\
S\text{IINC}_{4\text{PN}}={}&\Bigl(\frac{(d - 1)^2 (d^2 - 8 d + 8)^{}}{16 (d - 2)^4 (d + 4)} (\partial_t V)^2 V
 -  \frac{3 (d - 1)^3}{8 (d - 2)^3 (d + 4)} V^2 \partial_t^{2} V
 -  \frac{(d - 1)^3 (3 d - 2)^{}}{16 (d - 2)^4 (d + 4)} V^2 \partial_{a}V \partial_{a}V\nonumber\\
& -  \frac{3 (d - 1)^2}{2 (d - 2)^2 (d + 4)} V V_{a} \partial_t \partial_{a}V
 -  \frac{d (d - 1)^{}}{2 (d - 2)^2 (d + 4)} V_{a} \partial_t V \partial_{a}V
 + \frac{(d - 1)^2}{2 (d - 2)^2 (d + 4)} V \partial_t V_{a} \partial_{a}V\nonumber\\
& + \frac{1}{(d - 2) (d + 4)} \partial_t V_{a} \partial_t V_{a}
 -  \frac{(d - 3)^{} (d - 1)^{}}{(d - 2)^2 (d + 4)} V_{a} \partial_{a}V_{b} \partial_{b}V
 + \frac{(d - 5)^{} (d - 1)^{}}{(d - 2)^2 (d + 4)} V_{a} \partial_{b}V_{a} \partial_{b}V\nonumber\\
& + \frac{3 (d - 1)^{}}{2 (d - 2) (d + 4)} V \partial_{a}V_{b} \partial_{b}V_{a}
 + \frac{(d - 1)^{}}{2 (d - 2) (d + 4)} V \partial_{b}V_{a} \partial_{b}V_{a}
 + \frac{(d - 1)^{}}{4 (d - 2) (d + 4)} \partial_t V \partial_t \hat{W}\nonumber\\
& -  \frac{(d - 1)^{} (2 d - 3)^{}}{8 (d - 2)^2 (d + 4)} \hat{W} \partial_{a}V \partial_{a}V
 -  \frac{(d - 1)^2}{2 (d - 2)^2 (d + 4)} V \partial_{a}\hat{W} \partial_{a}V
 -  \frac{(d - 1)^2}{2 (d - 2)^2 (d + 4)} V \hat{W}_{ab} \partial_{ba}V\nonumber\\
& -  \frac{(d - 1)^{}}{4 (d - 2)^2 (d + 4)} \hat{W}_{ab} \partial_{a}V \partial_{b}V
 + \frac{(d - 1)^{}}{(d - 2) (d + 4)} \partial_t \hat{W}_{ab} \partial_{b}V_{a}
 -  \frac{1}{4 (d - 2) (d + 4)} \partial_{a}\hat{W} \partial_{a}\hat{W}\nonumber\\
& + \frac{1}{2 (d - 2) (d + 4)} \partial_{i}\hat{W}_{ab} \partial_{i}\hat{W}_{ab}
 -  \frac{(d - 1)^{}}{(d - 2) (d + 4)} \hat{R}_{a} \partial_t \partial_{a}V
 + \frac{2 (d - 1)^{}}{(d - 2) (d + 4)} \partial_{a}V_{b} \partial_{b}\hat{R}_{a}\nonumber\\
& + \frac{2 (d - 3)^{}}{(d - 2) (d + 4)} \partial_{b}V_{a} \partial_{b}\hat{R}_{a}
 -  \frac{(d - 1)^2}{(d - 2)^2 (d + 4)} \partial_{a}\hat{X} \partial_{a}V
 -  \frac{(d - 1)^{}}{(d - 2) (d + 4)} \partial_{a}\hat{Z} \partial_{a}V\nonumber\\
& -  \frac{(d - 1)^{}}{(d - 2) (d + 4)} \hat{Z}_{ab} \partial_{ba}V
 -  \frac{(d - 4)^{} (d - 3)^{} (d - 1)^2}{4 (d - 2)^4 (d + 4)} \partial_t K \partial_t V
 + \frac{(d - 3)^{} (d - 1)^2}{4 (d - 2)^3 (d + 4)} V \partial_t^{2} K\nonumber\\
& + \frac{(d - 3)^{} (d - 1)^2}{4 (d - 2)^3 (d + 4)} K \partial_t^{2} V
 + \frac{(d - 3)^{} (d - 1)^2 (3 d - 2)^{}}{4 (d - 2)^4 (d + 4)} V \partial_{a}V \partial_{a}K\nonumber\\
& + \frac{(d - 3)^{} (d - 1)^2 (3 d - 2)^{}}{8 (d - 2)^4 (d + 4)} K \partial_{a}V \partial_{a}V
 -  \frac{(d - 3)^2 (d - 1)^2}{2 (d - 2)^4 (d + 4)} \partial_{a}K \partial_{a}K
 + \frac{(d - 3)^{} (d - 1)^{}}{(d - 2)^2 (d + 4)} V_{a} \partial_t \partial_{a}K\nonumber\\
& + \frac{(d - 3)^{} (d - 1)^{}}{2 (d - 2)^2 (d + 4)} \partial_{a}\hat{W} \partial_{a}K
 + \frac{(d - 3)^{} (d - 1)^{}}{2 (d - 2)^2 (d + 4)} \hat{W}_{ab} \partial_{ba}K\Bigr) r^2 \hat{x}_{ij}, \nonumber\\
S\text{IIINC}_{3\text{PN}}={}&- \frac{(d - 1)^2}{32 (d - 2)^2 (d + 4) (d + 6)} r^4 \hat{x}_{ij} \partial_{a}V \partial_{a}V,\\
S\text{IIINC}_{4\text{PN}}={}&\Bigl(\frac{(d - 4)^{} (d - 1)^2}{64 (d - 2)^3 (d + 4) (d + 6)} (\partial_t V)^2
 -  \frac{(d - 1)^2}{32 (d - 2)^2 (d + 4) (d + 6)} V \partial_t^{2} V\nonumber\\
& -  \frac{(d - 1)^2 (3 d - 2)^{}}{64 (d - 2)^3 (d + 4) (d + 6)} V \partial_{a}V \partial_{a}V
 -  \frac{(d - 1)^{}}{8 (d - 2) (d + 4) (d + 6)} V_{a} \partial_t \partial_{a}V\nonumber\\
& + \frac{(d - 1)^{}}{8 (d - 2) (d + 4) (d + 6)} \partial_{a}V_{b} \partial_{b}V_{a}
 + \frac{(d - 3)^{}}{8 (d - 2) (d + 4) (d + 6)} \partial_{b}V_{a} \partial_{b}V_{a}\nonumber\\
& -  \frac{(d - 1)^{}}{16 (d - 2) (d + 4) (d + 6)} \partial_{a}\hat{W} \partial_{a}V
 -  \frac{(d - 1)^{}}{16 (d - 2) (d + 4) (d + 6)} \hat{W}_{ab} \partial_{ba}V\nonumber\\
& + \frac{(d - 3)^{} (d - 1)^2}{8 (d - 2)^3 (d + 4) (d + 6)} \partial_{a}V \partial_{a}K\Bigr) r^4 \hat{x}_{ij}, \nonumber\\
S\text{IVNC}_{4\text{PN}}={}&- \frac{(d - 1)^2}{192 (d - 2)^2 (d + 4) (d + 6) (d + 8)} r^6 \hat{x}_{ij} \partial_{a}V \partial_{a}V, \nonumber\\
V\text{INC}_{2\text{PN}}={}&- \Bigl(\frac{(d - 1)^2 (d + 2)^{}}{4 (d - 2)^3 (d + 4)} \partial_t V \partial_{a}V
 + \frac{(d - 1)^2 (d + 2)^{}}{d (d - 2)^2 (d + 4)} \partial_{a}V_{b} \partial_{b}V\Bigr) \hat{x}_{ija},\\
V\text{INC}_{3\text{PN}}={}&\Bigl(- \frac{(d - 1)^3 (d + 2)^{}}{4 (d - 2)^4 (d + 4)} V \partial_t V \partial_{a}V
 -  \frac{(d - 1)^2 (d + 2)^{}}{d (d - 2)^2 (d + 4)} \partial_t V \partial_t V_{a}
 -  \frac{(d - 1)^2 (d + 2)^{}}{2 d (d - 2)^2 (d + 4)} V_{a} \partial_t^{2} V\nonumber\\
& + \frac{(d - 1)^2 (d + 2)^{}}{2 d (d - 2)^2 (d + 4)} V \partial_t^{2} V_{a}
 -  \frac{(d - 1)^2 (d + 2)^{}}{2 (d - 2)^3 (d + 4)} V_{b} \partial_{a}V \partial_{b}V
 + \frac{(d - 1)^3 (d + 2)^{}}{2 d (d - 2)^3 (d + 4)} V_{a} \partial_{b}V \partial_{b}V\nonumber\\
& + \frac{4 (d - 1)^{} (d + 2)^{}}{d (d - 2) (d + 4)} V_{b} \partial_t \partial_{b}V_{a}
 -  \frac{(d - 1)^2 (d + 2)^{}}{d (d - 2)^2 (d + 4)} \partial_t \hat{W}_{ab} \partial_{b}V
 + \frac{2 (d - 1)^{} (d + 2)^{}}{d (d - 2) (d + 4)} \hat{W}_{bi} \partial_{ib}V_{a}\nonumber\\
& + \frac{2 (d - 1)^{} (d + 2)^{}}{d (d - 2) (d + 4)} \partial_{a}\hat{W}_{bi} \partial_{i}V_{b}
 -  \frac{2 (d - 1)^{} (d + 2)^{}}{d (d - 2) (d + 4)} \partial_{b}\hat{W}_{ai} \partial_{i}V_{b}
 + \frac{(d - 3)^{} (d - 1)^2 (d + 2)^{}}{2 (d - 2)^4 (d + 4)} \partial_t V \partial_{a}K\nonumber\\
& + \frac{(d - 3)^{} (d - 1)^2 (d + 2)^{}}{2 (d - 2)^4 (d + 4)} \partial_t K \partial_{a}V
 + \frac{2 (d - 3)^{} (d - 1)^2 (d + 2)^{}}{d (d - 2)^3 (d + 4)} \partial_{a}V_{b} \partial_{b}K\Bigr) \hat{x}_{ija}\nonumber\\
& + \frac{(d - 1)^3 (d + 2)^{}}{2 (d - 2)^4 (d + 4)} \Psi_{ijb}^{\partial_{a} V} \partial_t \partial_{b}V \partial_{a}V
 + \frac{(d - 1)^3 (d + 2)^{}}{2 (d - 2)^4 (d + 4)} \Psi_{ija}^{\partial_{b} V} \partial_t V \partial_{ba}V\nonumber\\
& + \frac{2 (d - 1)^3 (d + 2)^{}}{d (d - 2)^3 (d + 4)} \Psi_{ijk}^{\partial_{b} V} \partial_{b}V_{a} \partial_{ka}V
 + \frac{2 (d - 1)^3 (d + 2)^{}}{d (d - 2)^3 (d + 4)} \Psi_{ijb}^{\partial_{k} V} \partial_{a}V \partial_{kb}V_{a},\\
V\text{INC}_{4\text{PN}}={}&\Bigl(- \frac{(d - 1)^2 (d + 2)^{}}{2 d (d - 2)^2 (d + 4)} (\partial_t V)^2 V_{a}
 -  \frac{(d - 1)^3 (d + 2)^{}}{d (d - 2)^3 (d + 4)} V \partial_t V \partial_t V_{a}
 -  \frac{(d - 1)^3 (d + 2)^{}}{2 d (d - 2)^3 (d + 4)} V V_{a} \partial_t^{2} V\nonumber\\
& + \frac{3 (d - 1)^3 (d + 2)^{}}{4 d (d - 2)^3 (d + 4)} V^2 \partial_t^{2} V_{a}
 -  \frac{(d - 1)^4 (d + 2)^{}}{d (d - 2)^4 (d + 4)} V V_{b} \partial_{a}V \partial_{b}V
 + \frac{2 (d - 1)^4 (d + 2)^{}}{d (d - 2)^4 (d + 4)} V V_{a} \partial_{b}V \partial_{b}V\nonumber\\
& + \frac{2 (d - 1)^4 (d + 2)^{}}{d (d - 2)^4 (d + 4)} V^2 \partial_{b}V_{a} \partial_{b}V
 + \frac{4 (d - 1)^2 (d + 2)^{}}{d (d - 2)^2 (d + 4)} V V_{b} \partial_t \partial_{b}V_{a}
 + \frac{2 (d - 1)^2 (d + 2)^{}}{d (d - 2)^2 (d + 4)} V_{b} \partial_t V_{a} \partial_{b}V\nonumber\\
& -  \frac{4 (d - 1)^2 (d + 2)^{}}{d (d - 2)^3 (d + 4)} V_{b} \partial_t V_{b} \partial_{a}V
 + \frac{2 (d - 4)^{} (d - 1)^2 (d + 2)^{}}{d (d - 2)^3 (d + 4)} V_{b} \partial_t V \partial_{a}V_{b}
 + \frac{4 (d - 1)^2 (d + 2)^{}}{d (d - 2)^2 (d + 4)} V_{a} \partial_t V_{b} \partial_{b}V\nonumber\\
& -  \frac{4 (d - 1)^{} (d + 2)^{}}{d (d - 2) (d + 4)} V_{b} \partial_{a}V_{i} \partial_{b}V_{i}
 -  \frac{4 (d - 1)^{} (d + 2)^{}}{(d - 2)^2 (d + 4)} V_{b} \partial_{a}V_{i} \partial_{i}V_{b}
 + \frac{4 (d - 1)^{} (d + 2)^{}}{d (d - 2) (d + 4)} V_{b} \partial_{b}V_{i} \partial_{i}V_{a}\nonumber\\
& + \frac{4 (d - 1)^{} (d + 2)^{}}{d (d - 2) (d + 4)} V_{a} \partial_{b}V_{i} \partial_{i}V_{b}
 -  \frac{2 (d - 1)^{} (d + 2)^{}}{d (d - 2) (d + 4)} V_{a} \partial_{i}V_{b} \partial_{i}V_{b}
 + \frac{(d - 1)^2 (d + 2)^{}}{2 (d - 2)^3 (d + 4)} \hat{W}_{ab} \partial_t V \partial_{b}V\nonumber\\
& -  \frac{(d - 1)^3 (d + 2)^{}}{d (d - 2)^3 (d + 4)} V \partial_t \hat{W}_{ab} \partial_{b}V
 -  \frac{2 (d - 1)^{} (d + 2)^{}}{d (d - 2) (d + 4)} \partial_t V_{a} \partial_t \hat{W}
 -  \frac{(d - 1)^{} (d + 2)^{}}{d (d - 2) (d + 4)} \hat{W} \partial_t^{2} V_{a}\nonumber\\
& + \frac{(d - 1)^{} (d + 2)^{}}{d (d - 2) (d + 4)} \hat{W}_{ab} \partial_t^{2} V_{b}
 + \frac{(d - 1)^{} (d + 2)^{}}{d (d - 2) (d + 4)} V_{b} \partial_t^{2} \hat{W}_{ab}
 -  \frac{(d - 1)^{} (d + 2)^{}}{d (d - 2) (d + 4)} V_{a} \partial_t^{2} \hat{W}\nonumber\\
& + \frac{2 (d - 1)^2 (d + 2)^{}}{d (d - 2)^2 (d + 4)} \hat{W}_{bi} \partial_{a}V_{i} \partial_{b}V
 -  \frac{2 (d - 1)^2 (d + 2)^{}}{d (d - 2)^2 (d + 4)} V_{b} \partial_{b}\hat{W}_{ai} \partial_{i}V
 + \frac{2 (d - 1)^2 (d + 2)^{}}{d (d - 2)^2 (d + 4)} V_{b} \partial_{i}\hat{W}_{ab} \partial_{i}V\nonumber\\
& + \frac{2 (d - 1)^2 (d + 2)^{}}{d (d - 2)^2 (d + 4)} \hat{W}_{ai} \partial_{b}V \partial_{i}V_{b}
 + \frac{(d - 1)^{} (d + 2)^{}}{d (d - 2) (d + 4)} \partial_t \hat{W}_{bi} \partial_{a}\hat{W}_{bi}
 -  \frac{2 (d - 1)^{} (d + 2)^{}}{d (d - 2) (d + 4)} \partial_t \hat{W}_{bi} \partial_{i}\hat{W}_{a}{}_{b}\nonumber\\
& -  \frac{2 (d - 1)^2 (d + 2)^{}}{d (d - 2)^2 (d + 4)} \partial_t \hat{R}_{a} \partial_t V
 + \frac{(d - 1)^2 (d + 2)^{}}{d (d - 2)^2 (d + 4)} V \partial_t^{2} \hat{R}_{a}
 -  \frac{(d - 1)^2 (d + 2)^{}}{d (d - 2)^2 (d + 4)} \hat{R}_{a} \partial_t^{2} V\nonumber\\
& -  \frac{(d - 1)^2 (d + 2)^{}}{(d - 2)^3 (d + 4)} \hat{R}_{b} \partial_{a}V \partial_{b}V
 -  \frac{2 (d - 1)^3 (d + 2)^{}}{d (d - 2)^3 (d + 4)} V \partial_{b}V \partial_{b}\hat{R}_{a}
 + \frac{8 (d - 1)^{} (d + 2)^{}}{d (d - 2) (d + 4)} V_{b} \partial_t \partial_{b}\hat{R}_{a}\nonumber\\
& + \frac{8 (d - 1)^{} (d + 2)^{}}{d (d - 2) (d + 4)} \hat{R}_{b} \partial_t \partial_{b}V_{a}
 + \frac{4 (d - 1)^{} (d + 2)^{}}{d (d - 2) (d + 4)} \hat{W}_{bi} \partial_{ib}\hat{R}_{a}
 + \frac{4 (d - 1)^{} (d + 2)^{}}{d (d - 2) (d + 4)} \partial_{a}\hat{W}_{bi} \partial_{i}\hat{R}_{b}\nonumber\\
& -  \frac{4 (d - 1)^{} (d + 2)^{}}{d (d - 2) (d + 4)} \partial_{b}\hat{W}_{ai} \partial_{i}\hat{R}_{b}
 -  \frac{(d - 1)^2 (d + 2)^{}}{(d - 2)^3 (d + 4)} \partial_t \hat{X} \partial_{a}V
 -  \frac{(d - 1)^2 (d + 2)^{}}{(d - 2)^3 (d + 4)} \partial_t V \partial_{a}\hat{X}\nonumber\\
& -  \frac{4 (d - 1)^2 (d + 2)^{}}{d (d - 2)^2 (d + 4)} \partial_{a}V_{b} \partial_{b}\hat{X}
 -  \frac{4 (d - 1)^2 (d + 2)^{}}{d (d - 2)^2 (d + 4)} \partial_t \hat{Z}_{ab} \partial_{b}V
 + \frac{8 (d - 1)^{} (d + 2)^{}}{d (d - 2) (d + 4)} \hat{Z}_{bi} \partial_{ib}V_{a}\nonumber\\
& + \frac{8 (d - 1)^{} (d + 2)^{}}{d (d - 2) (d + 4)} \partial_{a}\hat{Z}_{bi} \partial_{i}V_{b}
 -  \frac{8 (d - 1)^{} (d + 2)^{}}{d (d - 2) (d + 4)} \partial_{b}\hat{Z}_{ai} \partial_{i}V_{b}
 + \frac{(d - 3)^{} (d - 1)^3 (d + 2)^{}}{2 (d - 2)^5 (d + 4)} V \partial_t V \partial_{a}K\nonumber\\
& + \frac{(d - 3)^{} (d - 1)^3 (d + 2)^{}}{2 (d - 2)^5 (d + 4)} V \partial_t K \partial_{a}V
 + \frac{(d - 3)^{} (d - 1)^3 (d + 2)^{}}{2 (d - 2)^5 (d + 4)} K \partial_t V \partial_{a}V\nonumber\\
& -  \frac{(d - 3)^2 (d - 1)^2 (d + 2)^{}}{(d - 2)^5 (d + 4)} \partial_t K \partial_{a}K
 + \frac{2 (d - 3)^{} (d - 1)^2 (d + 2)^{}}{d (d - 2)^3 (d + 4)} \partial_t K \partial_t V_{a}\nonumber\\
& + \frac{(d - 3)^{} (d - 1)^2 (d + 2)^{}}{d (d - 2)^3 (d + 4)} V_{a} \partial_t^{2} K
 -  \frac{(d - 3)^{} (d - 1)^2 (d + 2)^{}}{d (d - 2)^3 (d + 4)} K \partial_t^{2} V_{a}\nonumber\\
& + \frac{(d - 3)^{} (d - 1)^2 (d + 2)^{}}{(d - 2)^4 (d + 4)} V_{b} \partial_{a}K \partial_{b}V
 + \frac{(d - 3)^{} (d - 1)^2 (d + 2)^{}}{(d - 2)^4 (d + 4)} V_{b} \partial_{a}V \partial_{b}K\nonumber\\
& -  \frac{2 (d - 3)^{} (d - 1)^3 (d + 2)^{}}{d (d - 2)^4 (d + 4)} V_{a} \partial_{b}V \partial_{b}K
 + \frac{2 (d - 3)^{} (d - 1)^2 (d + 2)^{}}{d (d - 2)^3 (d + 4)} \partial_t \hat{W}_{ab} \partial_{b}K\nonumber\\
& + \frac{4 (d - 3)^{} (d - 1)^2 (d + 2)^{}}{d (d - 2)^3 (d + 4)} \partial_{a}\hat{R}_{b} \partial_{b}K\Bigr) \hat{x}_{ija}
 + \frac{(d - 1)^4 (d + 2)^{}}{2 (d - 2)^5 (d + 4)} V \Psi_{ijb}^{\partial_{a} V} \partial_t \partial_{b}V \partial_{a}V\nonumber\\
& + \frac{(d - 1)^4 (d + 2)^{}}{2 (d - 2)^5 (d + 4)} \Psi_{ija}^{\partial_{b} V} \partial_t V \partial_{a}V \partial_{b}V
 + \frac{(d - 1)^4 (d + 2)^{}}{2 (d - 2)^5 (d + 4)} V \Psi_{ija}^{\partial_{b} V} \partial_t V \partial_{ba}V\nonumber\\
& -  \frac{2 (d - 1)^3 (d + 2)^{}}{d (d - 2)^3 (d + 4)} V \Psi_{ijb}^{\partial_{a} V} \partial_t^{2} \partial_{b}V_{a}
 -  \frac{2 (d - 1)^3 (d + 2)^{}}{d (d - 2)^3 (d + 4)} \Psi_{ija}^{\partial_{b} V} \partial_t^{2} V_{b} \partial_{a}V\nonumber\\
& + \frac{(d - 1)^3 (d + 2)^{}}{(d - 2)^4 (d + 4)} V_{a} \Psi_{ijk}^{\partial_{b} V} \partial_{b}V \partial_{ka}V
 + \frac{(d - 1)^3 (d + 2)^{}}{(d - 2)^4 (d + 4)} V_{b} \Psi_{ijk}^{\partial_{a} V} \partial_{b}V \partial_{ka}V\nonumber\\
& + \frac{(d - 1)^3 (d + 2)^{}}{(d - 2)^4 (d + 4)} \Psi_{ijk}^{\partial_{b} V} \partial_{a}V \partial_{b}V \partial_{k}V_{a}
 + \frac{2 (d - 1)^4 (d + 2)^{}}{d (d - 2)^4 (d + 4)} \Psi_{ijb}^{\partial_{k} V} \partial_{a}V_{k} \partial_{a}V \partial_{b}V\nonumber\\
& + \frac{2 (d - 1)^4 (d + 2)^{}}{d (d - 2)^4 (d + 4)} V \Psi_{ijk}^{\partial_{b} V} \partial_{a}V \partial_{ka}V_{b}
 + \frac{2 (d - 1)^4 (d + 2)^{}}{d (d - 2)^4 (d + 4)} V \Psi_{ijk}^{\partial_{a} V} \partial_{b}V_{a} \partial_{kb}V\nonumber\\
& -  \frac{8 (d - 1)^2 (d + 2)^{}}{d (d - 2)^2 (d + 4)} V_{a} \Psi_{ijk}^{\partial_{b} V} \partial_t \partial_{ka}V_{b}
 -  \frac{8 (d - 1)^2 (d + 2)^{}}{d (d - 2)^2 (d + 4)} \Psi_{ijb}^{\partial_{k} V} \partial_t \partial_{a}V_{k} \partial_{b}V_{a}\nonumber\\
& + \frac{2 (d - 1)^3 (d + 2)^{}}{d (d - 2)^3 (d + 4)} \Psi_{ijk}^{\partial_{b} V} \partial_t \partial_{k}\hat{W}_{a}{}_{b} \partial_{a}V
 + \frac{2 (d - 1)^3 (d + 2)^{}}{d (d - 2)^3 (d + 4)} \Psi_{ijb}^{\partial_{k} V} \partial_t \hat{W}_{a}{}_{k} \partial_{ba}V\nonumber\\
& -  \frac{4 (d - 1)^2 (d + 2)^{}}{d (d - 2)^2 (d + 4)} \hat{W}_{ab} \Psi_{ijk}^{\partial_{l} V} \partial_{kba}V_{l}
 -  \frac{4 (d - 1)^2 (d + 2)^{}}{d (d - 2)^2 (d + 4)} \Psi_{ijk}^{\partial_{l} V} \partial_{b}V_{a} \partial_{lk}\hat{W}_{ab}\nonumber\\
& + \frac{4 (d - 1)^2 (d + 2)^{}}{d (d - 2)^2 (d + 4)} \Psi_{ijl}^{\partial_{k} V} \partial_{b}V_{a} \partial_{la}\hat{W}_{b}{}_{k}
 -  \frac{4 (d - 1)^2 (d + 2)^{}}{d (d - 2)^2 (d + 4)} \Psi_{ijk}^{\partial_{l} V} \partial_{kb}V_{a} \partial_{l}\hat{W}_{ab}\nonumber\\
& + \frac{4 (d - 1)^2 (d + 2)^{}}{d (d - 2)^2 (d + 4)} \Psi_{ijk}^{\partial_{l} V} \partial_{a}\hat{W}_{b}{}_{l} \partial_{kb}V_{a}
 -  \frac{4 (d - 1)^2 (d + 2)^{}}{d (d - 2)^2 (d + 4)} \Psi_{ijl}^{\partial_{a} V} \partial_{kb}V_{a} \partial_{l}\hat{W}_{bk}\nonumber\\
& -  \frac{2 (d - 1)^2 (d + 2)^{}}{d (d - 2)^2 (d + 4)} \Psi_{ijb}^{\partial_{a} V} \partial_t^{2} \partial_{b}\hat{R}_{a}
 + \frac{4 (d - 1)^3 (d + 2)^{}}{d (d - 2)^3 (d + 4)} \Psi_{ijk}^{\partial_{b} V} \partial_{b}\hat{R}_{a} \partial_{ka}V\nonumber\\
& + \frac{4 (d - 1)^3 (d + 2)^{}}{d (d - 2)^3 (d + 4)} \Psi_{ijb}^{\partial_{k} V} \partial_{a}V \partial_{kb}\hat{R}_{a}
 -  \frac{(d - 3)^{} (d - 1)^3 (d + 2)^{}}{(d - 2)^5 (d + 4)} \Psi_{ijb}^{\partial_{a} V} \partial_t \partial_{b}V \partial_{a}K\nonumber\\
& -  \frac{(d - 3)^{} (d - 1)^3 (d + 2)^{}}{(d - 2)^5 (d + 4)} \Psi_{ijb}^{\partial_{a} V} \partial_t \partial_{b}K \partial_{a}V
 -  \frac{(d - 3)^{} (d - 1)^3 (d + 2)^{}}{(d - 2)^5 (d + 4)} \Psi_{ija}^{\partial_{b} V} \partial_t V \partial_{ba}K\nonumber\\
& -  \frac{(d - 3)^{} (d - 1)^3 (d + 2)^{}}{(d - 2)^5 (d + 4)} \Psi_{ija}^{\partial_{b} V} \partial_t K \partial_{ba}V
 -  \frac{4 (d - 3)^{} (d - 1)^3 (d + 2)^{}}{d (d - 2)^4 (d + 4)} \Psi_{ijk}^{\partial_{b} V} \partial_{b}V_{a} \partial_{ka}K\nonumber\\
& -  \frac{4 (d - 3)^{} (d - 1)^3 (d + 2)^{}}{d (d - 2)^4 (d + 4)} \Psi_{ijb}^{\partial_{k} V} \partial_{a}K \partial_{kb}V_{a}, \nonumber\\
S\text{VIINC}_{3\text{PN}}={}&\Bigl(- \frac{(d - 1)^2 (d + 2)^{}}{8 (d - 2)^3 (d + 4) (d + 6)} \partial_t V \partial_{a}V
 -  \frac{(d - 1)^2 (d + 2)^{}}{2 d (d - 2)^2 (d + 4) (d + 6)} \partial_{a}V_{b} \partial_{b}V\nonumber\\
& + \frac{(d - 1)^2 (d + 2)^{}}{2 d (d - 2)^2 (d + 4) (d + 6)} \partial_{b}V_{a} \partial_{b}V\Bigr) r^2 \hat{x}_{ija},\\
S\text{VIINC}_{4\text{PN}}={}&\Bigl(- \frac{(d - 1)^3 (d + 2)^{}}{4 (d - 2)^4 (d + 4) (d + 6)} V \partial_t V \partial_{a}V
 -  \frac{(d - 1)^2 (d + 2)^{}}{2 d (d - 2)^2 (d + 4) (d + 6)} \partial_t V \partial_t V_{a}\nonumber\\
& + \frac{(d - 1)^2 (d + 2)^{}}{2 d (d - 2)^2 (d + 4) (d + 6)} V \partial_t^{2} V_{a}
 -  \frac{(d - 1)^2 (d + 2)^{}}{2 d (d - 2)^3 (d + 4) (d + 6)} V_{b} \partial_{a}V \partial_{b}V\nonumber\\
& -  \frac{(d - 1)^3 (d + 2)^{}}{2 d (d - 2)^3 (d + 4) (d + 6)} V \partial_{a}V_{b} \partial_{b}V
 + \frac{(d - 1)^2 (d + 2)^{}}{4 (d - 2)^3 (d + 4) (d + 6)} V_{a} \partial_{b}V \partial_{b}V\nonumber\\
& + \frac{(d - 1)^3 (d + 2)^{}}{2 d (d - 2)^3 (d + 4) (d + 6)} V \partial_{b}V_{a} \partial_{b}V
 + \frac{2 (d - 1)^{} (d + 2)^{}}{d (d - 2) (d + 4) (d + 6)} V_{b} \partial_t \partial_{b}V_{a}\nonumber\\
& -  \frac{(d - 1)^2 (d + 2)^{}}{2 d (d - 2)^2 (d + 4) (d + 6)} \partial_t \hat{W}_{ab} \partial_{b}V
 + \frac{(d - 1)^{} (d + 2)^{}}{d (d - 2) (d + 4) (d + 6)} \partial_{b}\hat{W} \partial_{b}V_{a}\nonumber\\
& + \frac{(d - 1)^{} (d + 2)^{}}{d (d - 2) (d + 4) (d + 6)} \hat{W}_{bi} \partial_{ib}V_{a}
 + \frac{(d - 1)^{} (d + 2)^{}}{d (d - 2) (d + 4) (d + 6)} \partial_{a}\hat{W}_{bi} \partial_{i}V_{b}\nonumber\\
& -  \frac{(d - 1)^{} (d + 2)^{}}{d (d - 2) (d + 4) (d + 6)} \partial_{b}\hat{W}_{ai} \partial_{i}V_{b}
 -  \frac{(d - 1)^{} (d + 2)^{}}{d (d - 2) (d + 4) (d + 6)} \partial_{i}\hat{W}_{ab} \partial_{i}V_{b}\nonumber\\
& -  \frac{(d - 1)^2 (d + 2)^{}}{d (d - 2)^2 (d + 4) (d + 6)} \partial_{a}\hat{R}_{b} \partial_{b}V
 + \frac{(d - 1)^2 (d + 2)^{}}{d (d - 2)^2 (d + 4) (d + 6)} \partial_{b}V \partial_{b}\hat{R}_{a}\nonumber\\
& + \frac{(d - 3)^{} (d - 1)^2 (d + 2)^{}}{4 (d - 2)^4 (d + 4) (d + 6)} \partial_t V \partial_{a}K
 + \frac{(d - 3)^{} (d - 1)^2 (d + 2)^{}}{4 (d - 2)^4 (d + 4) (d + 6)} \partial_t K \partial_{a}V\nonumber\\
& + \frac{(d - 3)^{} (d - 1)^2 (d + 2)^{}}{d (d - 2)^3 (d + 4) (d + 6)} \partial_{a}V_{b} \partial_{b}K
 -  \frac{(d - 3)^{} (d - 1)^2 (d + 2)^{}}{d (d - 2)^3 (d + 4) (d + 6)} \partial_{b}V_{a} \partial_{b}K\Bigr) r^2 \hat{x}_{ija}, \nonumber\\
S\text{VIIINC}_{4\text{PN}}={}&\Bigl(- \frac{(d - 1)^2 (d + 2)^{}}{32 (d - 2)^3 (d + 4) (d + 6) (d + 8)} \partial_t V \partial_{a}V
 -  \frac{(d - 1)^2 (d + 2)^{}}{8 d (d - 2)^2 (d + 4) (d + 6) (d + 8)} \partial_{a}V_{b} \partial_{b}V\nonumber\\
& + \frac{(d - 1)^2 (d + 2)^{}}{8 d (d - 2)^2 (d + 4) (d + 6) (d + 8)} \partial_{b}V_{a} \partial_{b}V\Bigr) r^4 \hat{x}_{ija}, \nonumber\\
V\text{IVC}_{4\text{PN}}={}&- \frac{(d - 1)^{} (d + 2)^{}}{24 d (d - 2) (d + 4) (d + 6) (d + 8) (d + 10)} r^6 \hat{x}_{ija} \sigma_{a}, \nonumber\\
T\text{INC}_{2\text{PN}}={}&\frac{(d - 1)^2 (d + 2)^{}}{8 d (d - 2)^2 (d + 1) (d + 6)} \hat{x}_{ijab} \partial_{a}V \partial_{b}V, \nonumber\\
T\text{INC}_{3\text{PN}}={}&\Bigl(\frac{(d - 1)^2 (d + 2)^{}}{2 d (d - 2)^2 (d + 1) (d + 6)} (\partial_t V_{a} \partial_{b}V
 + \partial_t V_{b} \partial_{a}V)
 -  \frac{(d - 1)^{} (d + 2)^{}}{d (d - 2) (d + 1) (d + 6)} \partial_{i}V_{b} \partial_{i}V_{a}\nonumber\\
& -  \frac{(d - 1)^{} (d + 2)^{}}{d (d - 2) (d + 1) (d + 6)} \partial_{a}V_{i} \partial_{b}V_{i}
 + \frac{(d - 1)^{} (d + 2)^{}}{d (d - 2) (d + 1) (d + 6)} (\partial_{a}V_{i} \partial_{i}V_{b}
 + \partial_{b}V_{i} \partial_{i}V_{a})\nonumber\\
& -  \frac{(d - 3)^{} (d - 1)^2 (d + 2)^{}}{4 d (d - 2)^3 (d + 1) (d + 6)} (\partial_{a}K \partial_{b}V
 + \partial_{a}V \partial_{b}K)\Bigr) \hat{x}_{ijab},\\
T\text{INC}_{4\text{PN}}={}&\frac{(d - 1)^3 (d + 2)^{}}{2 d (d - 2)^3 (d + 1) (d + 6)} (\partial_t V_{a} \partial_{b}V
 + \partial_t V_{b} \partial_{a}V) V \hat{x}_{ijab}
 + \frac{(-1 + d)^2 (2 + d)^{}}{2 (-2 + d)^3 d (1 + d) (6 + d)} V_{a} \hat{x}_{ijab} \partial_t V \partial_{b}V\nonumber\\
& + \frac{(-1 + d)^2 (2 + d)^{}}{2 (-2 + d)^3 d (1 + d) (6 + d)} V_{b} \hat{x}_{ijab} \partial_t V \partial_{a}V
 + \frac{2 (d - 1)^{} (d + 2)^{}}{d (d - 2) (d + 1) (d + 6)} \hat{x}_{ijab} \partial_t V_{a} \partial_t V_{b}\nonumber\\
& + \frac{(d - 1)^2 (d + 2)^{}}{d (d - 2)^2 (d + 1) (d + 6)} (\partial_{a}V \partial_{i}V_{b}
 + \partial_{b}V \partial_{i}V_{a}) V_{i} \hat{x}_{ijab}
 -  \frac{(-1 + d)^2 (2 + d)^{}}{(-2 + d)^2 d (1 + d) (6 + d)} V_{a} \hat{x}_{ijab} \partial_{i}V_{b} \partial_{i}V\nonumber\\
& -  \frac{(-1 + d)^2 (2 + d)^{}}{(-2 + d)^2 d (1 + d) (6 + d)} V_{b} \hat{x}_{ijab} \partial_{i}V_{a} \partial_{i}V
 + \frac{(-1 + d)^2 (2 + d)^{}}{(-2 + d)^2 d (1 + d) (6 + d)} V_{a} \hat{x}_{ijab} \partial_{b}V_{i} \partial_{i}V\nonumber\\
& + \frac{(-1 + d)^2 (2 + d)^{}}{(-2 + d)^2 d (1 + d) (6 + d)} V_{b} \hat{x}_{ijab} \partial_{a}V_{i} \partial_{i}V
 + \frac{(d - 1)^2 (d + 2)^{}}{4 d (d - 2)^2 (d + 1) (d + 6)} \hat{W} \hat{x}_{ijab} \partial_{a}V \partial_{b}V\nonumber\\
& -  \frac{(-1 + d)^2 (2 + d)^{}}{2 (-2 + d)^2 d (1 + d) (6 + d)} \hat{W}_{ai} \hat{x}_{ijab} \partial_{b}V \partial_{i}V
 -  \frac{(-1 + d)^2 (2 + d)^{}}{2 (-2 + d)^2 d (1 + d) (6 + d)} \hat{W}_{bi} \hat{x}_{ijab} \partial_{a}V \partial_{i}V\nonumber\\
& -  \frac{(d - 1)^{} (d + 2)^{}}{d (d - 2) (d + 1) (d + 6)} (\partial_t \hat{W}_{ai} \partial_{b}V_{i}
 + \partial_t \hat{W}_{bi} \partial_{a}V_{i}) \hat{x}_{ijab}
 + \frac{(d - 1)^{} (d + 2)^{}}{d (d - 2) (d + 1) (d + 6)} (\partial_t \hat{W}_{ai} \partial_{i}V_{b}\nonumber\\
& + \partial_t \hat{W}_{bi} \partial_{i}V_{a}) \hat{x}_{ijab}
 + \frac{(d - 1)^{} (d + 2)^{}}{2 d (d - 2) (d + 1) (d + 6)} \hat{x}_{ijab} \partial_{a}\hat{W}_{ij} \partial_{b}\hat{W}_{ij}\nonumber\\
& -  \frac{(d - 1)^{} (d + 2)^{}}{d (d - 2) (d + 1) (d + 6)} (\partial_{a}\hat{W}_{ij} \partial_{j}\hat{W}_{b}{}_{i}
 + \partial_{b}\hat{W}_{ij} \partial_{j}\hat{W}_{a}{}_{i}) \hat{x}_{ijab}\nonumber\\
& + \frac{(d - 1)^{} (d + 2)^{}}{d (d - 2) (d + 1) (d + 6)} \hat{x}_{ijab} \partial_{i}\hat{W}_{bj} \partial_{j}\hat{W}_{a}{}_{i}
 + \frac{(d - 1)^{} (d + 2)^{}}{d (d - 2) (d + 1) (d + 6)} \hat{x}_{ijab} \partial_{j}\hat{W}_{bi} \partial_{j}\hat{W}_{a}{}_{i}\nonumber\\
& + \frac{(d - 1)^2 (d + 2)^{}}{d (d - 2)^2 (d + 1) (d + 6)} (\partial_t \hat{R}_{a} \partial_{b}V
 + \partial_t \hat{R}_{b} \partial_{a}V) \hat{x}_{ijab}
 + \frac{2 (d - 1)^{} (d + 2)^{}}{d (d - 2) (d + 1) (d + 6)} (\partial_{a}\hat{R}_{i} \partial_{i}V_{b}\nonumber\\
& + \partial_{b}\hat{R}_{i} \partial_{i}V_{a}) \hat{x}_{ijab}
 -  \frac{2 (d - 1)^{} (d + 2)^{}}{d (d - 2) (d + 1) (d + 6)} (\partial_{i}V_{a} \partial_{i}\hat{R}_{b}
 + \partial_{i}V_{b} \partial_{i}\hat{R}_{a}) \hat{x}_{ijab}\nonumber\\
& -  \frac{2 (d - 1)^{} (d + 2)^{}}{d (d - 2) (d + 1) (d + 6)} (\partial_{a}\hat{R}_{i} \partial_{b}V_{i}
 + \partial_{a}V_{i} \partial_{b}\hat{R}_{i}) \hat{x}_{ijab}
 + \frac{2 (d - 1)^{} (d + 2)^{}}{d (d - 2) (d + 1) (d + 6)} (\partial_{a}V_{i} \partial_{i}\hat{R}_{b}\nonumber\\
& + \partial_{b}V_{i} \partial_{i}\hat{R}_{a}) \hat{x}_{ijab}
 + \frac{(d - 1)^2 (d + 2)^{}}{2 d (d - 2)^2 (d + 1) (d + 6)} (\partial_{a}V \partial_{b}\hat{X}
 + \partial_{a}\hat{X} \partial_{b}V) \hat{x}_{ijab}\nonumber\\
& + \frac{(d - 3)^2 (d - 1)^2 (d + 2)^{}}{2 d (d - 2)^4 (d + 1) (d + 6)} \hat{x}_{ijab} \partial_{a}K \partial_{b}K
 -  \frac{(d - 3)^{} (d - 1)^2 (d + 2)^{}}{d (d - 2)^3 (d + 1) (d + 6)} (\partial_t V_{a} \partial_{b}K
 + \partial_t V_{b} \partial_{a}K) \hat{x}_{ijab}, \nonumber\\
T\text{IINC}_{3\text{PN}}={}&\frac{(d - 1)^2 (d + 2)^{}}{16 d (d - 2)^2 (d + 1) (d + 6) (d + 8)} r^2 \hat{x}_{ijab} \partial_{a}V \partial_{b}V,\\
T\text{IINC}_{4\text{PN}}={}&\Bigl(- \frac{(d - 3)^{} (d - 1)^2 (d + 2)^{}}{8 d (d - 2)^3 (d + 1) (d + 6) (d + 8)} (\partial_{a}K \partial_{b}V
 + \partial_{a}V \partial_{b}K)\nonumber\\
& + \frac{(d - 1)^2 (d + 2)^{}}{4 d (d - 2)^2 (d + 1) (d + 6) (d + 8)} (\partial_t V_{a} \partial_{b}V
 + \partial_t V_{b} \partial_{a}V)\nonumber\\
& -  \frac{(d - 1)^{} (d + 2)^{}}{2 d (d - 2) (d + 1) (d + 6) (d + 8)} \partial_{i}V_{b} \partial_{i}V_{a}
 -  \frac{(d - 1)^{} (d + 2)^{}}{2 d (d - 2) (d + 1) (d + 6) (d + 8)} \partial_{a}V_{i} \partial_{b}V_{i}\nonumber\\
& + \frac{(d - 1)^{} (d + 2)^{}}{2 d (d - 2) (d + 1) (d + 6) (d + 8)} (\partial_{a}V_{i} \partial_{i}V_{b}
 + \partial_{b}V_{i} \partial_{i}V_{a})\Bigr) r^2 \hat{x}_{ijab}, \nonumber\\
T\text{IIINC}_{4\text{PN}}={}&\frac{(d - 1)^2 (d + 2)^{}}{64 d (d - 2)^2 (d + 1) (d + 6) (d + 8) (d + 10)} r^4 \hat{x}_{ijab} \partial_{a}V \partial_{b}V.
\end{align*}}
\noindent
Finally we present the surface terms. The terms of the Laplacian type should be multiplied by the operator $\hat{x}_{ij}\Delta$ for the scalar (S) terms, and by $\hat{x}_{ija}\Delta$ for the vector (V) terms. The terms of the divergence type should be multiplied by a spatial derivative $\partial_a$.
{\footnotesize
\begin{align*}
S\text{ISL}_{1\text{PN}}={}&- \frac{(d - 1)^2}{8 (d - 2)^2} V^2,\\
S\text{ISL}_{2\text{PN}}={}&- \frac{(d - 1)^3}{12 (d - 2)^3} V^3
 + \frac{(d - 3)^{}}{2 (d - 2)} V_{a} V_{a}
 -  \frac{(d - 1)^{}}{4 (d - 2)} V \hat{W}
 + \frac{(d - 3)^{} (d - 1)^2}{2 (d - 2)^3} K V,\\
S\text{ISL}_{3\text{PN}}={}&- \frac{(d - 1)^4}{24 (d - 2)^4} V^4
 + \frac{(d - 3)^{} (d - 1)^{}}{(d - 2)^2} V V_{a} V_{a}
 -  \frac{(d - 1)^2}{4 (d - 2)^2} V^2 \hat{W}
 -  \frac{1}{4 (d - 2)} \hat{W}^2
 + \frac{1}{2 (d - 2)} \hat{W}_{ab} \hat{W}_{ab}\nonumber\\
& + \frac{2 (d - 3)^{}}{d - 2} \hat{R}_{a} V_{a}
 -  \frac{(d - 1)^2}{(d - 2)^2} V \hat{X}
 -  \frac{(d - 1)^{}}{d - 2} V \hat{Z}
 + \frac{(d - 3)^{} (d - 1)^3}{2 (d - 2)^4} K V^2
 -  \frac{(d - 3)^2 (d - 1)^2}{2 (d - 2)^4} K^2\nonumber\\
& + \frac{(d - 3)^{} (d - 1)^{}}{2 (d - 2)^2} K \hat{W},\\
S\text{ISL}_{4\text{PN}}={}&- \frac{(d - 1)^5}{60 (d - 2)^5} V^5
 + \frac{3 (d - 3)^{} (d - 1)^2}{2 (d - 2)^3} V^2 V_{a} V_{a}
 -  \frac{(d - 1)^{}}{4 (d - 2)} V \hat{W}^2
 -  \frac{(d - 1)^3}{6 (d - 2)^3} V^3 \hat{W}
 + \frac{2 (d - 3)^{}}{d - 2} \hat{R}_{a} \hat{R}_{a}\nonumber\\
& -  \frac{(d - 1)^{}}{d - 2} \hat{W} \hat{X}
 -  \frac{2}{d - 2} \hat{W} \hat{Z}
 + \frac{4}{d - 2} \hat{W}_{ab} \hat{Z}_{ab}
 + \frac{4 (d - 3)^{}}{d - 2} V_{a} \hat{Y}_{a}
 -  \frac{4 (d - 1)^2}{(d - 2)^2} \hat{T} V
 + \frac{4 (d - 1)^{}}{(d - 2)^2} \hat{M} V\nonumber\\
& + \frac{(d - 3)^{} (d - 1)^3 (3 d - 2)^{}}{12 (d - 2)^5} K V^3
 -  \frac{(d - 5)^{} (d - 3)^{} (d - 1)^{}}{(d - 2)^3} K V_{a} V_{a}
 + \frac{2 (d - 3)^{} (d - 1)^2}{(d - 2)^3} K \hat{X}\nonumber\\
& + \frac{2 (d - 3)^{} (d - 1)^{}}{(d - 2)^2} K \hat{Z}, \nonumber\\
V\text{ISL}_{2\text{PN}}={}&\frac{(d - 1)^2 (d + 2)^{}}{2 d (d - 2)^2 (d + 4)} V V_{a},\\
V\text{ISL}_{3\text{PN}}={}&\frac{(d - 1)^3 (d + 2)^{}}{4 d (d - 2)^3 (d + 4)} V^2 V_{a}
 + \frac{(d - 1)^{} (d + 2)^{}}{d (d - 2) (d + 4)} V_{a} \hat{W}
 -  \frac{(d - 1)^{} (d + 2)^{}}{d (d - 2) (d + 4)} V_{b} \hat{W}_{ab}
 + \frac{(d - 1)^2 (d + 2)^{}}{d (d - 2)^2 (d + 4)} \hat{R}_{a} V\nonumber\\
& -  \frac{(d - 3)^{} (d - 1)^2 (d + 2)^{}}{d (d - 2)^3 (d + 4)} K V_{a},\\
V\text{ISL}_{4\text{PN}}={}&\frac{(d - 1)^{} (d + 2)^{}}{(d - 2)^2 (d + 4)} V_{a} V_{b} V_{b}
 + \frac{(d - 1)^2 (d + 2)^{}}{d (d - 2)^2 (d + 4)} V V_{a} \hat{W}
 -  \frac{(d - 1)^2 (d + 2)^{}}{d (d - 2)^2 (d + 4)} V V_{b} \hat{W}_{ab}\nonumber\\
& + \frac{(d - 1)^3 (d + 2)^{}}{d (d - 2)^3 (d + 4)} \hat{R}_{a} V^2
 + \frac{2 (d - 1)^{} (d + 2)^{}}{d (d - 2) (d + 4)} \hat{R}_{a} \hat{W}
 -  \frac{2 (d - 1)^{} (d + 2)^{}}{d (d - 2) (d + 4)} \hat{R}_{b} \hat{W}_{ab}
 + \frac{2 (d - 1)^2 (d + 2)^{}}{d (d - 2)^2 (d + 4)} V_{a} \hat{X}\nonumber\\
& + \frac{4 (d - 1)^{} (d + 2)^{}}{d (d - 2) (d + 4)} V_{a} \hat{Z}
 -  \frac{4 (d - 1)^{} (d + 2)^{}}{d (d - 2) (d + 4)} V_{b} \hat{Z}_{ab}
 + \frac{2 (d - 1)^2 (d + 2)^{}}{d (d - 2)^2 (d + 4)} V \hat{Y}_{a}\nonumber\\
& -  \frac{(d - 3)^{} (d - 1)^3 (d + 2)^{}}{d (d - 2)^4 (d + 4)} K V V_{a}
 -  \frac{2 (d - 3)^{} (d - 1)^2 (d + 2)^{}}{d (d - 2)^3 (d + 4)} K \hat{R}_{a}, \nonumber\\
S\text{ISD}_{2\text{PN}}={}&\frac{(d - 1)^{}}{2 (d - 2)} (\Psi_{ai}^{\partial_{bk} V} \partial_{j}\hat{W}_{bk}
 -  \hat{W}_{bk} \partial_{a}\Psi_{ij}^{\partial_{bk} V}),\\
S\text{ISD}_{3\text{PN}}={}&- \frac{4 (d - 1)^{}}{d - 2} \Psi_{ai}^{\partial_{b}V_{k}} \partial_{kj}\hat{R}_{b}
 + \frac{4 (d - 1)^{}}{d - 2} \partial_{a}\Psi_{ij}^{\partial_{b}V_{k}} \partial_{k}\hat{R}_{b}
 + \frac{2 (d - 1)^{}}{d - 2} \Psi_{ai}^{\partial_{tb} V} \partial_{j}\hat{R}_{b}
 -  \frac{2 (d - 1)^{}}{d - 2} \hat{R}_{b} \partial_{a}\Psi_{ij}^{\partial_{tb} V}\nonumber\\
& + \frac{2 (d - 1)^{}}{d - 2} \Psi_{ai}^{\partial_{bk} V} \partial_{j}\hat{Z}_{bk}
 -  \frac{2 (d - 1)^{}}{d - 2} \hat{Z}_{bk} \partial_{a}\Psi_{ij}^{\partial_{bk} V},\\
S\text{ISD}_{4\text{PN}}={}&- \frac{8 (d - 1)^{}}{d - 2} \Psi_{ai}^{\partial_{b}V_{k}} \partial_{kj}\hat{Y}_{b}
 + \frac{8 (d - 1)^{}}{d - 2} \partial_{a}\Psi_{ij}^{\partial_{b}V_{k}} \partial_{k}\hat{Y}_{b}
 + \frac{4 (d - 1)^{}}{d - 2} \Psi_{ai}^{\partial_{tb} V} \partial_{j}\hat{Y}_{b}
 -  \frac{4 (d - 1)^{}}{d - 2} \hat{Y}_{b} \partial_{a}\Psi_{ij}^{\partial_{tb} V}\nonumber\\
& + \frac{4 (d - 1)^{}}{d - 2} \Psi_{ai}^{\partial_{bk} V} \partial_{j}\hat{M}_{bk}
 -  \frac{4 (d - 1)^{}}{d - 2} \hat{M}_{bk} \partial_{a}\Psi_{ij}^{\partial_{bk} V}, \nonumber\\
V\text{ISD}_{3\text{PN}}={}&\frac{2 (d - 1)^2 (d + 2)^{}}{d (d - 2)^2 (d + 4)} (\Psi_{aib}^{\partial_{k} V} \partial_{kj}\hat{R}_{b}
 -  \partial_{a}\Psi_{ijb}^{\partial_{k} V} \partial_{k}\hat{R}_{b}),\\
V\text{ISD}_{4\text{PN}}={}&\frac{4 (d - 1)^2 (d + 2)^{}}{d (d - 2)^2 (d + 4)} (\Psi_{aib}^{\partial_{k} V} \partial_{kj}\hat{Y}_{b}
 -  \partial_{a}\Psi_{ijb}^{\partial_{k} V} \partial_{k}\hat{Y}_{b}).
\end{align*}}

\bibliography{ListeRef_Quad4PN}

\end{document}